\documentclass[a4paper,fleqn,usenatbib]{mnras}
\usepackage{newtxtext,newtxmath}

\usepackage[T1]{fontenc}
\usepackage{ae,aecompl}

\usepackage{graphicx}	
\usepackage{amsmath}	
\usepackage{amssymb}	

\usepackage{natbib}
\usepackage{float}
\usepackage{hyperref}
\usepackage{multicol}

\usepackage{etoolbox}
\makeatletter
\patchcmd\@combinedblfloats{\box\@outputbox}{\unvbox\@outputbox}{}{%
  \errmessage{\noexpand\@combinedblfloats could not be patched}%
}%
\makeatother

\usepackage{xspace}

\newcommand{\Msun}{\rm M_{\odot}}

\newcommand{\art}{ART${^2}$\xspace}
\newcommand{\lya}{\ifmmode {\rm Ly}\alpha \else Ly$\alpha$\fi\xspace}

\newcommand{\cii}{[CII]\xspace}

\newcommand{\oi}{[OI]\xspace}
\newcommand{\oii}{[OII]\xspace}
\newcommand{\oiii}{[OIII]\xspace}

\newcommand{\nii}{[NII]\xspace}

\newcommand{\Gizmo}{{\small\sc Gizmo}}

\newcommand{\Gadget}{{\small\sc Gadget}}

\graphicspath{{./}}

\title[\art: A 3D Parallel Multi-wavelength Radiative Transfer Code for Continuum and Lines]{\art: A 3D Parallel Multi-wavelength Radiative Transfer Code for Continuum and Atomic and Molecular Lines}

\author[Y. Li et al.]{
Yuexing Li$^{1, 2, 3}$\thanks{E-mail: yul20@psu.edu}, 
Ming F. Gu$^{4}$,
Hidenobu Yajima$^{5}$, 
Qirong Zhu$^{6}$ 
and Moupiya Maji$^{7}$
\vspace{0.2cm}\\
$^{1}$ Department of Astronomy \& Astrophysics, The Pennsylvania State University, University Park, PA 16802, USA\\
$^{2}$ Institute for Cosmology and Gravity, The Pennsylvania State University, University Park, PA 16802, USA\\
$^{3}$ Department of Physics, Sapienza University of Rome, Rome, 00185, Italy \\
$^{4}$ Space Science Laboratory, University of California, Berkeley, CA 94720, USA \\
$^{5}$ Center for Computational Sciences University of Tsukuba, Tsukuba, Ibaraki 305-8577, Japan\\
$^{6}$ Department of Physics, Carnegie Mellon University, 5000 Forbes Avenue, Pittsburgh, PA 15213, USA \\
$^{7}$ Department of Astronomy, University of Geneva, Switzerland
}

\date{Accepted XXX. Received YYY; in original form ZZZ}
\pubyear{2019}

\begin{document}
\label{firstpage}
\pagerange{\pageref{firstpage}--\pageref{lastpage}} 
\maketitle

\begin{abstract}
\art is a 3D multi-wavelength Monte Carlo radiative transfer (RT) code that couples continuum and emission lines to track the propagation of photons and their interactions with the ISM. The original \art has been extensively applied to hydrodynamics simulations to study panchromatic properties of galaxies and ISM. Here, we describe new implementations of non-local thermodynamic equilibrium RT of molecular and atomic fine structure emission lines, and the parallelization of the code using a number of novel methods. The new \art can efficiently and self-consistently produce a full spectrum that includes both continuum and lines such as \cii, \nii, \oiii, \lya, and CO. These essential features, together with the multi-phase ISM model and the adaptive grid, make \art a multi-purpose code to study multi-wavelength properties of a wide range of astrophysical systems from planetary disks to large-scale structures. 

To demonstrate the capability of the new \art, we applied it to two hydrodynamics simulations: the zoom-in Milky Way Simulation to obtain panchromatic properties of individual galaxies, and the large-scale IllustrisTNG100 Simulation to obtain global properties such as the line intensity mappings. These products are vital for a broad array of studies. By enabling direct comparison between numerical simulations and multi-band observations, \art provides a crucial theoretical framework for the understanding of existing and future surveys, and the synergy between multi-band galaxy surveys and line intensity mappings. Therefore, \art is a powerful and versatile tool to bridge the gap between theories and observations of cosmic structures.
\end{abstract}

\begin{keywords}
radiative transfer: multi-wavelength, continuum,  line, non-local thermodynamic equilibrium -- dust: extinction, re-emission -- interstellar medium -- intergalactic medium -- galaxy: formation -- galaxy: evolution -- astrophysics
\end{keywords}

\section{Introduction}
\label{sec:intro}

The past decade has witnessed a ``golden age" of panchromatic astronomy thanks to an impressive array of multi-wavelength instruments, from NASA's Great Observatories (Hubble, Chandra, and Spitzer), to small-scale missions such as GALEX, SWIFT, FERMI, and NuSTAR, to European missions such as Herschel, Planck, XMM, and to ground-based telescopes as SDSS, Subaru and ALMA (e.g., see a recent review by \citealt{Megeath2019}). These observatories span the full electromagnetic spectrum from radio, to infrared /optical /UV, and to X-ray and gamma-ray. These facilities have led to major discoveries in observational cosmology from multi-band surveys using both continuum and emission lines. These surveys have measured the cosmic star formation history  \citep[e.g.,][]{Madau2014, Goto2019, Wilkins2019}, the black hole growth history \citep[e.g.,][]{Hickox2018, Aird2019}, the evolution of galaxy luminosity functions \citep[e.g.,][]{Bouwens2014, Koprowski2017, Park2019}, the evolution of interstellar medium (ISM)  \citep[e.g.,][]{Scoville2017, Decarli2018, Riechers2019}, and a full-spectrum extragalactic background light \citep[e.g.,][]{Hill2018}. 

In recent years, a new technique called Line Intensity Mapping (LIM) has emerged as a promising method to study the evolution of the Universe \citep[e.g.,][]{Lidz2011, Visbal2011, Pullen2014, Kovetz2017, Fonseca2017, Moradinezhad2019, Karkare2018, Chung2019, Sun2019, Bernal2019}.  In contrast to the aforementioned single-object telescopes which require high resolutions, the LIM technique uses low-resolution instruments to measure the integrated emission of atomic and molecular lines  from  galaxies  and  the  intergalactic  medium  (IGM) such as \lya, \cii, CO and 21-cm line. This technique can probe a wide array of topics such as the epoch of reionization, cosmic star formation history, and evolution of ISM and IGM. 

Looking ahead, the next decade will be a new era in astrophysics with a host of planned or proposed ambitious facilities from radio to x-ray, such as SKA, JWST, WFIRST, LSST, Athena and Lynx, and LIM instruments such as CONCERTO,  SPHEREx and Origins (e.g., see recent reviews by \citealt{Cooray2019} and \citealt{Kovetz2019}). These developments suggest that full-spectrum surveys that include both continuum and lines will be the name of the game, and that a synergy between galaxy surveys and LIMs is on the horizon to provide a complete picture of our cosmos. 

However, in order to understand the existing observations and the underlying physical processes, and prepare for the future surveys, we need a comprehensive theoretical framework that can simultaneously provide both the physical and the panchromatic properties of galaxies, active galactic nuclei, ISM and IGM across cosmic time. A desired approach is to combine state-of-the-art cosmological simulations with state-of-the-art multi-wavelength radiative transfer calculations. Radiation at different wavelengths are often intertwined due to the interaction between photons and baryonic matter in different forms, such as ionized gas, molecular clouds, and dusts. Therefore, detailed comparison of observations and theoretical models must include radiative transfer calculations that couples photons of different wavelengths and the physical states of matter \citep[e.g.,][]{Steinacker2013,  Kewley2019}.

Recently, a number of impressive cosmological simulations have been performed, including large-scale ones Illustris \citep[precursor of IllustrisTNG,][]{Vogelsberger2014, Genel2014}, EAGLE \citep{Schaye2015, Crain2015}, Horizon-AGN \citep{Dubois2014}, Romulus \citep{Tremmel2017}, Simba \citep{Dave2019}, Magneticum \citep{Dolag2016}, and IllustrisTNG~\citep[][]{Pillepich2018b, Springel2018, Marinacci2018, Naiman2018, Nelson2018a, Nelson2019a, Nelson2019b, Pillepich2019},  as well as high-redshift simulations such as BlueTides \citep{Feng2016} and  Sphinx \citep{Rosdahl2018}, and zoom-in ones such as MilkyWay \citep{Zhu2016}, FIRE-2 \citep{Hopkins2018}, and Auriga \citep{Grand2017}. 

On the other hand, a number of radiative transfer codes have been developed in recent years, for examples, RADMC-3D~\cite{Dullemond2012} and SKIRT~\citep{Camps2015} for dust continuum; CLOUDY~\citep{Ferland2017}, DESPOTIC \citep{Krumholz2014}, LIME~\citep{Brinch2010},  MAIHEM~\citep{Gray2015} and MAPPINGS~\citep{Allen2008} for emission lines from ionized ISM; MOLLIE~\citep{Keto2010} and Turtlebeach~\citep{Narayanan2009} for molecular lines, CRASH for ionization~\citep{Hariharan2017}, as well as   IGMtransfer~\citep{Laursen2011},  McLya~\citep{Schaerer2011} and RASCAS~(Michel-Dansac et al., in prep) for \lya line and other calculations ~\citep[e.g.,][]{Zheng2010, Zheng2011, Dijkstra2016}. 

Over the last decade, we have developed a 3D Monte Carlo radiative transfer code, All-Wavelength Radiative Transfer with Adaptive Refinement Tree ({\art}). {\art} originally included  continuum transfer of dust scattering and absorption, which produced a continuum spanning 7 orders of magnitude in wavelength from X-ray to sub-millimeter. The code  was first used to model broad-band properties of high-redshift quasars \citep{Li2008}. The ionization of neutral Hydrogen, and resonant scattering of \lya photons were later added as additional modules \citep{Yajima2012}. This version of \art has been extensively applied to multi-scale cosmological hydrodynamic simulations to study multi-band properties of galaxies and quasars, such as the 21-cm signals of the first galaxies and quasars, high-redshift \lya emitters, and progenitors of the Milky Way \citep{Yajima2012b,  Yajima2013, Yajima2014, Yajima2015,  Yajima2018, Arata2019, Arata2020}. 

In the present paper, we further extend the scope of \art to make it more versatile, by improving the algorithms of existing modules, increasing its efficiency by parallelization, and adding new modules for the non-local thermal equilibrium (non-LTE) molecular and atomic fine structure line transfer process, in order to provide a comprehensive framework for both continuum and useful lines such as \lya, \cii, \nii, \oiii,  and  CO as commonly employed in multi-wavelength surveys of galaxies and ISM.
 
The paper is organized in three parts: the code, the applications, and the conclusions. In Section~\ref{sec:art}, we describe the improvements to the original \art, and new implementations of the non-LTE molecular and atomic line transfer processes, their verifications and the parallelization in the following order: dust continuum in Section~\ref{sec:dustrt}; the Hydrogen ionization and \lya resonant line transfer in Section~\ref{sec:ionlya}; the non-LTE molecular line transfer in Section~\ref{sec:co}; the non-LTE atomic fine structure line transfer in Section~\ref{sec:afs};  and method for RT through a sub-grid multi-phase interstellar medium model in Section~\ref{sec:multiphase}. In Sections~\ref{sec:apps}, we demonstrate the power  of the new \art by applying it to two hydrodynamics simulations: the zoom-in Milky Way Simulation to obtain multi-band properties of individual galaxies in Sections~\ref{sec:mw}, and the large-scale IllustrisTNG100 Simulation to obtain global properties of the entire system such as the line intensity mappings in Sections~\ref{sec:tng}. Finally, in Section~\ref{sec:summary} we discuss and summarize the essential features of \art and its potential contributions to astrophysics. 

\section{\art: The State-of-the-art Radiative Transfer Code for Continuum and Lines}
\label{sec:art}

\art is a 3D, fully-parallel, multi-wavelength Monte Carlo RT code that can self-consistently and efficiently calculate a full spectrum that includes a continuum from X-ray to sub-millimeter, atomic lines such as \lya, [CII], [NII] and [OIII], and molecular lines such as CO and HCN. In addition, \art  adopts a multi-phase ISM model, which ensures an appropriate prescription of the ISM physics in case hydrodynamics simulations have insufficient resolution to resolve the multi-phase ISM, and it employs an adaptive grid scheme, which can handle arbitrary geometry and cover a large dynamical range of gas densities in hydrodynamical simulations. These essential features make \art a flexible, reliable and versatile tool to study the multi-wavelength properties of a wide range of astrophysical systems, from planetary disks, to star forming regions, to galaxies, and to large-scale structures of the Universe.

\art was first developed with continuum only to study the dust properties of high-redshift quasars \citep{Li2008}. Since then, significant developments and improvements have been made to it over the years. We added modules of ionization of neutral hydrogen and $\lya$ emission in \cite{Yajima2012a}, and recently we have implemented the non-LTE radiative transfer of molecular and atomic fine structure lines, and have parallelized all the RT processes. 

The main objective of \art is to seek the solution of the radiative transfer equation \citep{Rybicki1986}:
\begin{equation}\\
  \frac{dI_{\nu}}{ds} = -\alpha_{\nu}I_{\nu} + j_{\nu},
\end{equation}
or equivalently,
\begin{equation}\\
  \frac{dI_{\nu}}{d\tau_{\nu}} = -I_{\nu} + S_{\nu},
\end{equation}
where $I_{\nu}$ is the specific intensity of the radiation, $\alpha_{\nu}$ is the total opacity, $j_{\nu}$ is the emissivity, $\tau_{\nu} = \alpha_{\nu}ds$ is the differential optical depth, and $S_{\nu} = j_{\nu}/\alpha_{\nu}$ is the source function. 

This is a multi-dimensional equation as the radiation depends on the positions $(x, y, z)$, the viewing angles $(\Theta, \Phi)$, the photon frequency $(\nu)$, and in some cases also the time $(t)$. To solve such a complex problem, \art uses the flexible Monte Carlo method to robustly follow the propagation of the photons and their interactions with ISM and IGM with rigorous random samplings, as described in~\citet{Li2008}.

The type of radiative transfer problem is prescribed by specific coupling between the $S_{\nu}$ and the local state of the matter responsible for absorption,
scattering, and emission. In the case of dust continuum transfer, the emissivity function depends on the dust temperature, which in turn is determined by the local radiative field through the radiative equilibrium. For the transfer of ionizing radiation, the opacity function depends on the thermal and ionization state of
various atomic species, which must be determined through the thermal and ionization equilibrium involving the radiation field through photoionization. For the transfer of molecular or atomic lines, the source function depends on the statistical populations of molecular or atomic levels, which are also coupled with the radiation field through photo-excitation. The $\lya$ resonance scattering is the only process in which the radiation is not explicitly coupled with the matter state. However, $\lya$ scattering presents its own unique challenges due to the often extremely large scattering optical depth. 

In the following sections, we describe the numerical techniques implemented in \art for these specific problems. In particular, we highlight the new improvements over the original implementations, and the novel approaches for the non-LTE radiative transfer of molecular and atomic fine structure lines, and the parallelization of each RT process.

\subsection{Dust Continuum Radiative Transfer}
\label{sec:dustrt}

The dust RT implemented in the original \art is based on the
algorithm of \citet{Bjorkman2001}, which exploits the fact that the dust
opacity does not depend on its temperature, and obtains the solution of
radiative equilibrium and emergent spectrum simultaneously without
iterations. The central ingredient of this algorithm is the immediate
re-emission of the photon packet by the dust following an absorption event. For
each absorption and re-emission step, the radiative equilibrium is maintained
by updating the dust temperature so that the re-emitted luminosity is the same
as the absorbed one, while the re-emitted photon frequency is sampled from a
spectrum proportional to: 
\begin{equation}\\
\label{eq:reemitsed}
\Delta j_{\nu} = \kappa_{\nu}\rho\left[B_{\nu}(T_i)-B_{\nu}(T_{i-1})\right],
\end{equation}
where $\kappa_{\nu}$ is the dust opacity, $\rho$ is the gas density of the
cell where the photon is absorbed, $B_{\nu}$ is the Planck function, and
$T_{i}$ is the dust temperature after the absorption of the $i$-th photon
packet while maintaining the radiative equilibrium. 

After all photons have been traced, it is evident that the ensemble of all re-emitted photons are
sampled from the spectrum $\kappa_{\nu}\rho B_{\nu}(T_d)$, where
$T_d$ is the final dust temperature in radiative equilibrium.

While this algorithm is very efficient, it has two drawbacks. First, all
photon packets are completely destroyed and re-emitted in cells where the
absorptions occur. This leads to large statistical sampling errors. Second,
the immediate re-emission step with temperature update leads to dependency of
successively traced photons, making it difficult to parallelize the
algorithm. In the present work, we implement a modified version of this
algorithm that avoids these two problems while retaining its non-iterative
nature.

Instead of absorbing and re-emitting the photon packets completely at a single
site, we first trace all photon packets without re-emission and dust
temperature update. The photon packets are not absorbed statistically in any
single cell. Instead we record the average absorbed luminosity along the
photon path for the entire batch of photons. Dust scattering, however, is
still followed statistically at every interaction site at the time of the scattering, because it would be rather
difficult to re-emit the scattered luminosity at a later time or after all the interactions, if the scattering phase function
is not isotropic, as is often the case with dust. 

After tracing all photon packets, we then update the dust temperature according the total absorbed
luminosity at each cell, and re-emit them with a spectrum corresponding to
$\kappa_{\nu}\rho B_{\nu}(T_0)$. The re-emitted thermal photons are then traced
again, whose absorption results in further update of the dust temperature to
$T_1$. The absorption and re-emission processes are then repeated until the total
absorbed luminosity becomes negligible. Denoting the dust temperature after
the $i$-th step as $T_i$, the frequencies of the re-emitted photons in each
step are sampled from the spectrum of Equation~\ref{eq:reemitsed}. 

After convergence, we obtain both the radiative equilibrium solution and the
emergent spectrum as in the original algorithm. However, in this modified
method, the photon packets traced in each step are
uncorrelated. It is therefore straightforward to distribute the ray tracing
workload over a large number of processors. Data communication is only
required for updating the dust temperature, where the total absorbed
luminosity is collected from individual processors. 

\begin{figure}
\includegraphics[width=3.2in]{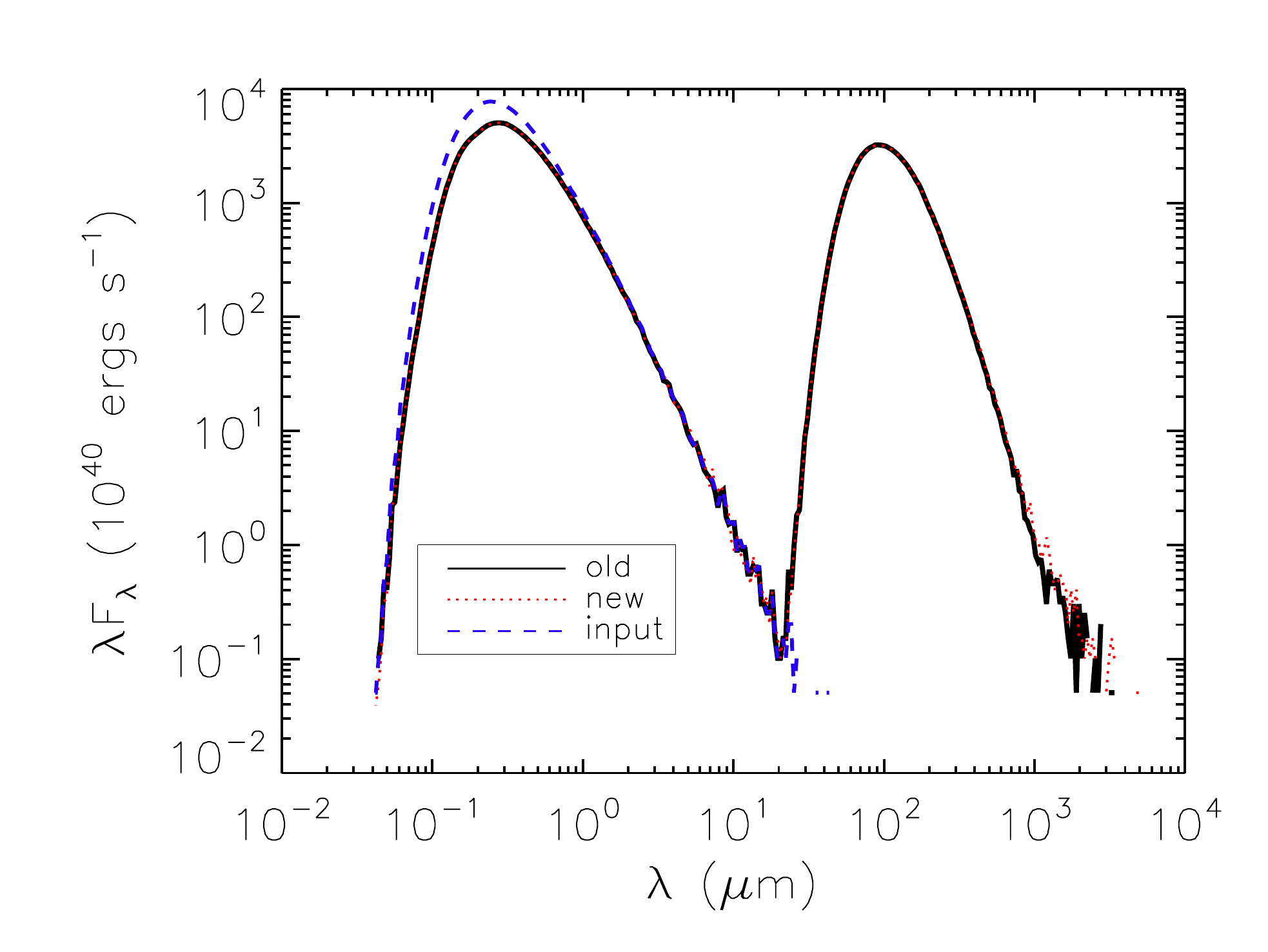}
\caption{Test of the new dust continuum algorithm with a disk galaxy model. The emergent continuum from the disk galaxy is calculated using the new and the old~\citep{Li2008} algorithms. The spectra produced by both algorithms are nearly identical. }
\label{fig:jaffe}
\end{figure}

To test this new algorithm and the parallelization, we run the same test problem of dust radiative transfer through a model disk as in \citet{Li2008}. A comparison of the emergent spectrum from the new algorithm  with that from the original one is shown in Figure~\ref{fig:jaffe}.  The reprocessed spectra from the two implementations are virtually indistinguishable, except at frequencies where statistical sampling error becomes dominant, thus verifying the new algorithm for dust continuum RT and the parallelization.

\subsection{Ionization and \lya Scattering}
\label{sec:ionlya}

The ionization and \lya scattering components implemented in \art were described in detail in \citet{Yajima2012}. In this Section, we only describe the improvements to the original implementation in the new \art.   

Similar to the continuum dust transfer, we modified the implementation of the ionization process, by not treating absorption as a single statistical
event, but by following the average absorbed luminosity along the paths of photon packets. However, there is an important distinction between the dust
and the ionization RT. Due to the tight coupling between the gas absorption opacity and its ionization fraction, one must update the ionization states of
the gas cell as the photons pass through it. This creates dependency in successively traced photons, making parallelization difficult. 

To circumvent this issue, we have developed a two-step procedure to evolve the gas ionization fraction. In the first step, individual photon packets are traced, and the ionization states are updated as the ray tracing proceeds. When multiple, e.g., $n_p$, processors are used for this step, each processor works independently by tracing an equal number of photons to advance the ionization evolution for a reasonable time step, $\delta t$. This provides $n_p$ independent estimates of the ionizing flux within each cell. In the second step, these independent estimates are collected and averaged to provide a single improved estimate, which is then
used to evolve the ionization states for the same time step. This two-step process is then repeated to advance the ionization state evolution until convergence is reached.

In some situations, one can ignore the time evolution of the ionization states but only needs the final equilibrium. For this purpose we implement another mode to solve the equilibrium ionization fractions iteratively. In each iteration, the ray tracing procedure is similar to that used in advancing the ionization evolution for one time step. When an estimate of the ionizing flux is available from the ray tracing step, we solve the steady state statistical equilibrium equation, instead of evolving the ionization fractions at this time step.

\begin{figure}
\includegraphics[width=3.2in]{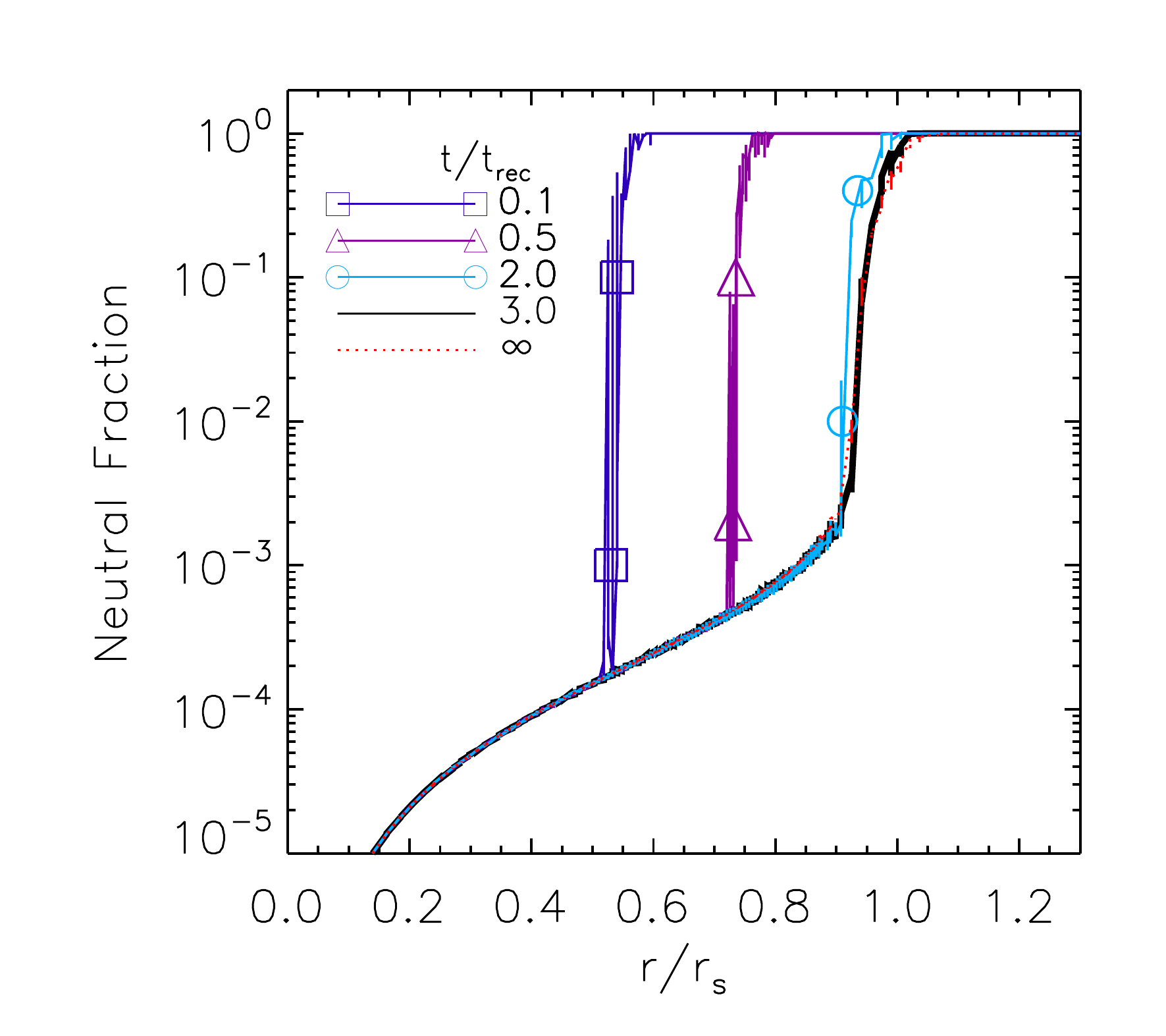}
\caption{Test of the new ionization algorithm with a dust free Str\"{o}mgren sphere. The neutral H fraction profile of a dust free Str\"{o}mgren sphere is calculated at
  different times. The profile at $t/t_{\rm rec} = 3$ successfully converges to the final solution obtained with the equilibrium ionization mode, as shown in red curve. } 
\label{fig:ssev}
\end{figure}

We test the new implementation of the ionization process with a dust-free Str\"{o}mgren sphere model at a temperature of $10^4$~K. We calculate the neutral H fraction profile at different times  relative to the recombination time scale, as shown in Figure~\ref{fig:ssev}.  The ionization front reaches the Stro\"{o}mgren sphere radius at  $t/t_{\rm rec}=3$, and the neutral fraction profile successfully converges to the final solution obtained with the equilibrium ionization mode as expected.

\begin{figure}
\includegraphics[width=3.2in]{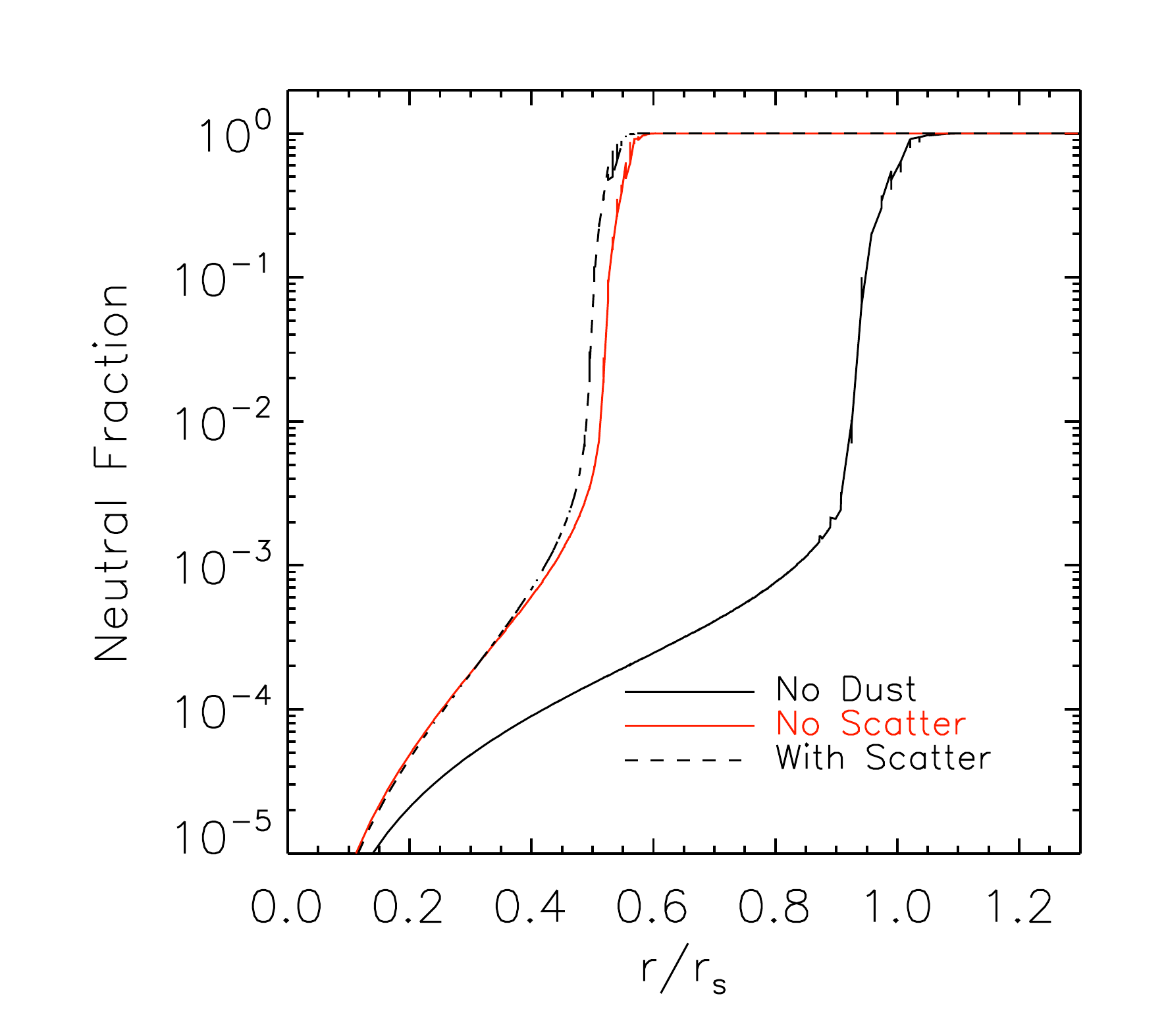}
\caption{A comparison of dust-free and dusty Str\"{o}mgren sphere calculated with the equilibrium ionization mode. A dusty sphere has a smaller ionization radius than a dust-free one due to dust absorption and scattering of the ionizing photons. }
\label{fig:sseq}
\end{figure}

We also compare the equilibrium neutral fraction profiles of a dust-free and a dusty Str\"{o}mgren sphere calculated with the equilibrium ionization mode, as shown in Figure~\ref{fig:sseq}.The dusty sphere has an absorption optical depth of 4.0 at the Stro\"{o}mgren sphere radius. The location of the ionization front for the dusty sphere agrees with the analytic solution of \citet{Spitzer1978}. Figure~\ref{fig:sseq} shows that dusty sphere has a smaller ionization radius than dust-free medium due to absorption of the ionizing photons by dust, and that dust scattering has a small effect on the equilibrium neutral fraction profile and shrinks the ionized sphere.

\begin{figure}
\includegraphics[width=3.2in]{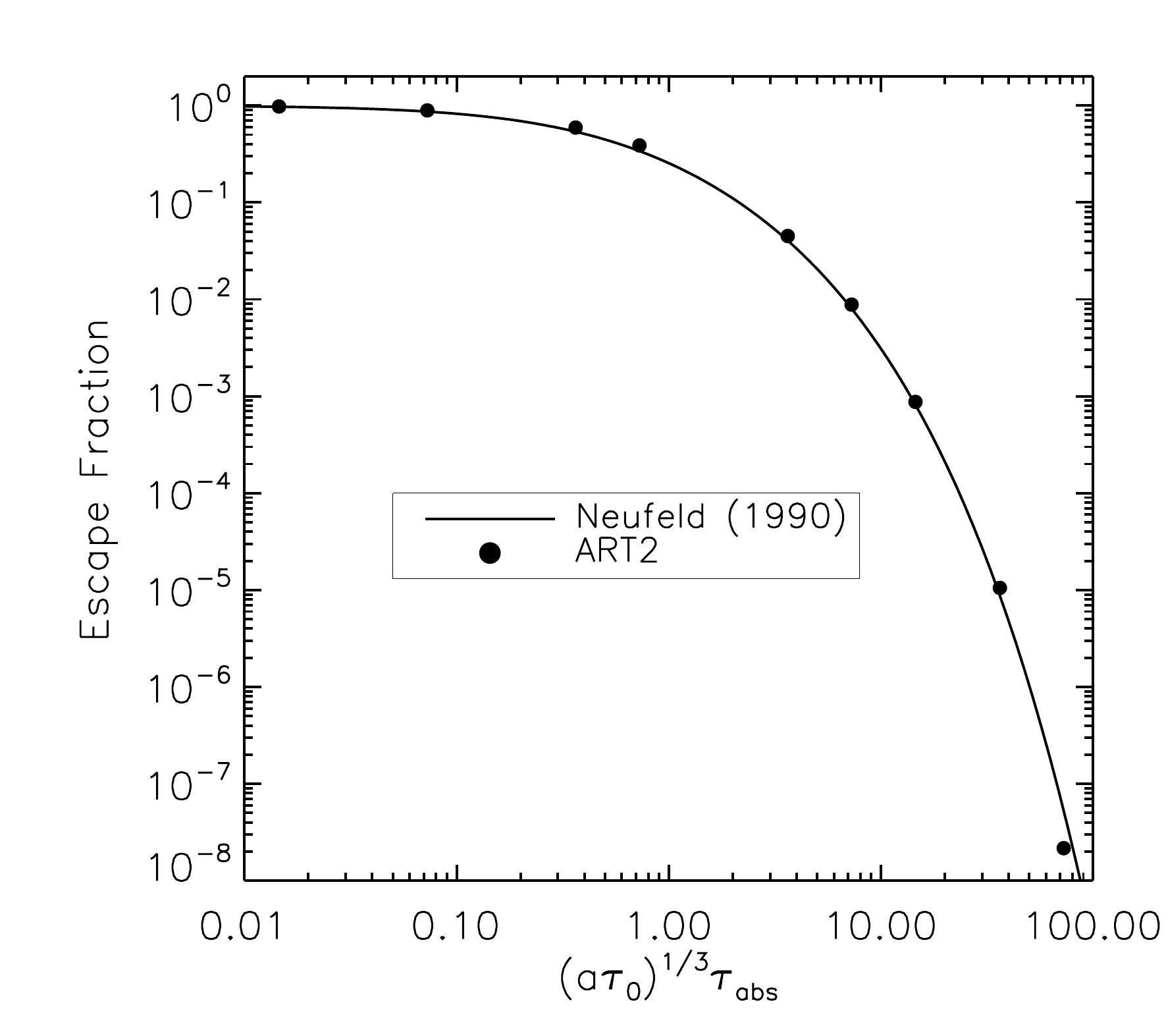}
\caption{Test of the new \lya algorithm with a dusty slab. The escape fraction of \lya photons from a source at the center of a dusty slab is calculated at different absorption optical depths. The parameter $\rm{(a\tau_0)^{1/3}\tau_{abs}}$ is a scaled optical depth combining the optical depth at the line center $\tau_0$ and the total absorption optical depth $\tau_{\rm abs}$, as defined in \citet{Neufeld1990}. Our result agrees with the analytic solution of \citet{Neufeld1990} very well.}
\label{fig:lyad}
\end{figure}

The algorithm for the resonant scattering of \lya photons is the same as that described in \citet{Yajima2012}, with one improvement for the handling of dust absorption. Similar to the implementations of the dust continuum and ionization RT,  the dust absorption is followed as the average luminosity of the entire batch of photons gradually diminishing  along the photon path, instead of a single statistical event that destroys the entire photon. Figure~\ref{fig:lyad} shows the standard Neufeld \citep{Neufeld1990} test for the escape fraction of \lya photons from a source placed at the center of a dusty slab. The comparison demonstrates good agreement between our new \lya calculation and the analytic solution.

\subsection{Molecular Line Radiative Transfer}
\label{sec:co}

The excitation of many molecules is non-LTE because the gas density is usually below the high threshold required for local thermal equilibrium. The line transfer is further complicated by the closely coupled relation between source function, level population and radiation field. Our implementation of the non-LTE molecular line transfer in \art is based on the accelerated Monte Carlo (AMC) method of \citet{Hogerheijde2000}, and the implementation of the AMC method on a 3D adaptive grid is similar to that of \citet{Rundle2010}. Here we give a brief description of the methods and highlight the improvements we made in \art.

The basic problem in non-LTE molecular line transfer is to seek the solution
of the coupled equations of statistical equilibrium of level populations $n_i$ and
radiative transfer,
\begin{equation}\\
\label{eq:se}
 \sum^N_{j\neq{i}}{n_jP_{ji}} - n_i\sum^N_{j\neq{i}}{P_{ij}} = 0,
\end{equation}
where the coefficients $P_{ij}$ are given by:
\begin{equation}\\
\label{eq:pij}
 P_{ij} = \left\{
  \begin{array}{l l}
   A_{ij} + B_{ij}\bar{J}_\nu + C_{ij} & \quad (i>j)\\
   B_{ij}\bar{J}_\nu + C_{ij} & \quad (i<j)\\
  \end{array}\right.
\end{equation}
where $A_{ij}$ and $B_{ij}$ are Einstein coefficients for transition from level i to j, and $C_{ij}$ are calculated as: 
\begin{equation}\\
 C_{ij} = n_{col}K_{ij}
\end{equation}
where $n_{col}$ is the number density of collision partner, which is often taken to be molecular hydrogen $\rm{H_2}$, and $K_{ij}$ are the collisional rate coefficients in unit of $\rm{cm^3~s^{-1}}$. 

The coupling to the radiation comes from the frequency integrated
mean intensity for a given transition $\bar{J}_\nu$, 
\begin{equation}\\
\label{eq:jnu}
\bar{J}_\nu=\int J_\nu\phi_\nu{d\nu}
      =\frac{1}{4\pi}\int{I_\nu\phi_\nu{d\nu}{d\Omega}},
\end{equation}
where $I_\nu$ is the specific intensity of the radiation field, and $\phi_\nu$
is the normalized line profile. 

In AMC, the Monte Carlo ray tracing used to obtain $\bar{J}_{\nu}$ takes a cell-centered point of view. A discrete set of rays passing through each cell with
different angles and frequencies are sampled to give an estimate of the double integral,
\begin{eqnarray}
\label{eq:jsplit}
\bar{J}_\nu &=& \bar{J}_{\nu}^{ext} + \bar{J}_{\nu}^{int} \nonumber\\
       &=&
       \frac{\sum_i{[I^i_\nu{e^{-\tau_i}}+S^i_\nu(1-e^{-\tau_i})]\phi_\nu}}{\sum_i{\phi_\nu}},
\end{eqnarray}
where the index $i$ denotes the $i$-th ray in the sample. 

In the original AMC method, the frequencies are uniformly sampled around the local systematic
velocity vector. $I^i_\nu$ is the incident radiation on  
the edge of cell for $i$-th ray, which is obtained by integrating the
radiative transfer equation from the cell edge to infinity. $S^i_\nu$ is the
source function in the cell. $\tau_i$ corresponds to the distance $ds_i$
between the spot where the ray contributes to the cell and the point where the
ray intersects the boundary of the cell. The convenient split of $\bar{J}_{\nu}$ into the
external part, $\bar{J}_{\nu}^{ext}$, and the internal part,
$\bar{J}_{\nu}^{int}$, is the key to the 
acceleration scheme used in AMC. In each iteration to solve
Equation~\ref{eq:se}, an inner iteration is inserted to obtain the level
populations of the local cell while fixing $\bar{J}_{\nu}^{ext}$. The benefit
of this inner iteration is substantial when the optical depth of the local
cell is relatively large. 

In the \art implementation, we make two modifications to
Equation~\ref{eq:jsplit}. First, the frequencies are not randomly sampled,
instead, we estimate the integration over $\nu$ by a five-point Gauss
quadrature formula. This is uniquely suitable as the line profile is
often close to Gaussian function. Second, the external and internal
contributions to mean intensity are taken to be the average along the photon
path within the cell, instead of the value at a random
point. Equation~\ref{eq:jsplit} may be rewritten as,
\begin{equation}\\
\label{eq:jnew}
\bar{J}_{\nu}=\Sigma_{ik}w_k\phi_k\left[\beta_{ik} I^{ik} + (1-\beta_{ik})S^{ik}\right],
\end{equation}
where the additional index $k$ denotes the Gauss quadrature points in the
frequency space, $w_k$ are the quadrature weights, and $\beta_{ik}$ is an
escape probability given by,
\begin{equation}\\
\label{eq:beta}
\beta_{ik} = \frac{1-e^{-\tau_{ik}}}{\tau_{ik}},
\end{equation}
where $\tau_{ik}$ is the total optical depth of the $i$-th ray at $k$-th frequency within the local cell. 

This recasts $\bar{J}_\nu$ into a familiar form encountered in, e.g., the escape probability formalism, or the large velocity gradient (LVG) approximation \citep{DeJong1975,
  Goldreich1974}. The source function $S^{ik}$ generally includes contributions from the molecular line and dust. The line contribution depends on the level populations, which may make the inner iteration unstable if the cell optical depth is large enough so that the internal part of the mean intensity dominates, or equivalently, if $\beta_{ik}$ is very small. In such cases,  it is more convenient to rewrite the source function $S^{ik}$ in Equation~\ref{eq:jnew} to include the line source function $S^0_{ul}$ and a residual term  $S^{ik}_{\rm res}$: 
\begin{eqnarray}
S^{ik} &=& (S^{ik}-S^0_{ul})+S^0_{ul} = S^{ik}_{\rm res} + S^0_{ul} \nonumber\\
S^0_{ul} &=& \frac{2h\nu_{ul}^3}{c^2}\left[\frac{n_lg_u}{n_ug_l}-1\right],
\end{eqnarray}
where $S^0_{ul}$ is the line source function for transition from $u$ to
$l$, which depends on the level populations $n$ and the statistical weight $g$  of level $u$ and $l$, respectively. 

As in the LVG approximation, we can drop the $S^0_{ul}$ term in the mean intensity, and substitute $S^{ik}$ in Equation~\ref{eq:jnew} with $S^{ik}_{\rm res}$. As suggested in \cite{Goldreich1974}, the combined effect of dropping $S^0_{ul}$ on the level population  equation is equivalent to modifying the Einstein coefficients $A_{ul}$ to $\Sigma_{ik}\beta_{ik}w_k\phi_kA_{ul}$ in Equation~\ref{eq:pij}. Now the inner iteration to solve the level populations is
much more stable when Equation~\ref{eq:se} is recast in this form. Note that since the molecular and the atomic fine structure lines we consider here do not have severe population inversion such as strong masing, so a simplified approach of limiting the negative optical depth can be used to solve the statistical equilibrium equations and obtain level population convergence. Once the level population $n$ and $S^{ik}_{\rm res}$ are solved, we can obtain the source function $\beta_{ik}$, and then solve the general radiative transfer equation.

Finally, the improved AMC method implemented in \art is parallelized by distributing the
level population equation for individual cells to different processors. The
solutions for individual cells are essentially independent if only the results
from the previous iteration is used. Data communication is therefore only
needed at the end of iteration to synchronize the level population solutions
on the entire grid cross different processors. For the construction of images
and spectra, the parallelization is achieved by distributing ray tracing to
different processors.

\begin{figure}
\includegraphics[width=3.2in]{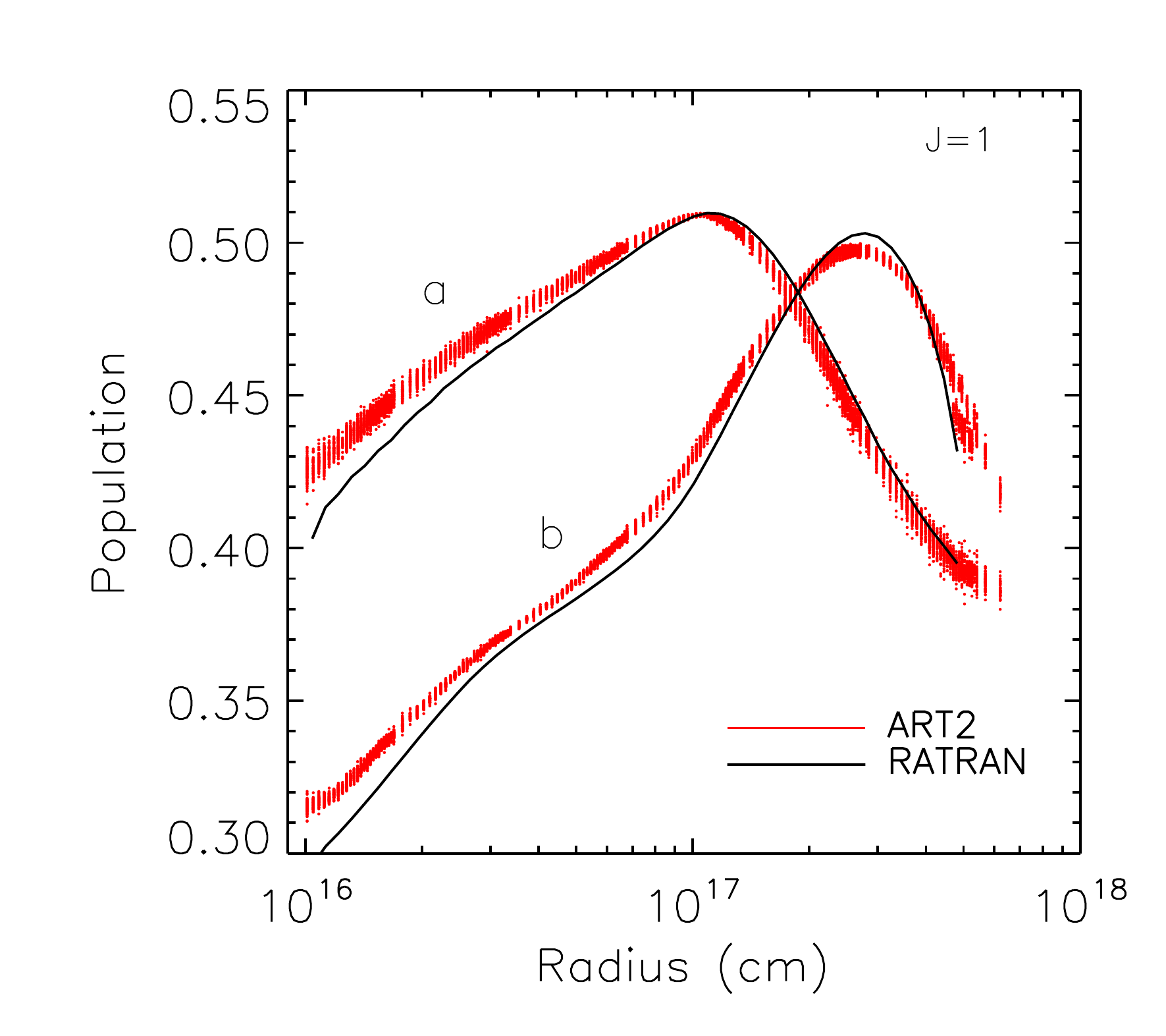}
\includegraphics[width=3.2in]{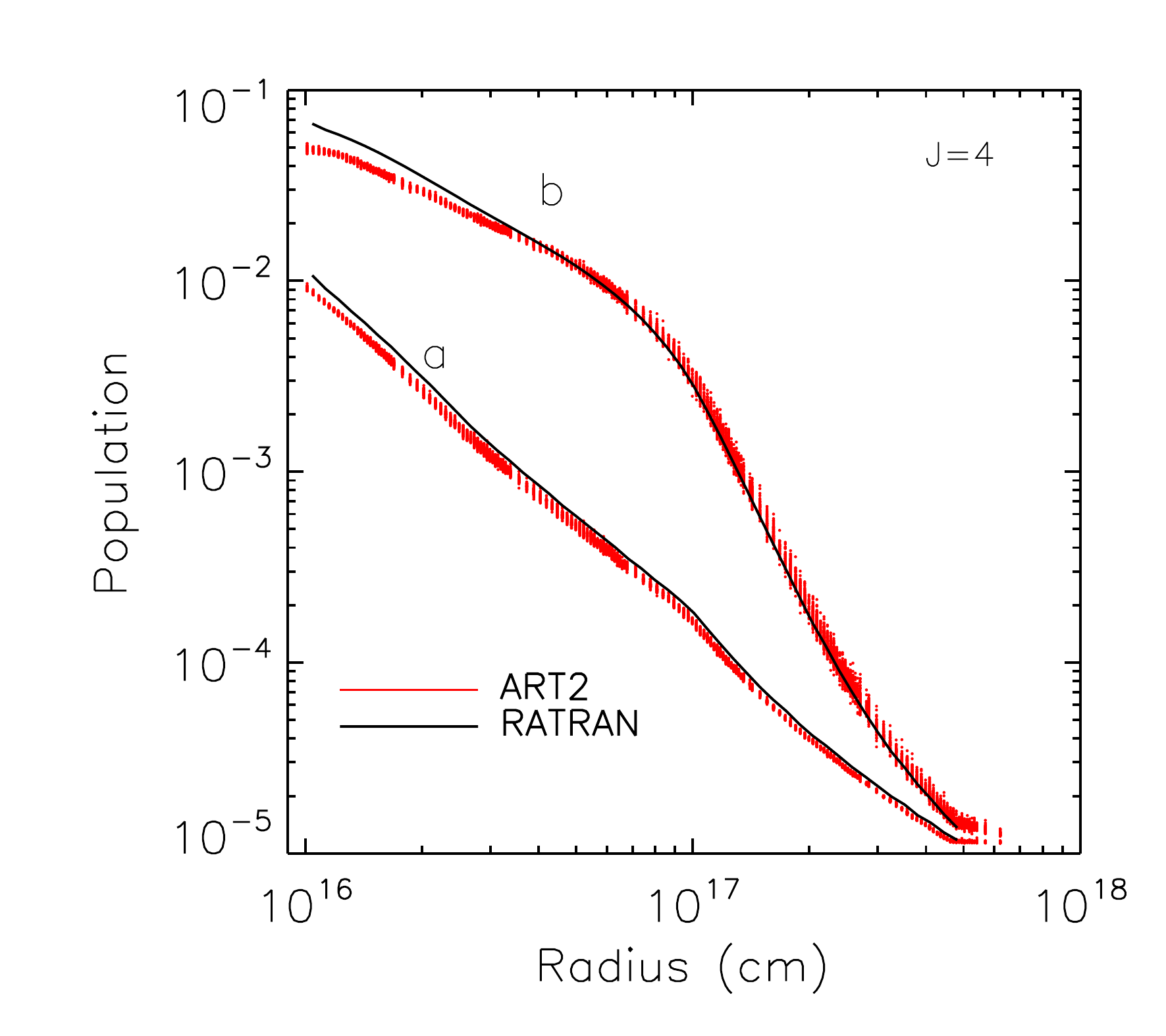}
\caption{Test of the AMC implementation in \art with a collapsing cloud model for HCO+ molecule for optically thin (a) and thick (b) cases, in comparison with Ratran. The top panel shows the relative population for $\rm J=1$ level, while the bottom panel shows the same but for $\rm J=4$ level, respectively. The comparisons show good agreement between \art and Ratran.}
\label{fig:j1j4}
\end{figure}

\begin{figure*}
\includegraphics[width=3.0in]{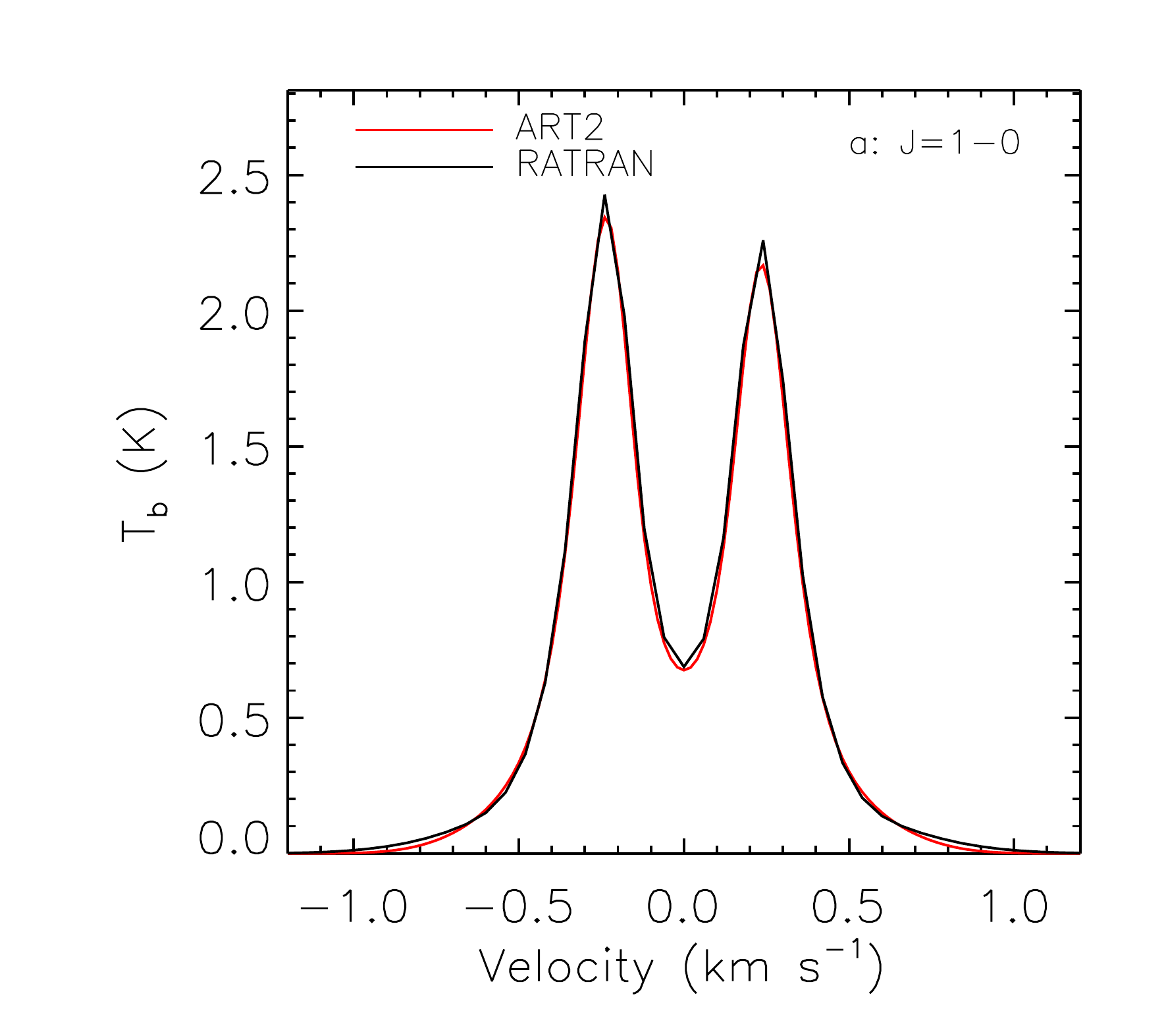}
\includegraphics[width=3.0in]{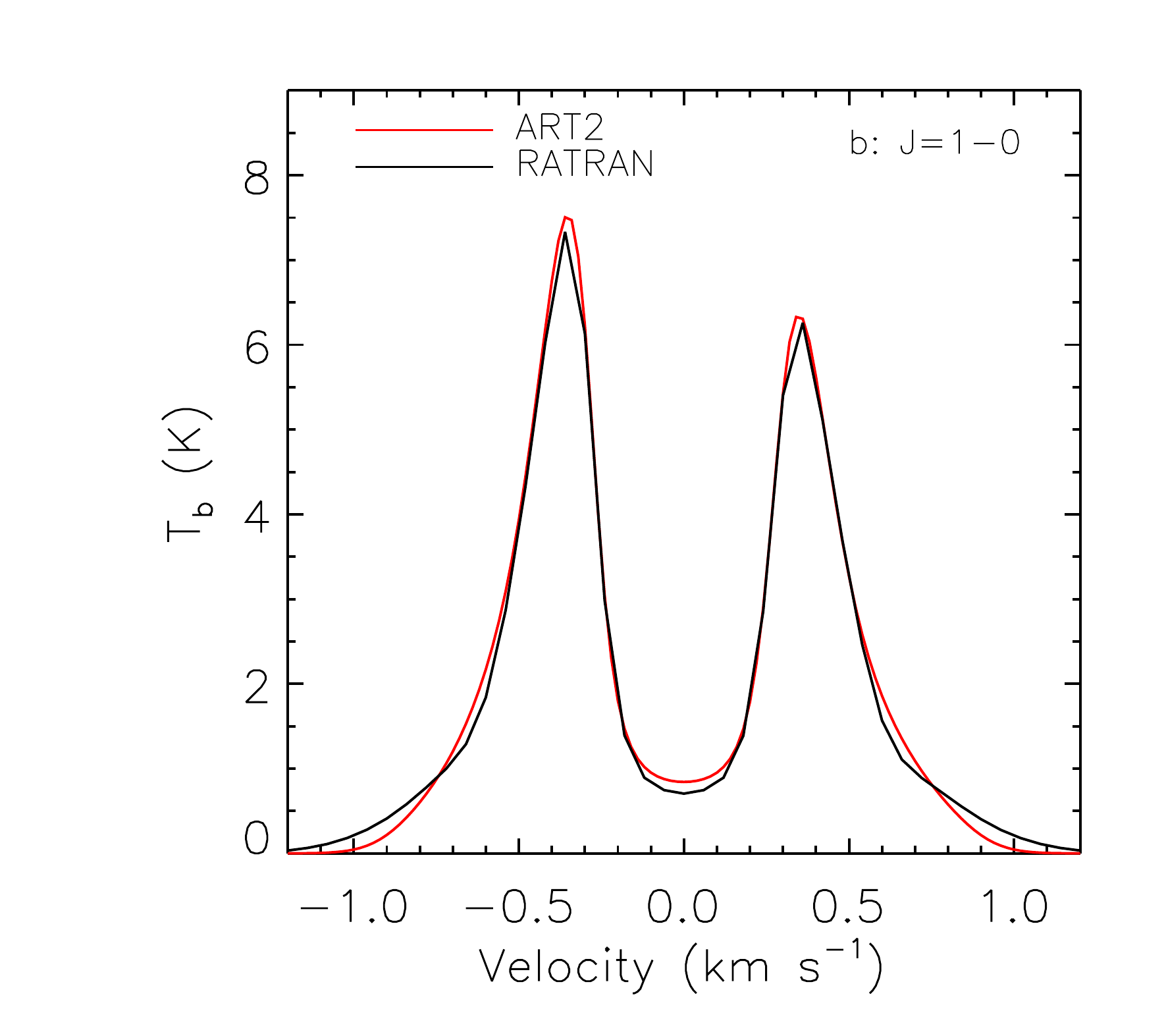} \\
\includegraphics[width=3.0in]{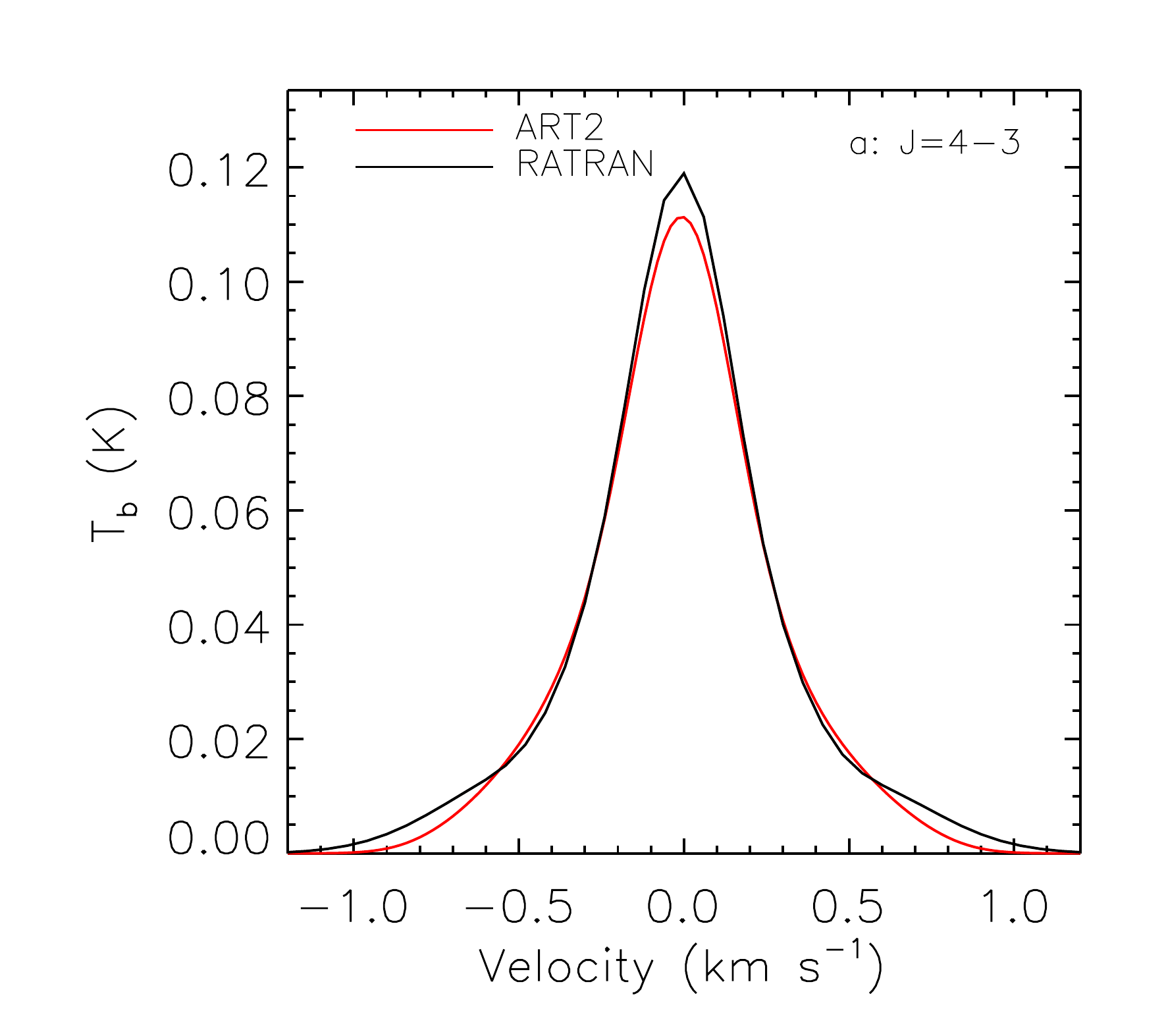} 
\includegraphics[width=3.0in]{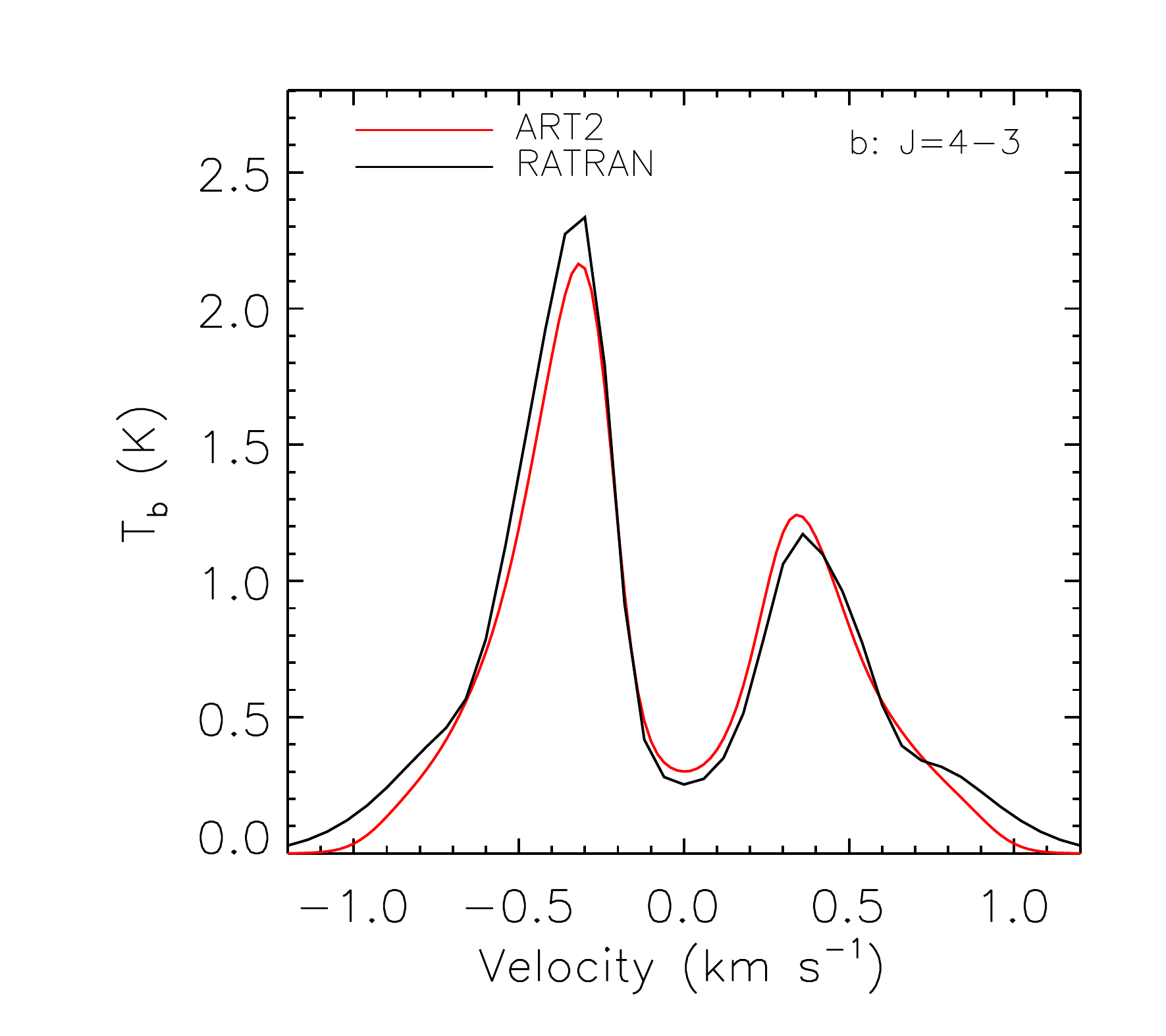}
\caption{Test of the new molecular RT algorithm in \art with a collapsing cloud model for HCO+ molecule for optically thin (a) and thick (b) cases, in comparison with Ratran. The top panels show the brightness temperature profiles for $\rm J=1-0$ transition, while the bottom panels show the same but for $\rm J=4-3$. The comparisons show good agreement between \art and Ratran.} 
\label{fig:s0134}
\end{figure*}

To verify our implementation of the AMC method, we run the test problem of a collapsing cloud for ${\rm HCO}^{+}$ molecule, as presented in \citet{vanZadelhoff2002}, with \art for both optically thin (a) and optical thick (b) cases. We have also run the same models using the one-dimensional code Ratran~\citep{Hogerheijde2000b}, which implements the original AMC method~\citep{Hogerheijde2000}. The resulting relative populations for the $\rm J=1$ and $\rm J=4$ levels from \art are shown in Figures~\ref{fig:j1j4}, in comparisons with results from Ratran. These comparisons show good agreement between  \art and Ratran.

The resulting brightness temperature profiles for $\rm J=1-0$ and $4-3$ transitions are shown in Figures~\ref{fig:s0134}, which again show reasonable agreements between \art and Ratran. 

\subsection{Atomic Fine Structure Line Transfer}
\label{sec:afs}

The non-LTE atomic fine structure line transfer includes two steps: First, we solve the ionization equilibrium of individual atomic species and the thermal equilibrium of the gas. After that, we solve the coupled equations of statistical equilibrium of level populations and radiative transfer of fine structure emission lines of individual ionic species using the same method of the molecular line RT in Section~\ref{sec:co}. 

In order to treat the fine structure emission lines from atomic ions, the
ionization module is extended to include elements other than H. The ionization
equilibrium of individual atomic species and the thermal equilibrium of the
gas are determined by solving a set of non-linear equations,
\begin{eqnarray}
  (C^a_i+R^a_i)n^a_i &=& C^a_{i-1}n^a_{i-1} + R^a_{i+1}n^a_{i+1} \nonumber\\
  n_e &=& \sum_{a,i} n^a_i q_i \nonumber\\
  n^a &=& \sum_i n^a_i \nonumber\\
  C(n_e,T) &=& H(n_e,T)
\end{eqnarray}
where $n^a_i$ represents the number density of the charge state $i$ of atomic
species $a$; $n^a$ is the total density of the atomic species $a$;
$n_e$ is the electron density; $C^a_i$ and $R^a_i$ are the total
ionization and recombination rates of the ion, $q_i$ is the ion charge;
$C(n_e,T)$ and $H(n_e,T)$ are the total cooling and heating rates,
respectively. 

The ionization processes include both photoionization and
collisional ionization by electrons, and the recombination processes include
both radiative and dielectronic recombinations. 
Charge exchange reactions between ions and neutral H and He
are also included in the ionization and
recombination processes. The ionization and recombination rate coefficients of relevant ions needed to compute the
ionization equilibrium are taken from the photoionization code CLOUDY version
17.00 \citep{Ferland2017}, while the local ionizing photon flux are determined
in our Monte Carlo ray tracing procedure. The cooling and heating
rates are also calculated self consistently. The heating sources include
photoionization and Compton scattering with high energy photons. The cooling
sources include Bremsstrahlung, collisional ionization, radiate and dielectronic
recombination, collisional excitation of fine structure lines, and
Comptonization with the low energy photons, including cosmic microwave
background.

\begin{figure}
\includegraphics[width=3.2in]{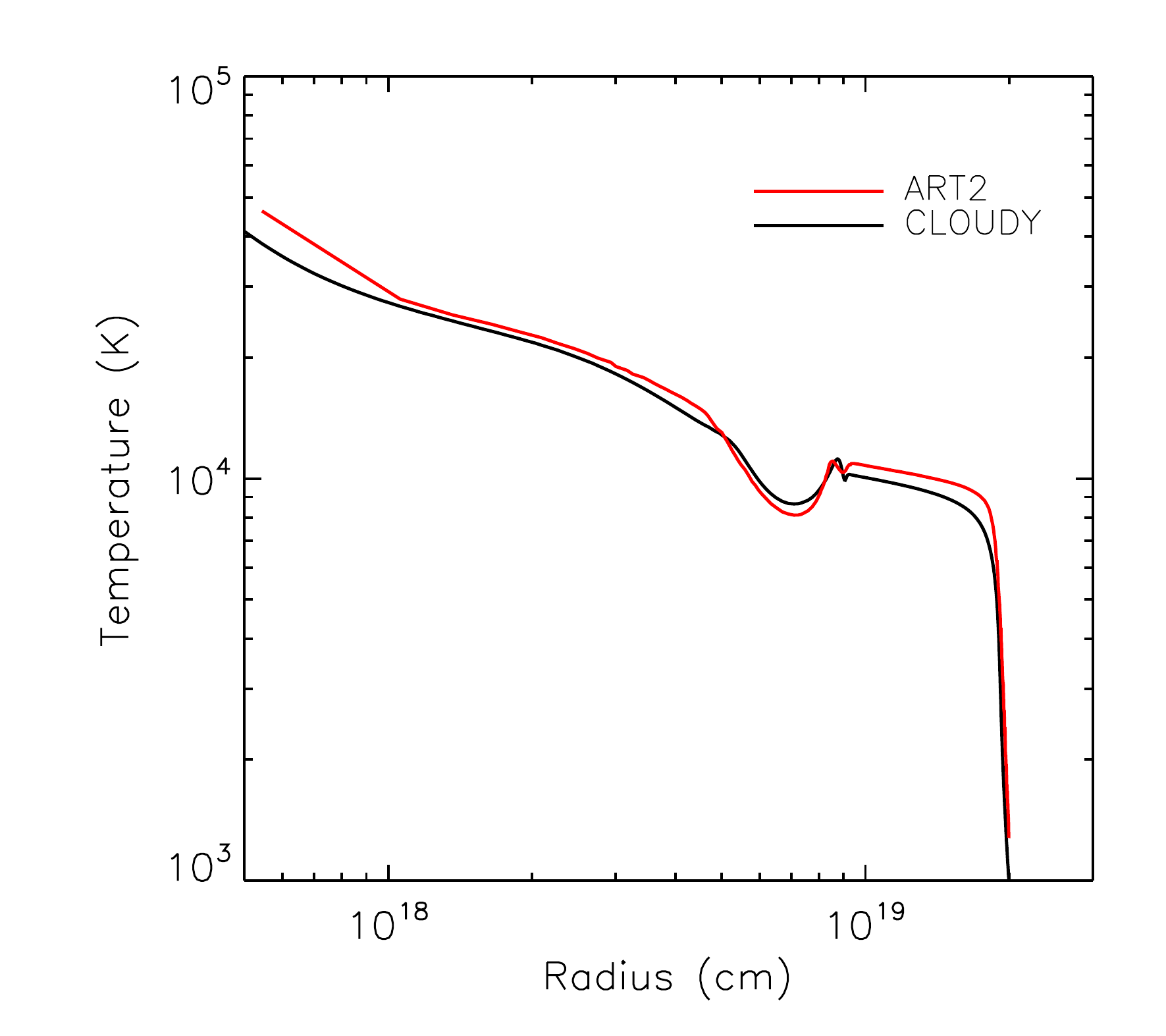}
\includegraphics[width=3.2in]{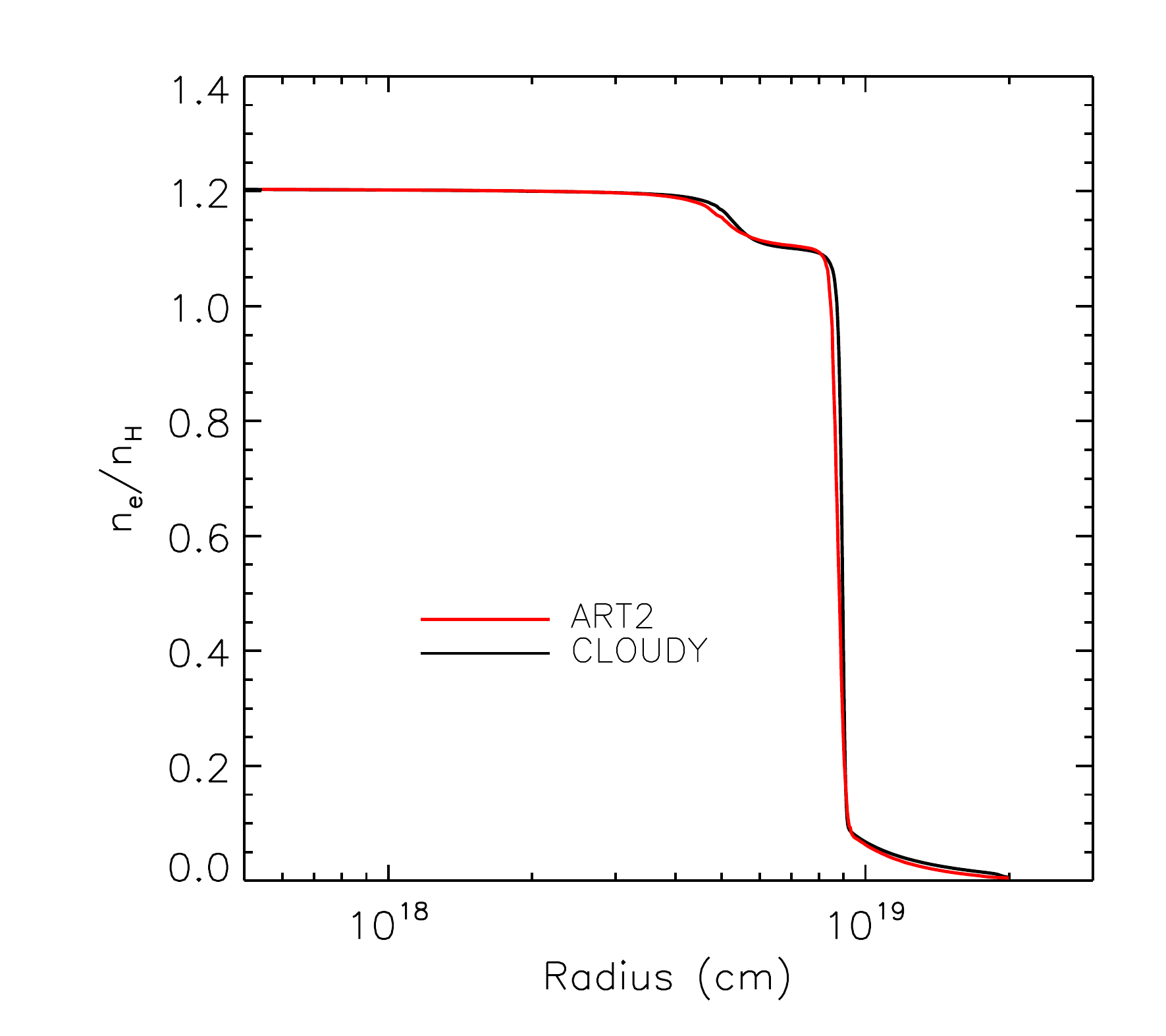}
\caption{Test of the new atomic fine structure line RT in \art  with a spherical cloud. The radial profile of the gas temperature (top) and the electron density relative to the H atoms (bottom) calculated by \art are in good agreements with CLOUDY. }
\label{fig:temp}
\end{figure}

\begin{figure}
\includegraphics[width=3.2in]{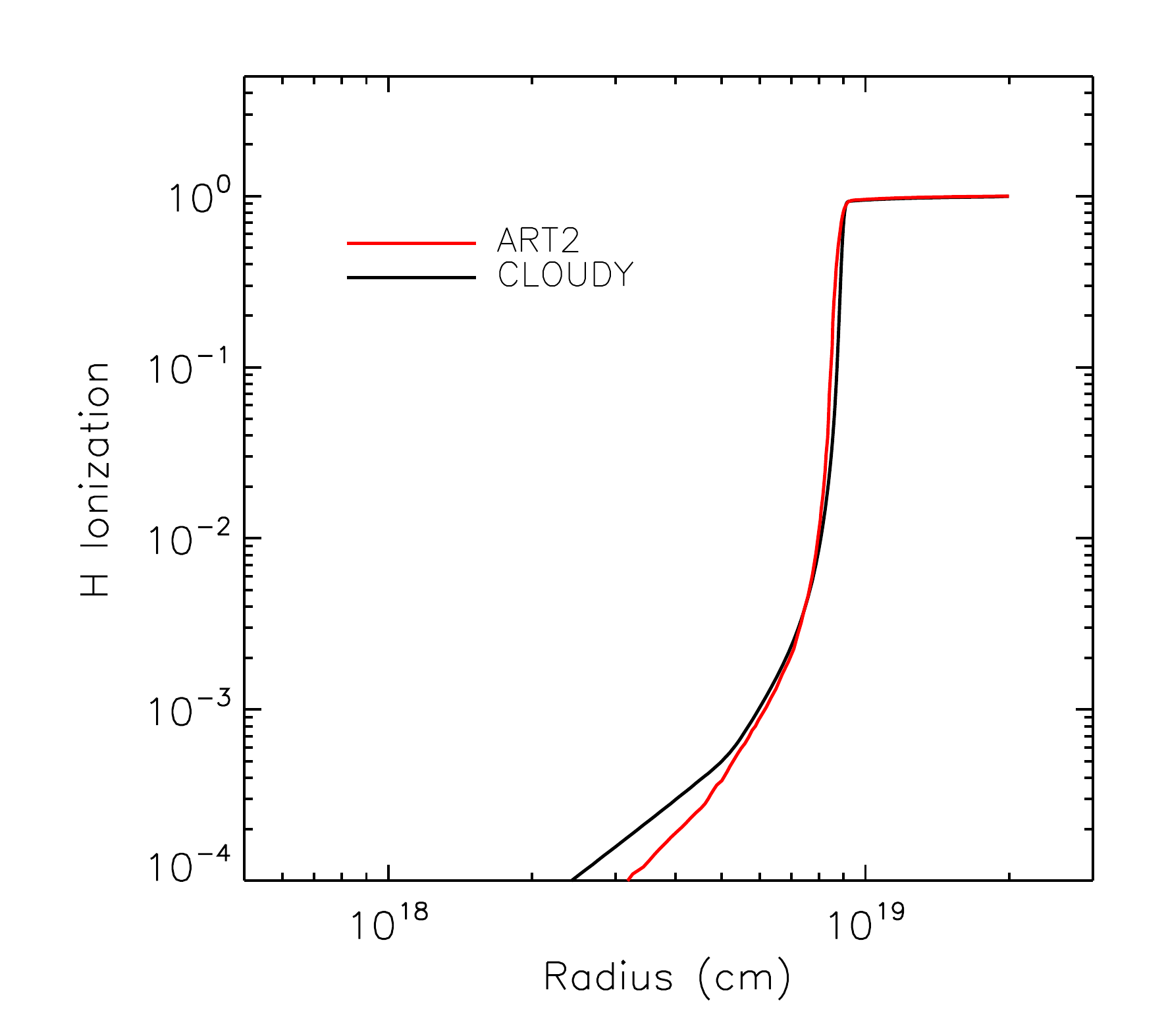}
\caption{A comparison of the HI fraction of a spherical cloud model calculated with \art
  and CLOUDY.}
\label{fig:xh}
\end{figure}

\begin{figure}
\includegraphics[width=3.2in]{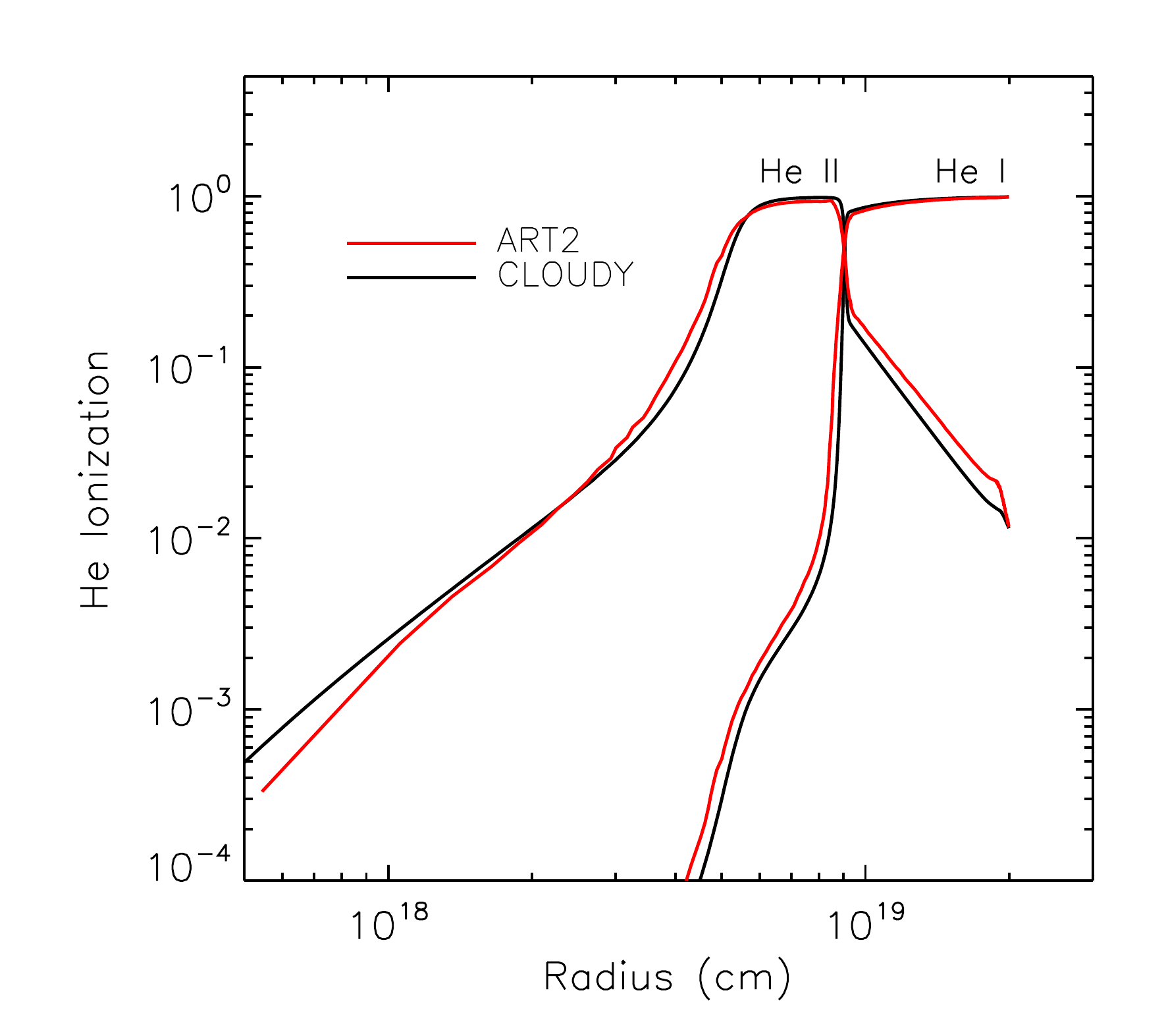}
\caption{A comparison of the HeI and HeII fractions of a spherical cloud model calculated with \art
 and CLOUDY.}
\label{fig:xhe}
\end{figure}

\begin{figure}
\includegraphics[width=3.2in]{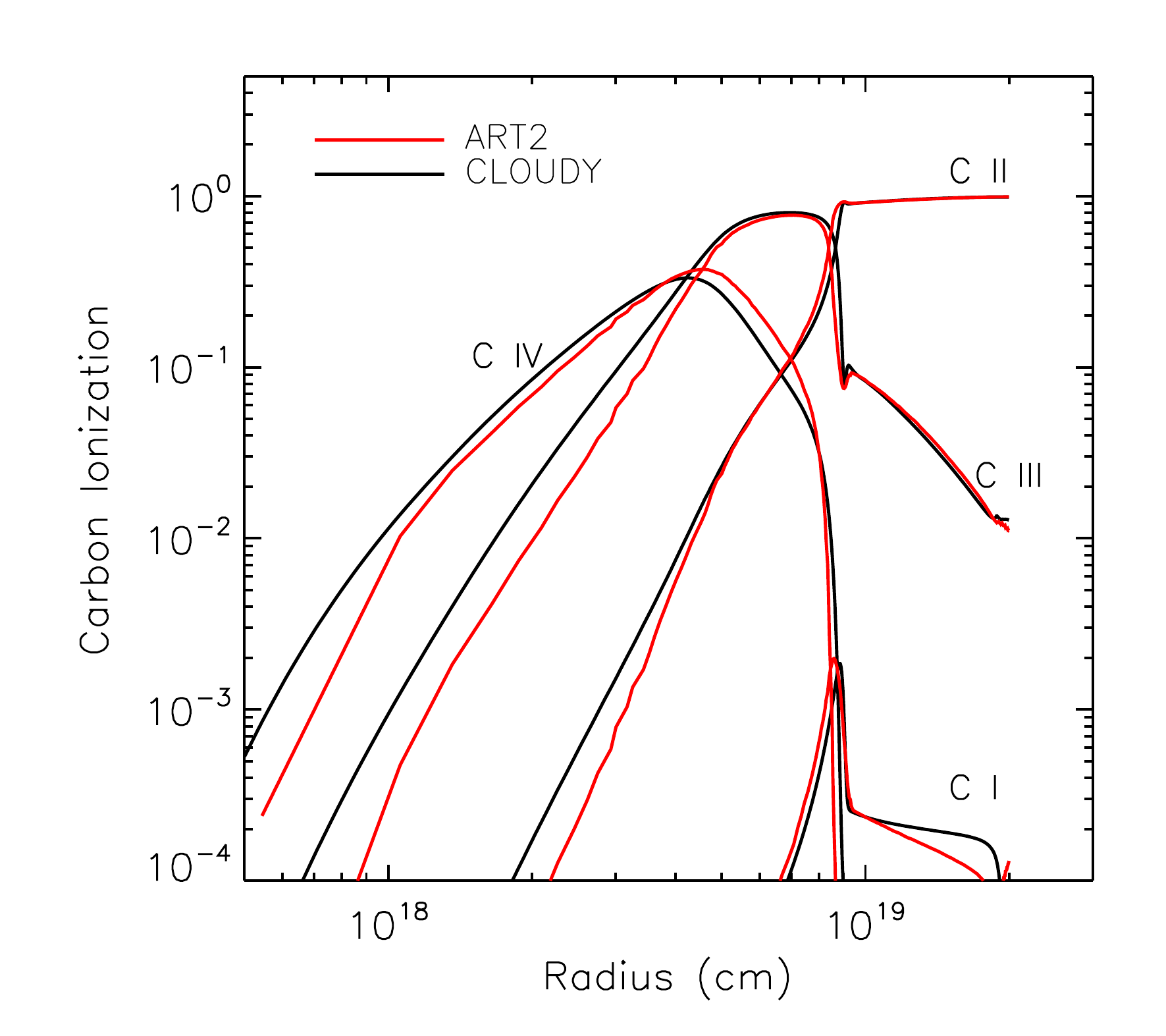}
\caption{A comparison of the CI -- CIV fractions of a spherical cloud model calculated
  with \art and CLOUDY.}
\label{fig:xc}
\end{figure}

\begin{figure}
\includegraphics[width=3.2in]{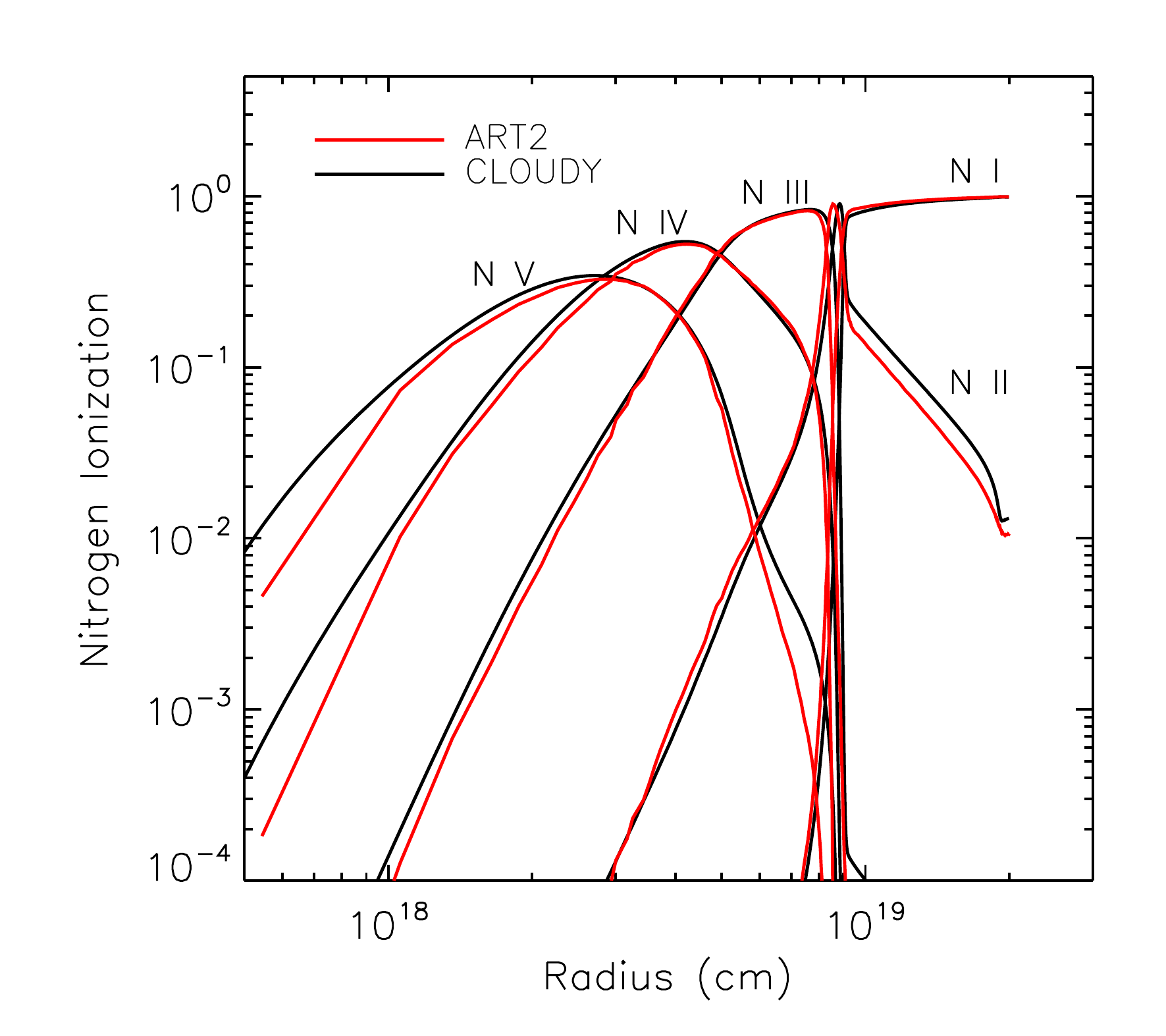}
\caption{A comparison of the NI -- NV fractions of a spherical cloud model calculated
  with \art and CLOUDY.}
\label{fig:xn}
\end{figure}

\begin{figure}
\includegraphics[width=3.2in]{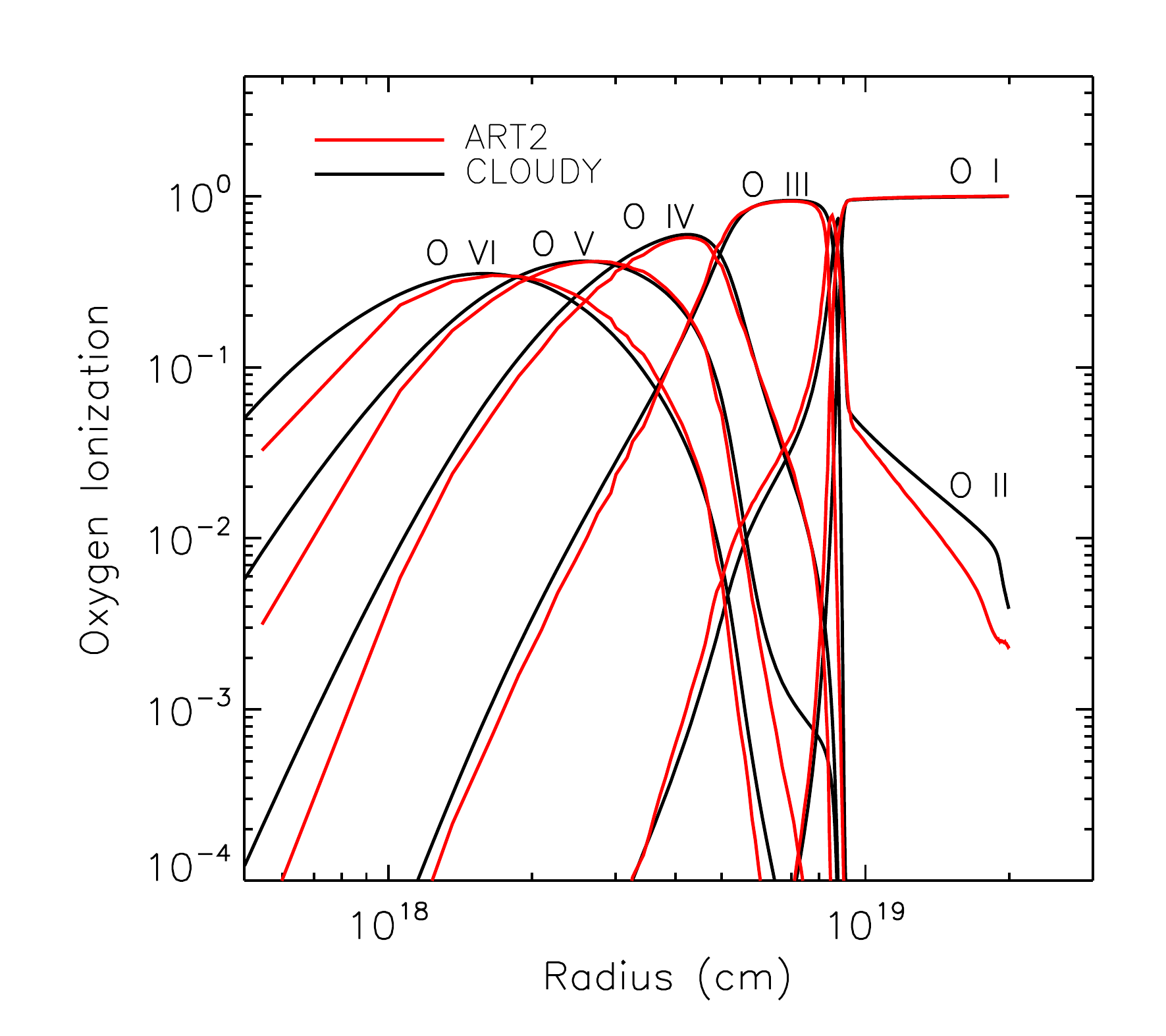}
\caption{A comparison of the OI -- OVI fractions of a spherical cloud model
  calculated  with \art and CLOUDY.}
\label{fig:xo}
\end{figure}

\begin{figure}
\includegraphics[width=3.2in]{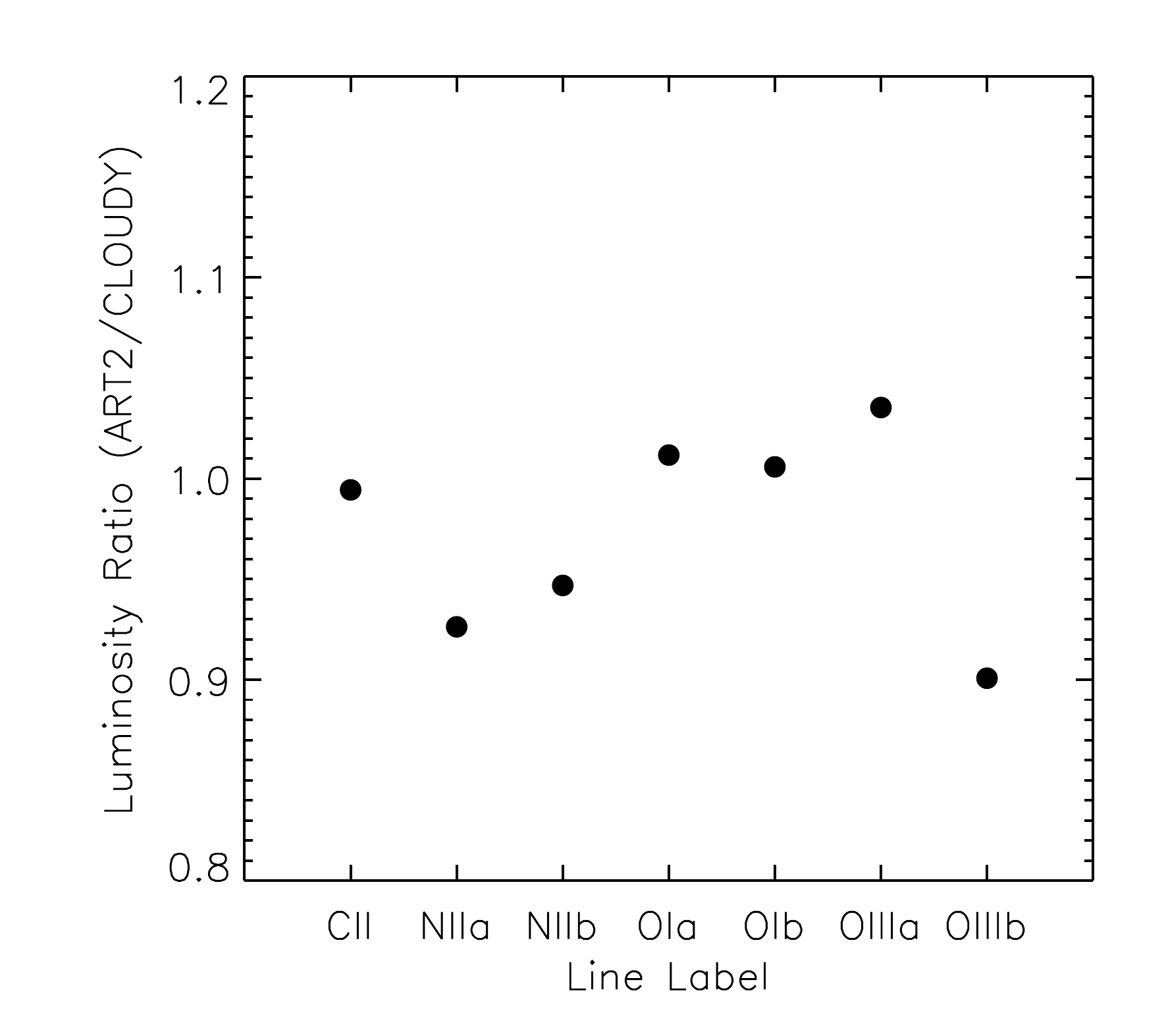}
\caption{A comparison of the atomic fine structure line luminosity calculated with \art and CLOUDY, 
  for the \cii, \nii, \oi and  \oiii lines. The comparison shows reasonable agreement between \art and CLOUDY to within 10\%.}
\label{fig:lum}
\end{figure}

After the ionization and thermal equilibrium are solved, the radiative transfer
of fine structure emission lines of individual ionic species are calculated
using the same method as the molecular line RT in Section~\ref{sec:co}. We use the same equations~(1) - (9) in Section \ref{sec:co}, but instead of solving the populations of molecular rotational levels, the fine structure level populations of a given ion are solved. The excitation mechanisms of the atomic
lines also differ significantly from those for molecular
lines. Collisional excitation with electrons and protons are typically the
dominant processes populating the fine structure levels in atomic ions. In our
implementation, we include excitations by electron, proton, and neutral H
atom. The rate coefficients of collisional processes, transition energies, and
spontaneous transition rates are taken from the CLOUDY database.

To verify our implementation, we compare calculations by \art of a spherical cloud model with a central ionizing source with those by the 1D photoionization code CLOUDY. The gas sphere has a uniform hydrogen density of 100~cm$^{-3}$, and a radius of  $2\times 10^{19}$~cm, which is about twice the Str\"{o}mgren radius of the cloud at the temperature of $10^4$~K. The ionizing spectrum is a power law with index of 1.5 in the energy range between 10 and 1000 Rydberg, and the total photon rate is $10^{49}$~s$^{-1}$. The atomic species included in the calculation are all ions of H, He, C, N, and O. The relative elemental abundances are set at the solar values. 

We first calculate the ionization and thermal structures of the cloud, and then the atomic fine structure lines of all species in the gas sphere, with both \art and CLOUDY. Figure~\ref{fig:temp} shows the gas temperature and the electron density as a function of radius as calculated by the two codes, and Figures~\ref{fig:xh} to \ref{fig:xo} show  the comparisons of the ionization fractions of H, He, C, N, and O computed, respectively. These comparisons show that the thermal and ionization structures calculated by the non-LTE atomic line RT in our 3D \art are consistent with the 1D calculations by CLOUDY.

Finally, we compute the radiative transfer for the prominent fine structure \cii line at 158~$\mu$m, the \nii lines at 122 and 205~$\mu$m, the \oi lines at 63 and 145~$\mu$m, and the  \oiii lines at 52 and 88~$\mu$m. In Figure~\ref{fig:lum}, we compare the total line luminosities calculated with \art and CLOUDY. It shows that the luminosities calculated with the two codes agree with each other to within 10\% for the 7 strongest lines.

\subsection{Radiative Transfer Through a Sub-grid Multi-phase ISM Model}
\label{sec:multiphase}

The RT processes in \art described in previous sections are comprehensive and applicable to situations with vastly different
physical scales, such as planetary disks, star forming regions, galaxies and large-scale structures. However, numerical simulations in
the cosmological context usually have limited spatial resolution to resolve molecular clouds where stars form. These simulations often employ sub-grid
recipes to capture the essence of star formation physics. For example, the Smoothed Particle Hydrodynamics (SPH) code \Gadget~ \citep{Springel2005} implements a sub-grid multi-phase ISM model to describe the star formation processes \citep{Springel2003}. When post processing such simulations with RT codes, it is important that a consistent sub-grid multi-phase ISM model is followed. 

The propagation of photons through a multi-phase ISM was first considered by \cite{Neufeld1991}, who calculated the escape of \lya photons from a clumpy, dusty ISM by modeling the medium as a collection of spherical clouds, which were treated as large particles capable of scattering or absorbing photons. Subsequently,  \cite{Hobson1993} named this approach the ``Mega-grain approximation", and they found good agreement between the approximation and calculations using  Markov processes. Recently, the ``Mega-grain approximation" was adopted by \cite{Hansen2006} to calculate the \lya RT  using a Monte Carlo method. We use the same approximation and follow the  approach of  \cite{Hansen2006} in our treatment of the multi-phase ISM. 

In the original \art, we implemented the multi-phase RT for dust continuum transfer by assuming that the cold phase consists of randomly distributed molecular clouds with a given mass-radius relation and a power-law mass distribution function, and statistically sample these clouds in the ray tracing process \citep{Li2008}. The physical properties such as density and volume filling factor of the cold phase were derived from the \Gadget  simulation output. It was shown in that work that the existence of hot and cold dust components are essential to reproduce both near and far infrared observations of high redshift quasars. In the present work, we adopt a similar  approach with some modifications to accommodate the new absorption handling and extend it to the RT components other than the dust continuum.

First, to simplify the handling, we do not impose a specific power-law mass distribution function in a single cell like in the original implementation. Instead, all clouds are assumed to have the same mass, $M_c$, and radius, $R_c$, within a single cell that satisfy the mass-radius relation of the form $M_c=AR_c^\alpha$, where $\alpha$ has a value between 2 and 2.5, as most observations suggest. In principle, the normalization $A$ could vary from galaxy to galaxy, but a value of a few hundred is appropriate for galaxies like Milky Way \citep{Solomon1987}. When a photon passes through a cell that contains cold clouds, the total opacity is given by:  
\begin{equation}\\
\alpha^t = (1-f_v)\alpha^t_h + f_v\beta_c\alpha^t_c,
\end{equation}
where $f_v$ is the volume filling factor of the cold phase in the cell, the
index $t$ is either $s$ for scattering and $a$ for absorption, while
$\alpha^t_h$ and $\alpha^t_c$ are the respective opacities of hot and cold
phases, and $\beta_c$ is the escape probability of individual cold clouds,
which we model as 
\begin{equation}\\
\beta_c = \frac{1-e^{-2f_g\alpha_cR_c}}{2f_g\alpha_cR_c},
\end{equation}
where $\alpha_c=\Sigma_t \alpha^t_c$, is the total opacity of the cold
cloud. 

The appearance of $\beta_c$ in the weighting factor for the cold phase
is due to the fact that photons do not completely penetrate the cold clouds if
its optical depth is high, with an average penetration depth reduced by the
factor $\beta_c$. Note that the definition of $\beta_c$ includes a geometric
factor of order unity, $f_g$. Since our treatment of the multi-phase RT is
approximate, and likely has larger uncertainties elsewhere, we have take
$f_g=1$, appropriate for a one-dimensional cloud with photon path length of
$2R_c$, even though $R_c$ is derived from the simulation output assuming
spherical clouds. 

When the absorption albedo per interaction,
$\epsilon=\sigma_a/(\sigma_a+\sigma_s)$, of the gas in 
cold clouds is small, as is the case with \lya scattering, the effective cloud
absorption albedo, $\epsilon_c$, is enhanced due to multiple scattering within
the cloud before the photon escapes \citep{Hansen2006}: 
\begin{equation}\\
\label{eq:albedo}
\epsilon_c = \frac{2\sqrt{\epsilon}}{1+\sqrt{\epsilon}}.
\end{equation}
where $\epsilon_c  \approx 2\sqrt{\epsilon}$ when $\epsilon <<1$,  which is similar to the power law originally found by \cite{Neufeld1991}.

Since we do not track multiple scattering inside the optically thick cold
clouds explicitly, the absorption opacity of the cold clouds are enhanced
according to Equation~\ref{eq:albedo} to account for this effect.

Second, to obtain the ionization structure and \lya emission
from the cold clouds, we must allow a thin layer of the cold cloud to be
heated and ionized by the external stellar and quasar radiation. The thickness
and temperature of this layer is determined self-consistently during the
ionization RT calculation, in a fashion similar to determining the
Str\"{o}mgren sphere radius for a central source in the cloud. In adopting
this model, we assume that the stellar radiation does not originate from the
center of molecular clouds, but uniformly distributed in cells where stars
form. Therefore, the escape fractions of ionizing or \lya photons under this
model is likely a lower limit, since some stellar sources may be surrounded by
cold gas of higher volume filling factor than the cell average and suffer
additional absorption. However, such effects are rather difficult to include,
as the location where the stars are born is not followed in the simulations
with sub-grid multi-phase ISM model, and the distribution or geometry of
these surrounding cold phase gases is unknown. Our treatment essentially
assumes a single cold phase volume filling factor in each grid cell,
regardless of where stars are actually born.

Finally, to take into account dust destruction processes in the hot phase
medium, we model the dust to gas ratio as a function of the gas temperature,
\begin{equation}\\
r = r_1\left[1-\left(1-r_0\right)\exp\left(-\left[\frac{T_c}{T}\right]^a\right)\right],
\end{equation}
where $r_1$ represents the dust to gas ratio ignoring any dust destruction
processes, and $r_0$ gives a residual dust content relative to $r_1$ when $T
>> T_c$. 

In the cosmological applications we present below, $r_1$ is taken to
be the Milky-Way dust to gas ratio, $r_0=0.01$, $T_c=5\times10^5$~K, and
$a=1.5$. These parameters are chosen because the resulting infrared
dust emission spectra agree with observations for the modeled galaxies
reasonably well. 

For the molecular line transfer, we assume molecular H$_2$ number density to
be proportional to the neutral H density resulting from the ionization RT
calculation. For the cold clouds, essentially all of the H is assumed to be in
the molecular form. To calculate the molecular line emissivity of the cold
clouds, we assume they are in a state of pressure free infall, and treated in
the LVG approximation with the velocity gradient given as in
\citet{Goldreich1974}. The LVG approximation is originally motivated by a
model cloud under pressure free collapse. However it can also be applied to
turbulence driven or virialized clouds, with appropriate reinterpretation of
the velocity gradient. The kinetic temperature of the H$_2$ is assumed to be
proportional to the dust temperature resulting from the dust continuum RT
calculation. For simplicity, we choose the proportional constant to be unity
for the example below, assuming that the energy transfer between the molecular
gas and dust particles due to collision is efficient enough, even though such
an assumption might not be entirely appropriate at low gas densities. 

We also assume the trace molecular number density relative to the H$_2$ is constant,
and for CO, we choose the canonical value of $2\times 10^{-4}$ for the example
presented in the next section. The assumptions regarding molecular gas
temperature, H$_2$ and CO abundance fractions adopted here are the most
simplistic for the demonstration purpose. A more complex and
self-consistent model can be built by following the thermal balance and energy
exchange between dust and gas, and molecular abundance fractions
can be made dependent on the physical conditions of clouds
\citep{Narayanan2011}.

In addition to the multi-phase ISM model, \art employs an adaptive octree grid scheme, as described in \cite{Li2008}. Each cartesian cell is adaptively refined by dividing it into $2^3$ sub-cells until a predefined maximum refinement level is reached, or if the total number of particles in the cell reaches the threshold. After the grid is constructed, the gas properties are calculated in each cell. This adaptive grid enables \art to efficiently handle arbitrary geometry and a large dynamical range of gas density in the simulations.

\section{Applications of \art}
\label{sec:apps}

\begin{figure*}
\includegraphics[width=2.2in]{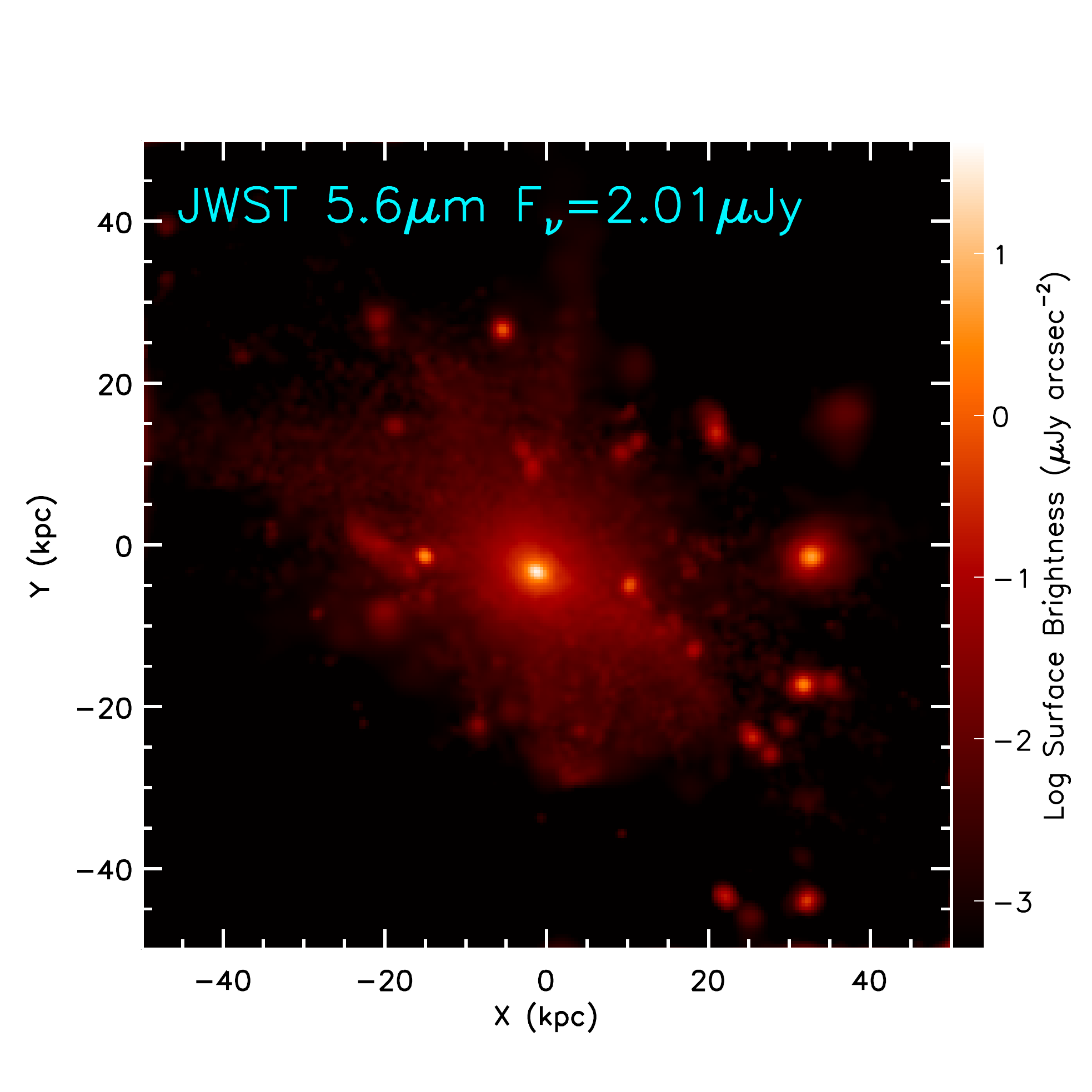}
\includegraphics[width=2.2in]{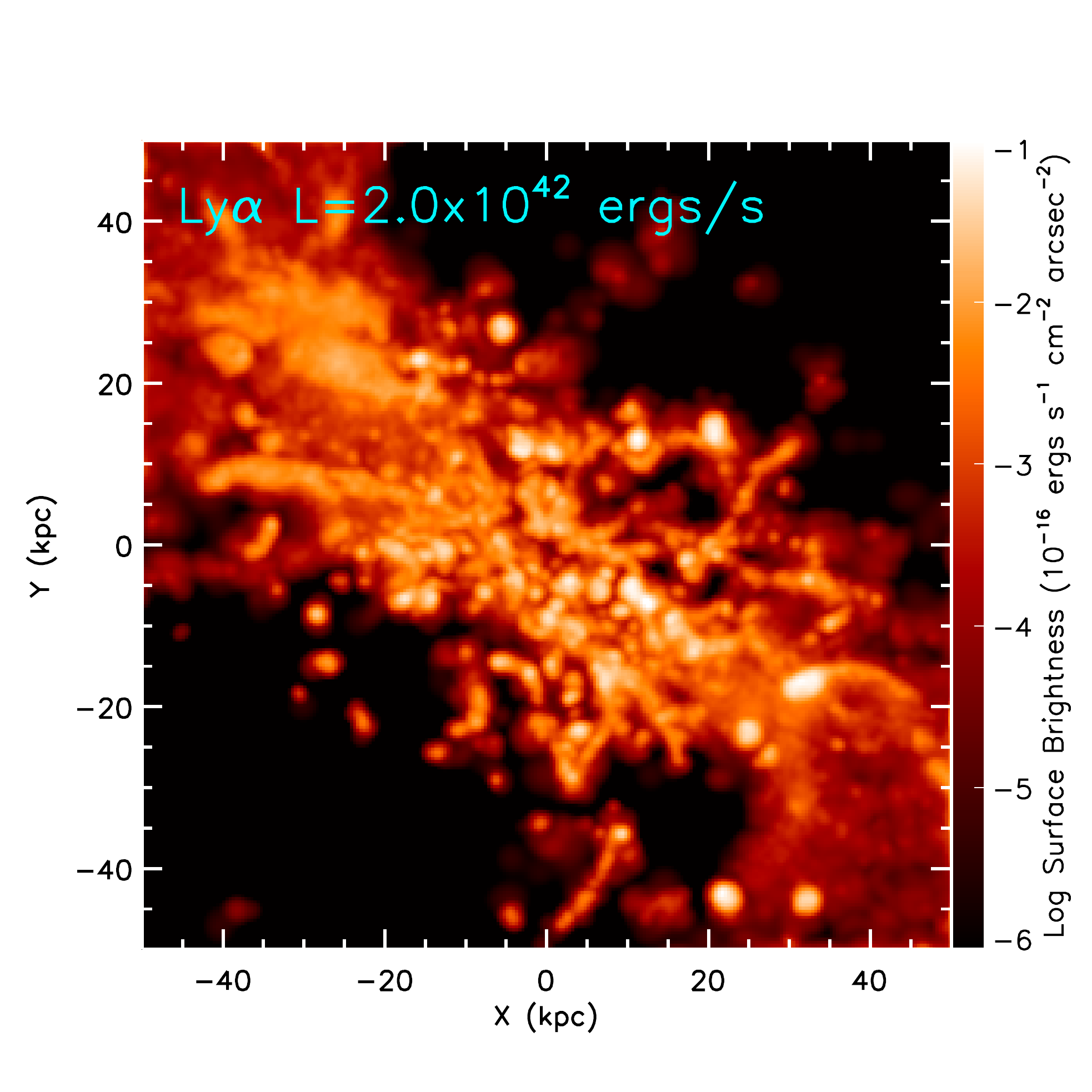}
\includegraphics[width=2.2in]{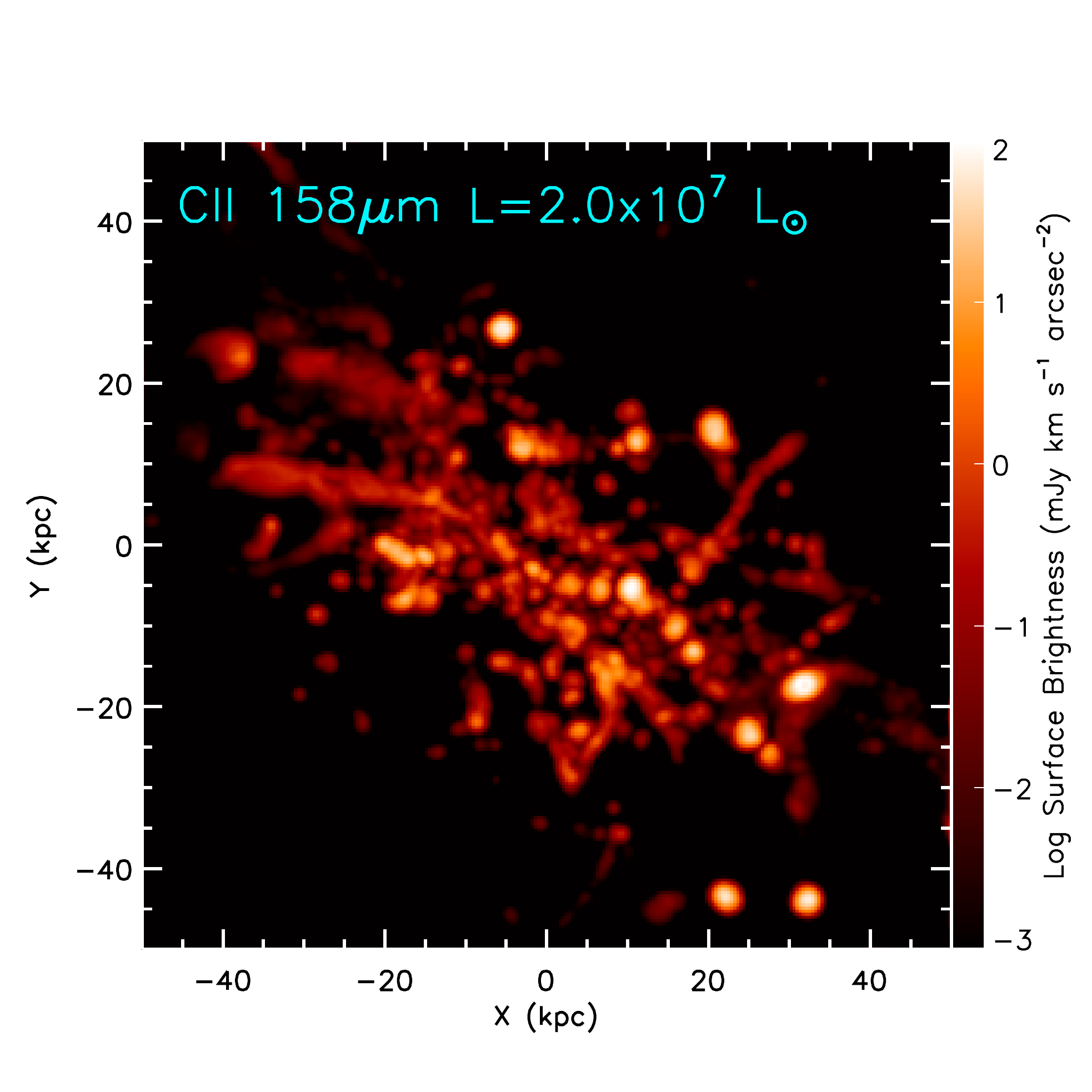} \\
\includegraphics[width=2.2in]{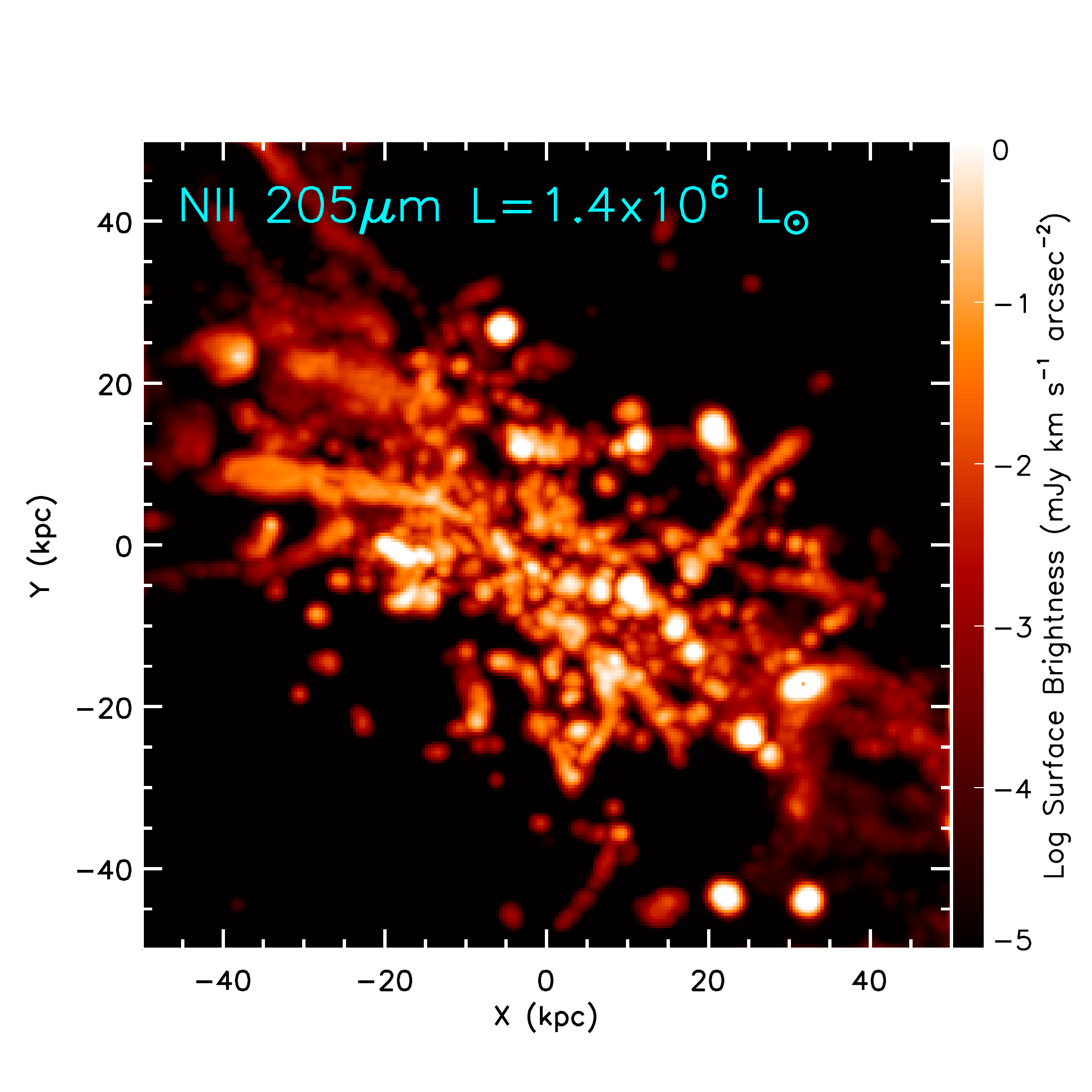}
\includegraphics[width=2.2in]{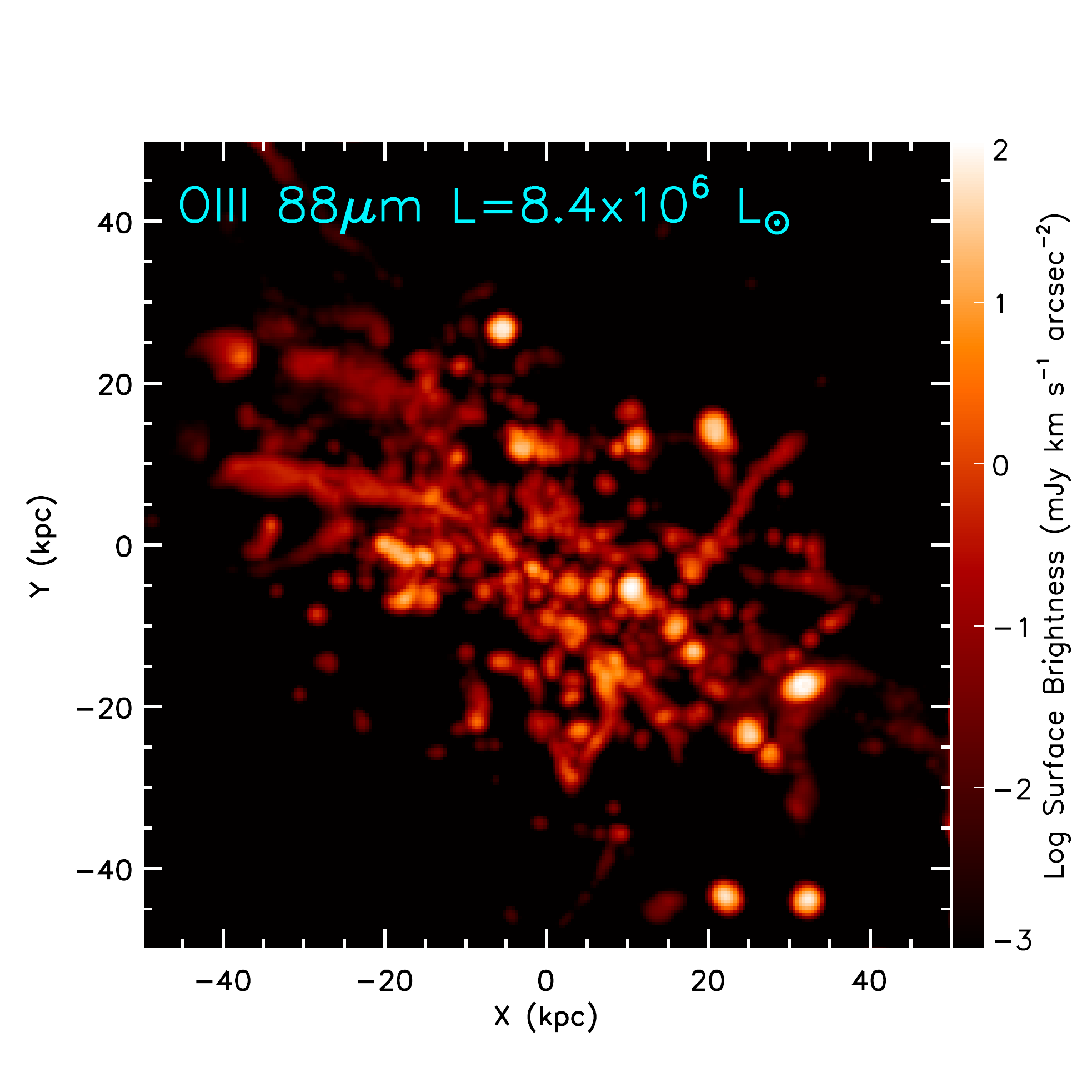}
\includegraphics[width=2.2in]{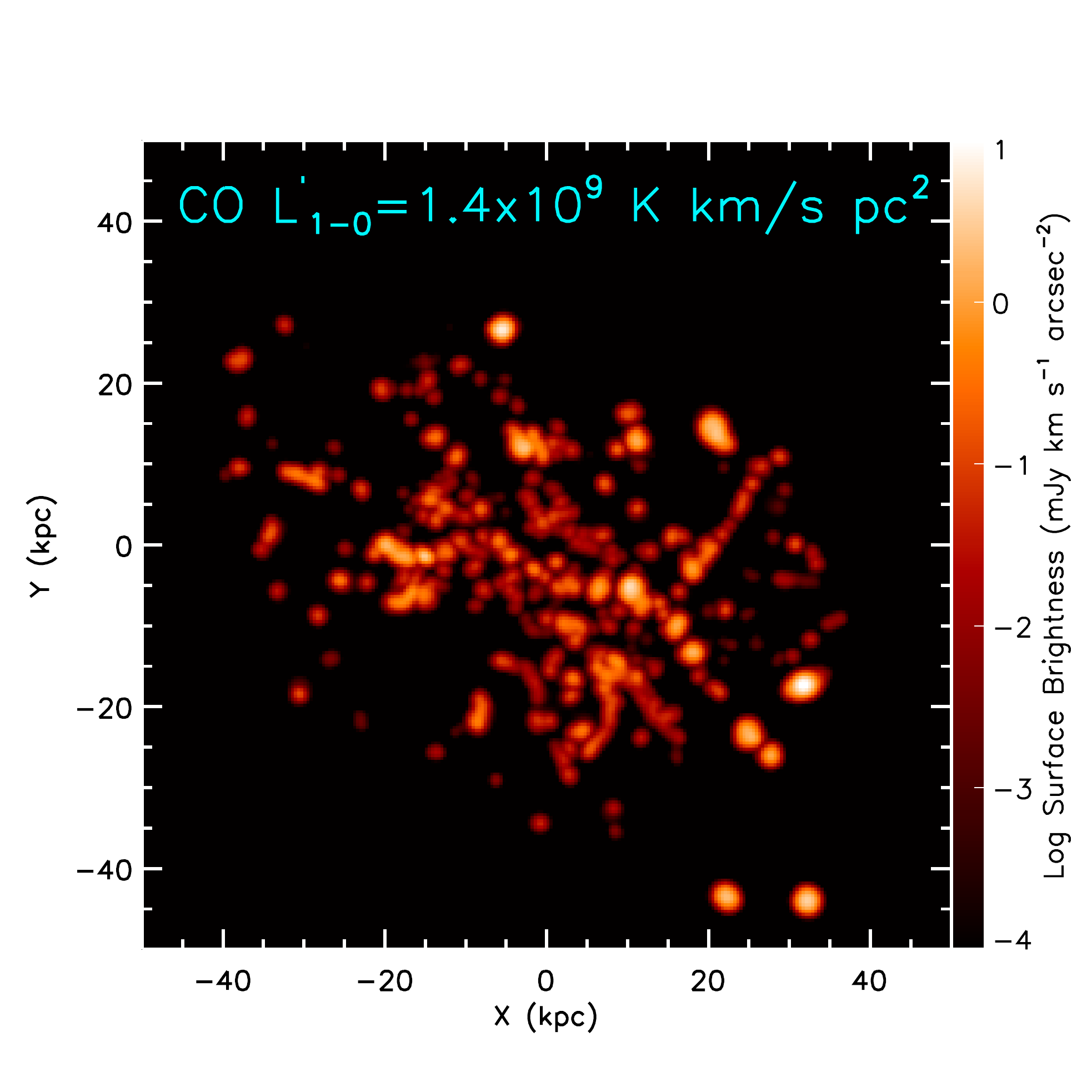}
\caption{Surface brightness map of the most massive galaxy at redshift z=3.1 from the Milky Way Simulation post processed by \art, in JWST F560W filter, \lya, \cii, \nii, \oiii, and CO line, respectively. The legend in the image indicates the total flux from the entire field of view of the JWST filter, and total line luminosities for the
  \lya \cii, \nii, \oiii and the CO 1-0 transition, respectively.}.
\label{fig:img}
\end{figure*}

The \art code can be applied to most hydrodynamics simulations to obtain multi-wavelength properties of the simulated system. There are two approaches for the post processing: individual objects for detailed properties of discrete targets, and whole snapshot for statistical properties of the entire system. To demonstrate these two approaches, we apply \art to two astrophysical simulations: the small-scale, zoom-in Milky Way Simulation~\citep{Zhu2016}, and the large-scale, full-box IllustrisTNG Simulation~\citep[][]{Pillepich2018b, Springel2018, Marinacci2018, Naiman2018, Nelson2018a, Nelson2019a, Nelson2019b, Pillepich2019}. 

For the first application, we obtain the output for each of the targeted galaxies, which contains the full spectral energy distribution (SED) in continuum  from X-ray to far-infrared, fine structure lines from atoms and ions such as H, He, C, N, O, and other elements, resonant scattering \lya line, and molecular lines such as CO, as well as images at different bands defined by filters of existing and future observatories such as HST and JWST. This data set is ideal to study of individual and statistical  properties of galaxies at different redshift, luminosity functions, dust and gas properties, photon escape fractions, and relations between photometric properties and galaxy properties such as galaxy mass, star formation rate (SFR) and metallicity.

For the second application, we obtain the output for the entire system, which is ideal to study the global properties of the simulation box, such as ionization and intensity mappings. In this example, we focus on the intensity mappings of emission lines \lya, \cii, \nii, \oiii, and CO. 

Since \art is a Monte Carlo code, we use photons to trace the continuum RT and the resonant scattering of the \lya photons, and we use rays to trace the non-LTE molecular and atomic line transfer. The number of continuum and \lya photons varies between $10^6$ to $10^8$ depending on the number of grid cells generated from the hydrodynamics snapshots. For the non-LTE molecular and atomic fine structure lines, each finest cell in the adaptive grids initially has 48 rays, and the number of rays increases by a factor of 4 for each iteration until the level population reaches convergence. For a snapshot with $\sim 1 \times10^6$ cells, the typical total number of rays is $\sim 5 \times 10^7$ for each of the line calculations.

In addition to the galaxy and gas properties from the hydrodynamics simulations, \art requires the following additional inputs and assumptions. (1)  The intrinsic stellar radiation is calculated with the stellar population synthesis code StarBurst99~\citep{Leitherer1999, Leitherer2010, Leitherer2014} on a grid of stellar age and metallicity. (2)  The black hole radiation is calculated using the accretion rate from the simulation assuming a double power law for spectral template of active galactic nuclei. (3)  For the continuum RT, we assume a Milky Way dust opacity curve \citep{Draine2003} and a dust-to-metal ratio of $10^{-2}$. (4) For CO line RT, a molecular fraction of $2\times 10^{-4}$ relative to H$_2$ is typically used, while the size of the molecular core of the cold clouds in our multi-phase ISM is calculated based on the local ionizing radiation flux. 

The computational cost varies with application as different astrophysical system has different contents such as the number of objects and the distribution of gas density and radiation sources.  Among the RT processes we consider, the CO molecular line calculation is typically the most time consuming module. Including 20 rotational levels in the statistical population equilibrium, and for a galaxy with $2 \times 10^6$ cells, the molecular RT computation takes about 500 cpu-hours on the state-of-the-art CyberLAMP supercomputer cluster at Penn State. 

For the Milky Way application, the largest galaxy at redshift z=3.1 has $5 \times 10^6$  cells, the \art calculations on this galaxy with all the modules (continuum, ionization, Lyman-alpha, atomic fine structure lines, and CO lines) required $\sim 20$ gigabytes RAM memory, and it took about 15 hours to run on 64 cores on the CyberLAMP cluster. For the IllustrisTNG application, the snapshot size is $\sim1.7$ terabytes, so we selected a slice of 5 Mpc of the simulation box for the line intensity mapping, and it took \art about 8 hours on 64 cores of the CyberLAMP cluster to calculate the lines. 

We emphasize that these applications are intended to serve as proof-of-concept examples of the new \art code in this work, an exploration of other assumptions and parameters such as supernovae dust models \citep[e.g.,][]{Li2008, Marassi2019},  and an in-depth analysis of the results or topics, will be presented in upcoming papers.

\subsection{Application to the Zoom-in Milky Way Simulation}
\label{sec:mw}

\begin{figure*}
\includegraphics[width=6.8in]{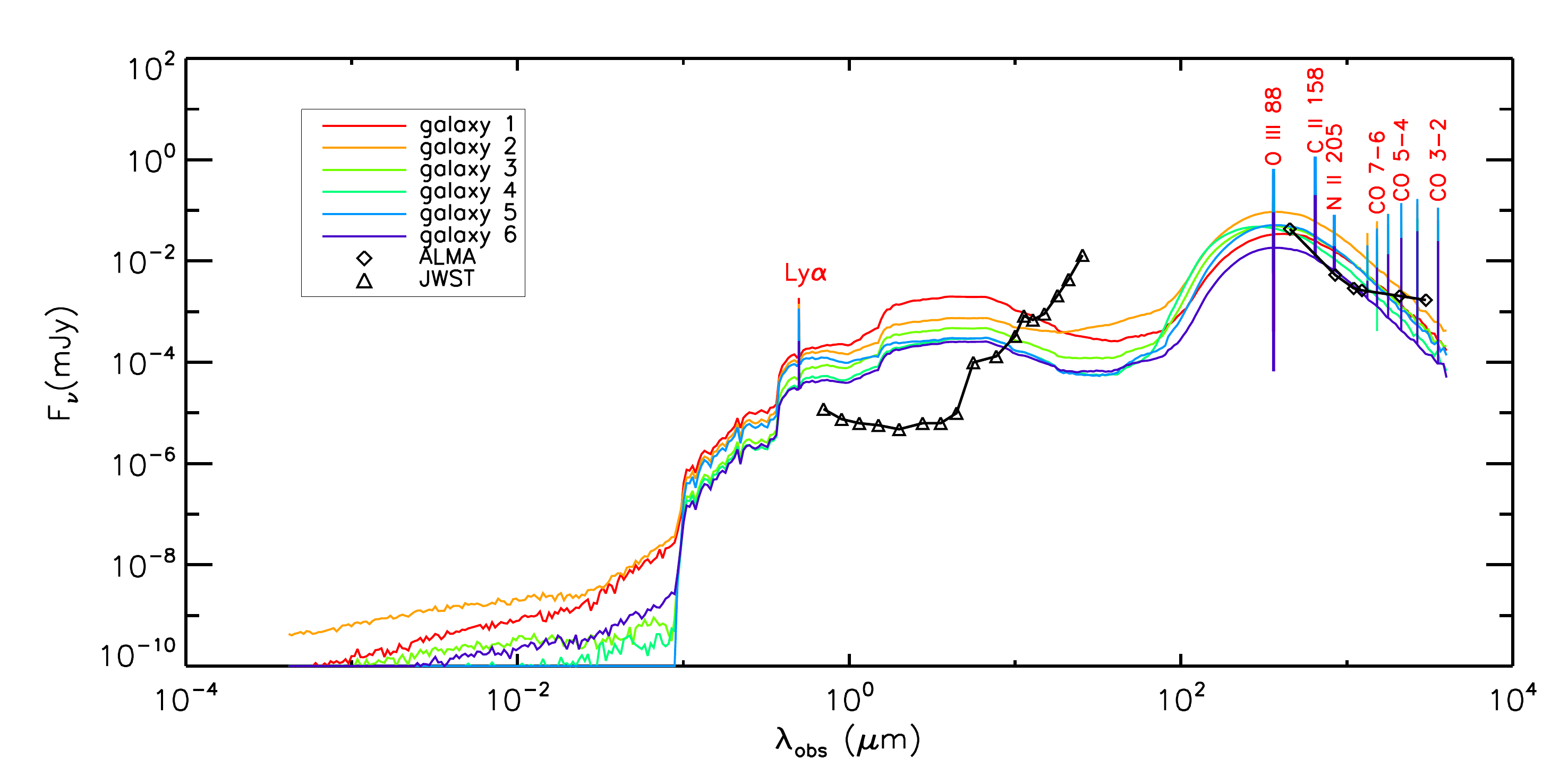}
\caption{Full spectral energy distribution with continuum and lines for the 6 most massive galaxies at redshift z=3.1 from the Milky Way Simulation post processed by \art. The detection limits of JWST and ALMA are shown for comparison. Note galaxy~\#6 in the plot has strong \oiii absorption line, while galaxy~\#4 has strong CO absorption lines.}
\label{fig:sed}
\end{figure*}

The Milky Way Simulation was described in detail in \cite{Zhu2016}, here we only give a brief summary of the simulation. It was a zoom-in hydrodynamics cosmological simulation run with the the Lagrangian Meshless Finite-Mass code {\Gizmo}  \citep{Hopkins2015}. A comprehensive list of physical processes was included in the simulation, including metal-dependent gas cooling, star formation, stellar evolution, chemical enrichment, and thermal and kinetic feedback from stars. This simulation did not include black holes.

The galaxy has a virial mass of $\sim 1.6\times10^{12}\, \Msun$. The mass resolutions of the simulation are  $\rm{m_{b}} = 4 \times10^5\, \rm{M_{\odot}}$ for gas and star particles  and  $\rm{m_{dm}} = 2.2 \times10^6\, \rm{M_{\odot}}$ for dark matter particles in the high resolution zoom-in region, and the gravitational softening length of gas particle is $\epsilon_{\rm {gas}}= 0.5\,\rm{kpc}$. It was run with the following cosmological parameters:  $\Omega_{\rm m}=0.25, \, \Omega_{\rm b}=0.04, \,  \Omega_\Lambda=0.75, \, \sigma_8=0.9, \, n_s=1$ and a Hubble constant  $\rm H=100\, h = 73\, \rm{km s^{-1} Mpc^{-1}}$. The simulation was evolved from redshift $z = 127$ to $z = 0$, and in each snapshot, galaxies are identified as ``groups" using the friend-of-friend group finder. 

As an example to demonstrate the capability of \art, we randomly choose the snapshot at redshift z=3.1 We first extract out the 50 largest halos from the snapshot, construct the adaptive grid, and run all RT components on these galaxies. The ionization RT is run with the equilibrium ionization option,and all RT runs use the sub-grid multi-phase ISM model. Molecular line RT is done for CO only.

\begin{figure}
\includegraphics[width=3.2in]{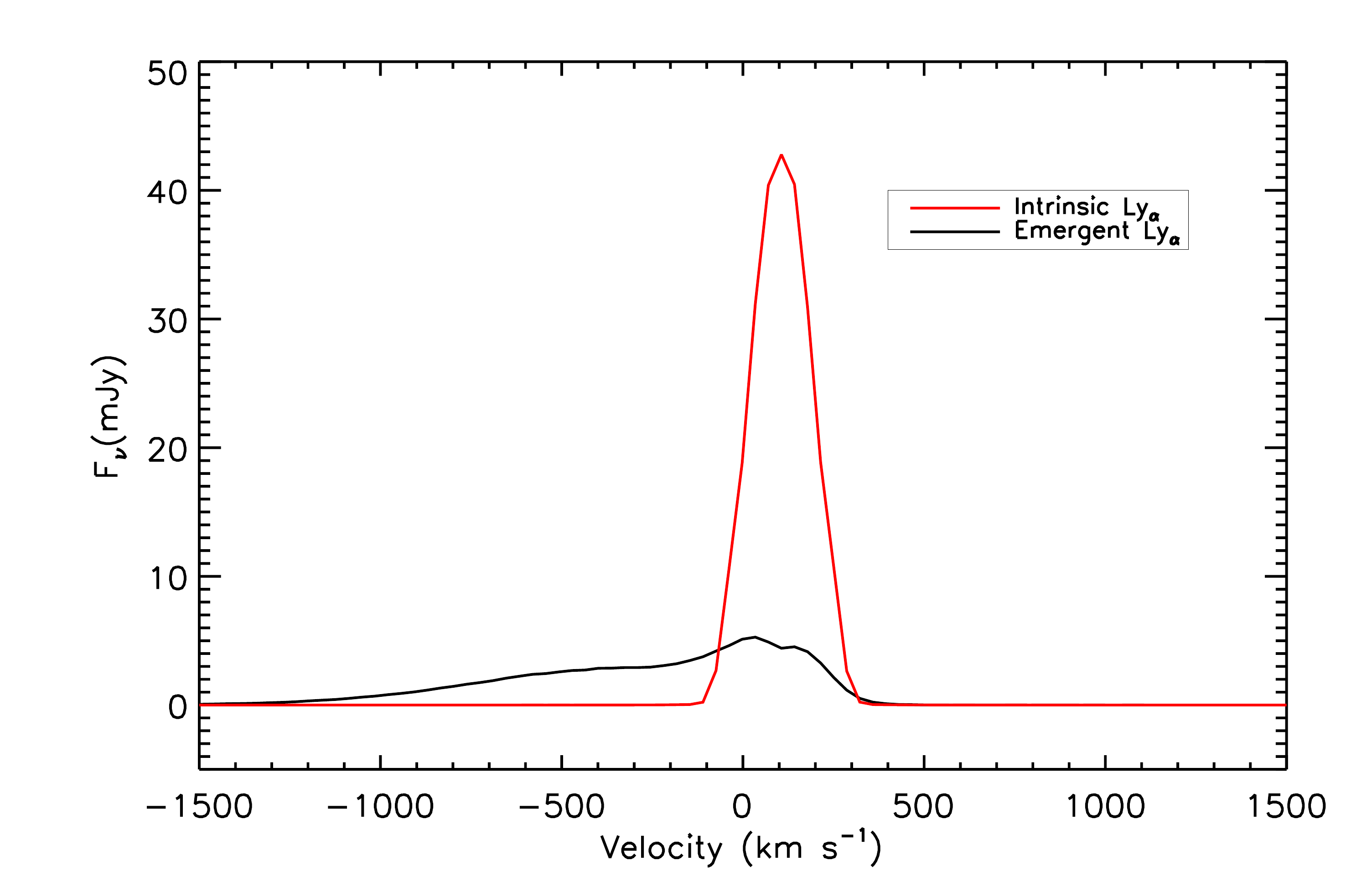}
\includegraphics[width=3.2in]{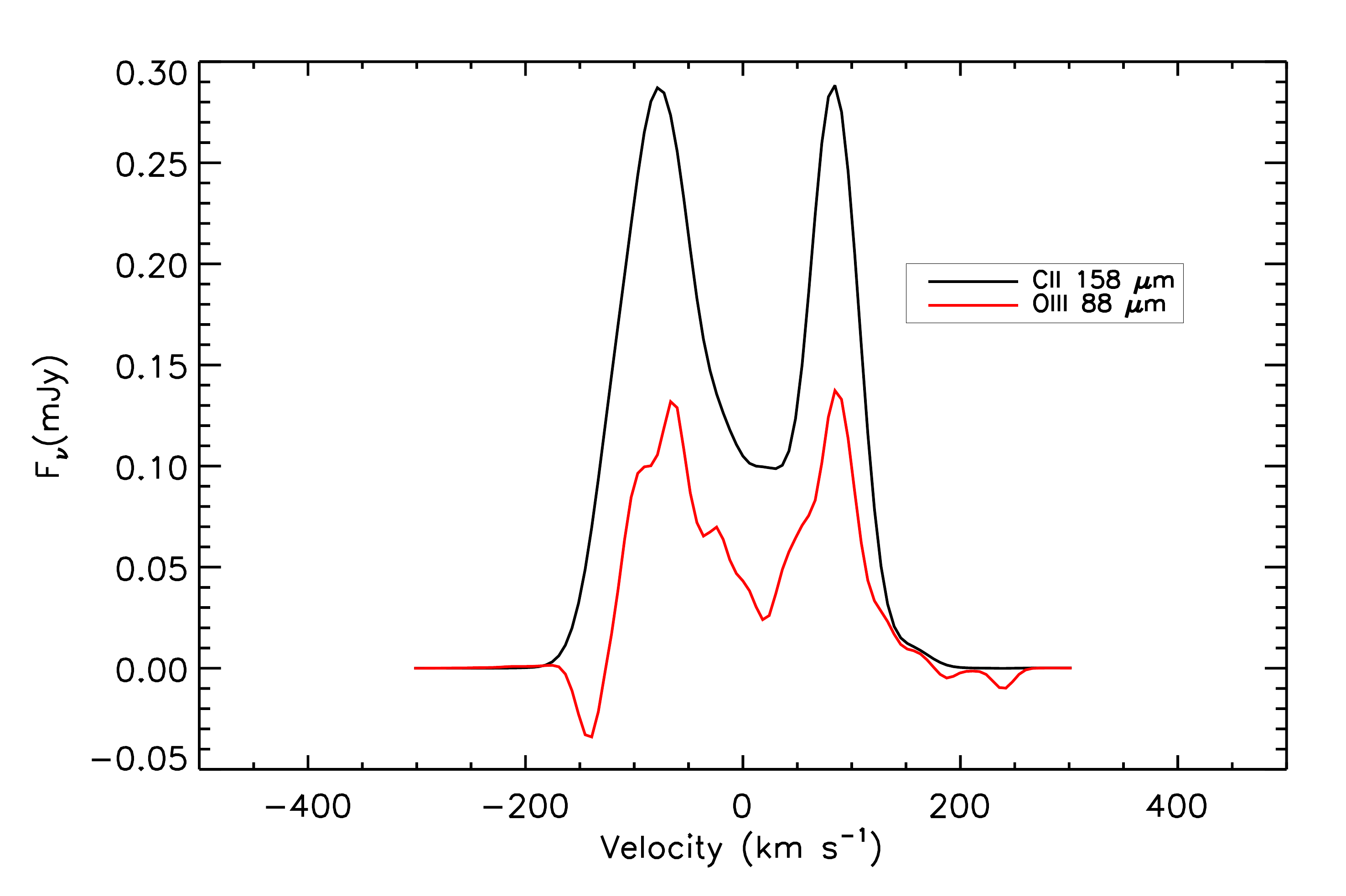}
\includegraphics[width=3.2in]{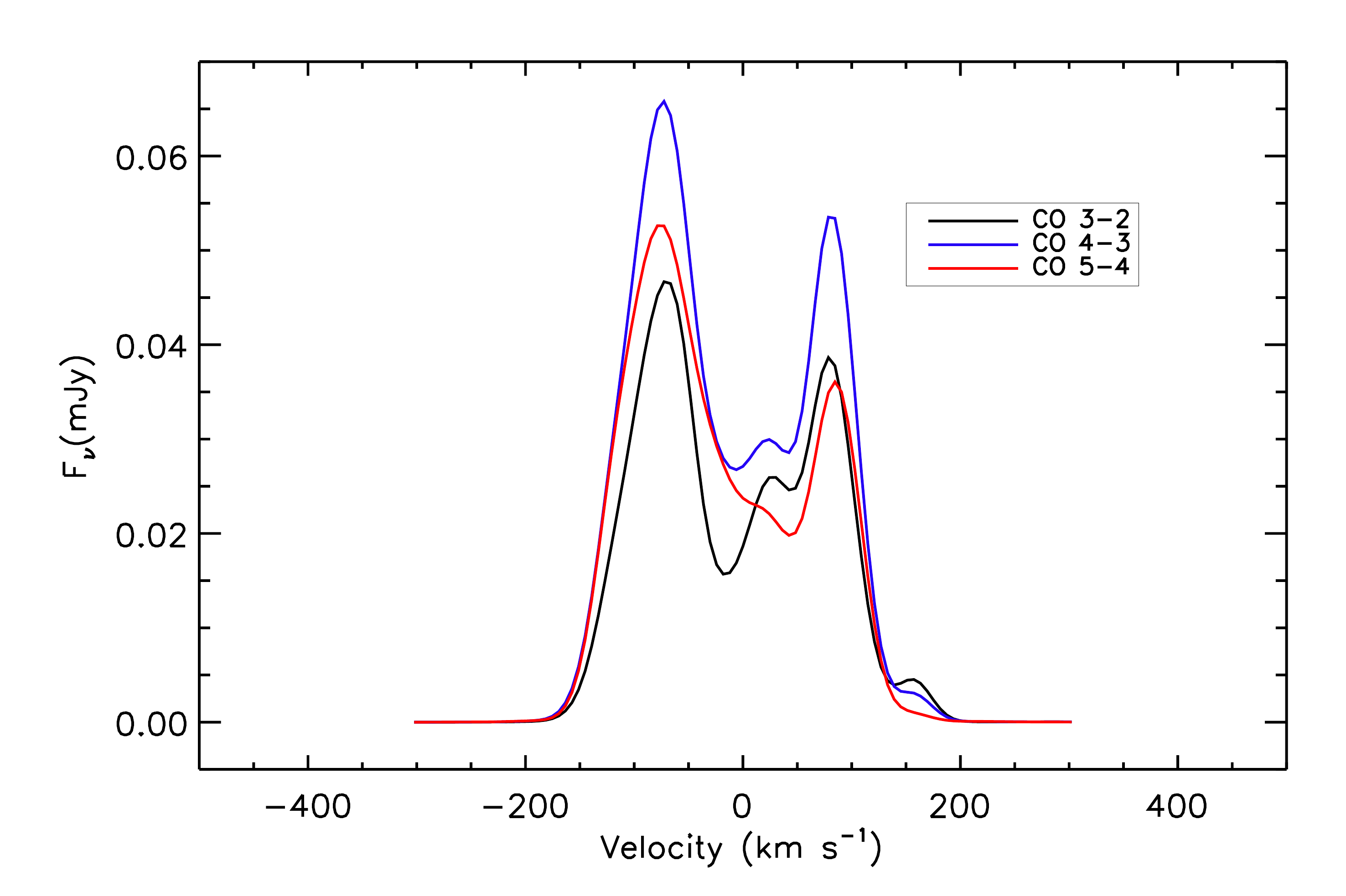}
\caption{Line profiles of resonant scattering line \lya, atomic fine structure lines \cii and \oiii, and molecular lines at transitions 5-4, 4-3, and 3-2, respectively, of the most massive galaxy at z=3.1 from the Milky Way Simulation.}
\label{fig:lines}
\end{figure}

\begin{figure}
\includegraphics[width=3.2in]{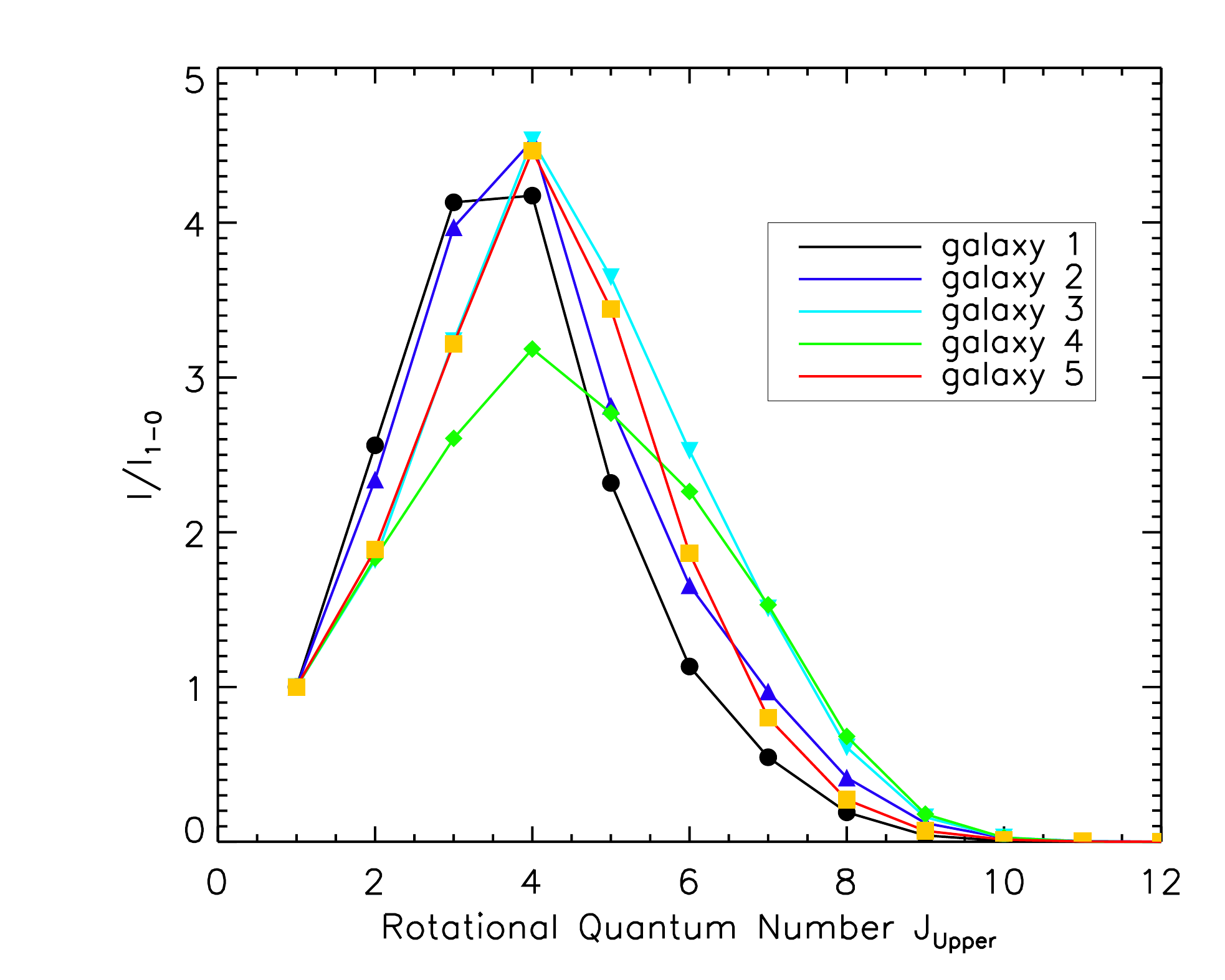}
\caption{The excitation ladder of CO transitions for the 5 most massive galaxies at z=3.1 from the Milky Way Simulation.}
\label{fig:ladder}
\end{figure}

\subsubsection{Multi-wavelength SEDs and Images of Galaxies}

The \art calculation can produce multi-wavelength spectral energy distribution, which includes both continuum and emission (or absorption) lines, and images of galaxies and ISM, as demonstrated in the following figures. Figure~\ref{fig:img} shows the images of the most massive galaxy as observed in the JWST 5.6~$\mu$m bands, \lya resonant line, atomic fine structure lines \cii, \nii and  \oiii, and the molecular line CO $J=$1-0 transition, respectively.  Figure~\ref{fig:sed} shows the overall SEDs, which include atomic and molecular lines and a continuum spanning 7 orders of magnitude in wavelength, of the 6 most massive galaxies at redshift z=3.1 of the Milky Way Simulation. The galaxies in this sample appear to be star-forming galaxies with strong lines of \lya, \oiii, \cii, \nii and CO,  with some having strong \oiii or CO absorption lines due to high gas density in the galaxies.

\subsubsection{Emission Line Properties of Galaxies}

\begin{figure}
\includegraphics[width=3.2in]{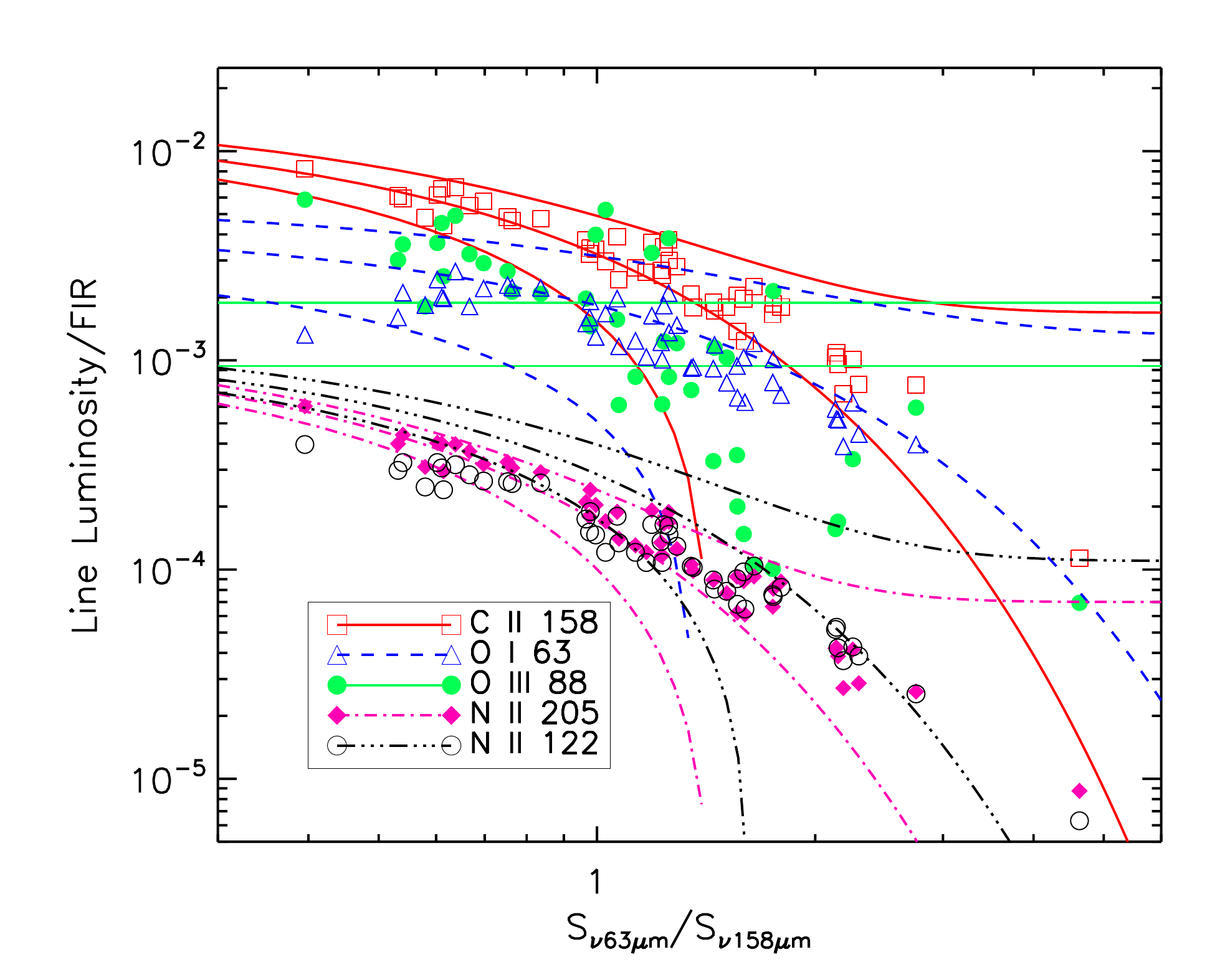}
\caption{The line luminosity of \cii 158~$\mu$m, \oi 63~$\mu$m,
  \oiii 88~$\mu$m, \nii 122~$\mu$m, and \nii 205~$\mu$m relative to the FIR
  luminosity between 8 and 1000~$\mu$m, as a function of the continuum flux
  ratio at 63 and 158 $\mu$m, $S_{\nu63\mu m}$/$S_{\nu158\mu m}$. The colored circles are the 50 most massive galaxies from the Milky Way Simulation, while the colored lines indicate the observed range of the relevant line from \citet{DiazSantos2017}. }
\label{fig:fslines}
\end{figure}

\begin{figure}
\includegraphics[width=3.2in]{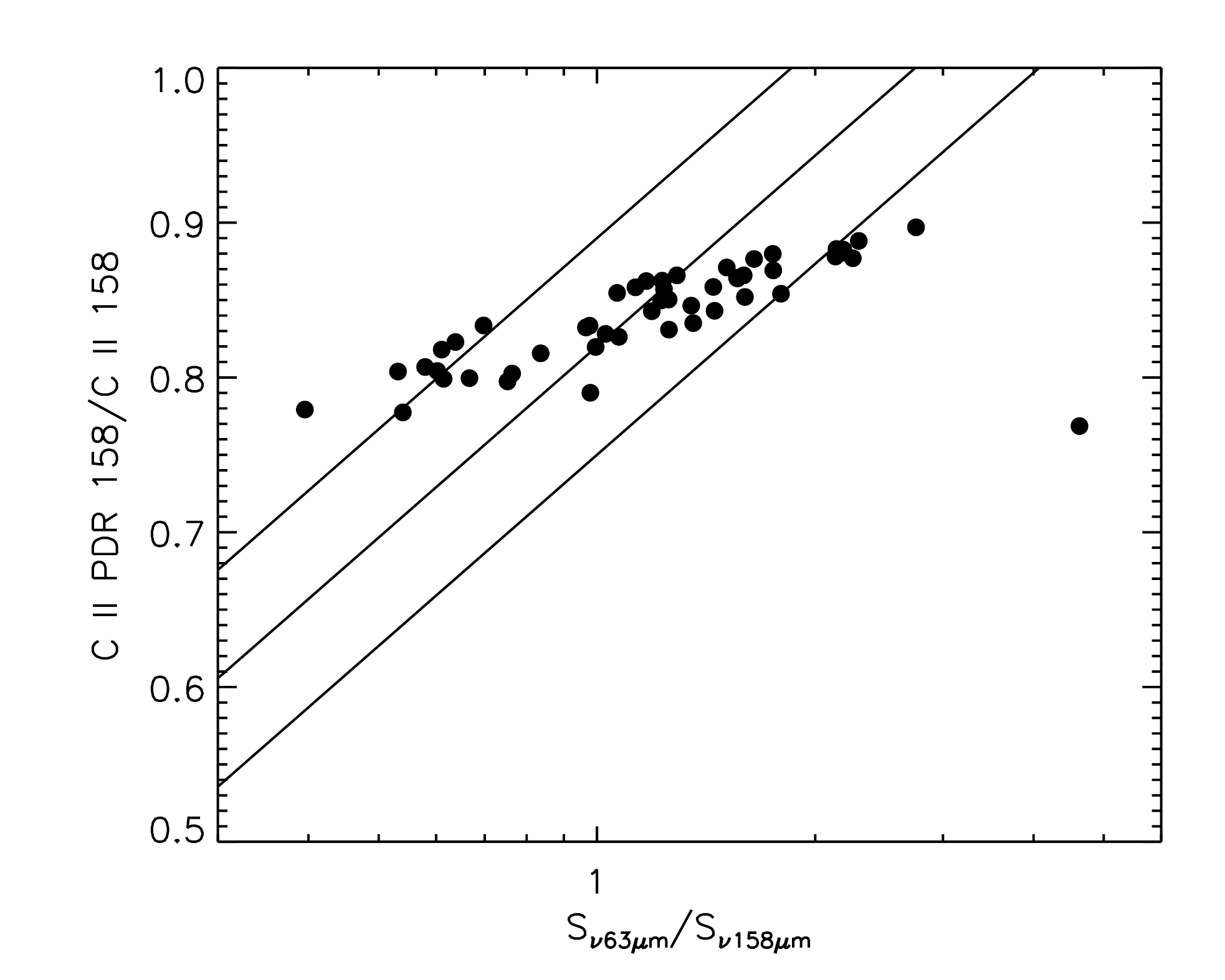}
\caption{The {\cii} emission from PDR relative to the total as a function of the
  continuum flux ratio at 63 and 158 $\mu$m for the 50 most massive galaxies from the Milky Way Simulation post processed by \art. The three straight lines
  represent the observed correlation and its scatter from \citet{DiazSantos2017}.}
\label{fig:ciir}
\end{figure}

The resonant scattering \lya line, and the atomic fine structure lines such as \cii and \oiii, and molecular lines such as CO are powerful lines to probe galaxies. Figure~\ref{fig:lines} shows the line profiles of  \lya,  \cii and \oiii, and molecular CO lines at transitions 5-4, 4-3, and 3-2 of the most massive galaxy at z=3.1. 
Figure~\ref{fig:ladder} showed the excitation ladder $\rm{I_{J - J-1}}/\rm{I_{1-0}}$ for the 5 most massive halos, which indicates the peak excitation is at $\rm J=4$. 

The emission lines survey of nearby infrared luminous galaxies by \citep{DiazSantos2017} shows an emission line deficit phenomenon in which the luminosity ratio between emission lines of  \cii, \nii, \oii and \oiii, and FIR emission decreases as dust temperature increases. 

The fine structure atomic emission lines of \cii, \nii,  \oii and \oiii are computed for the 50 most massive halos, and the line luminosities
relative the FIR luminosity  as a function of the continuum flux ratio at 63 and 158 $\mu$m, $S_{\nu63\mu m}$/$S_{\nu158\mu m}$, 
are plotted in Figure~\ref{fig:fslines}. The calculated luminosity ratios show the so called
emission line deficit effect, and are generally consistent with the
observations by \citet{DiazSantos2017}.

In an attempt to understand the origin of the \cii emission, \citet{DiazSantos2017} used the  \nii 220~$\mu$m emission as a proxy by  to divide the \cii 158~$\mu$m emission into two parts, one arising from the photodissociation region (PDR), and one arising from the ionized medium. The ionization emission is assumed to be 3 times that of the \nii 220~$\mu$m luminosity using photoionization calculations as a guide. With such a prescription, the \cii luminosity from PDR relative to the total is found to correlate with the dust temperature indicated by the flux ratio $S_{\nu63\mu m}$/$S_{\nu158\mu m}$.   

In Figure~\ref{fig:ciir} we compare this relation between our calculations and the observation. The slope of the calculated relation is somewhat smaller than the observed. However, given the larger scatter in the observed correlation, the calculation and the observation are still in broad agreement.

\subsubsection{Statistical Properties and Correlations}

\begin{figure}
\includegraphics[width=3.2in]{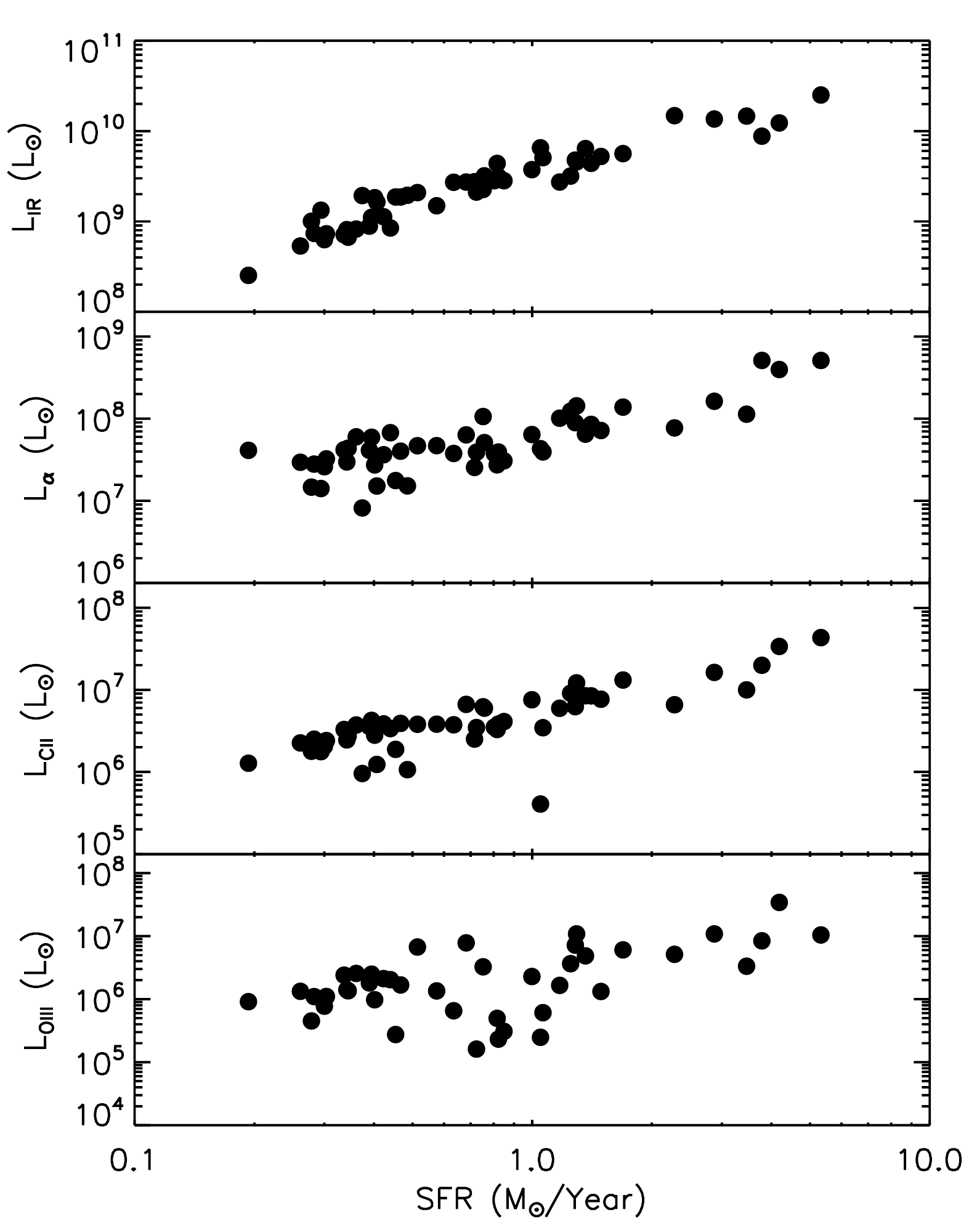}
\caption{Correlations between luminosities of FIR, \lya, \cii and \oiii and star formation rate of the 50 most massive galaxies at z=3.1 from the Milky Way Simulation. }
\label{fig:lum_sfr}
\end{figure}

\begin{figure}
\includegraphics[width=3.2in]{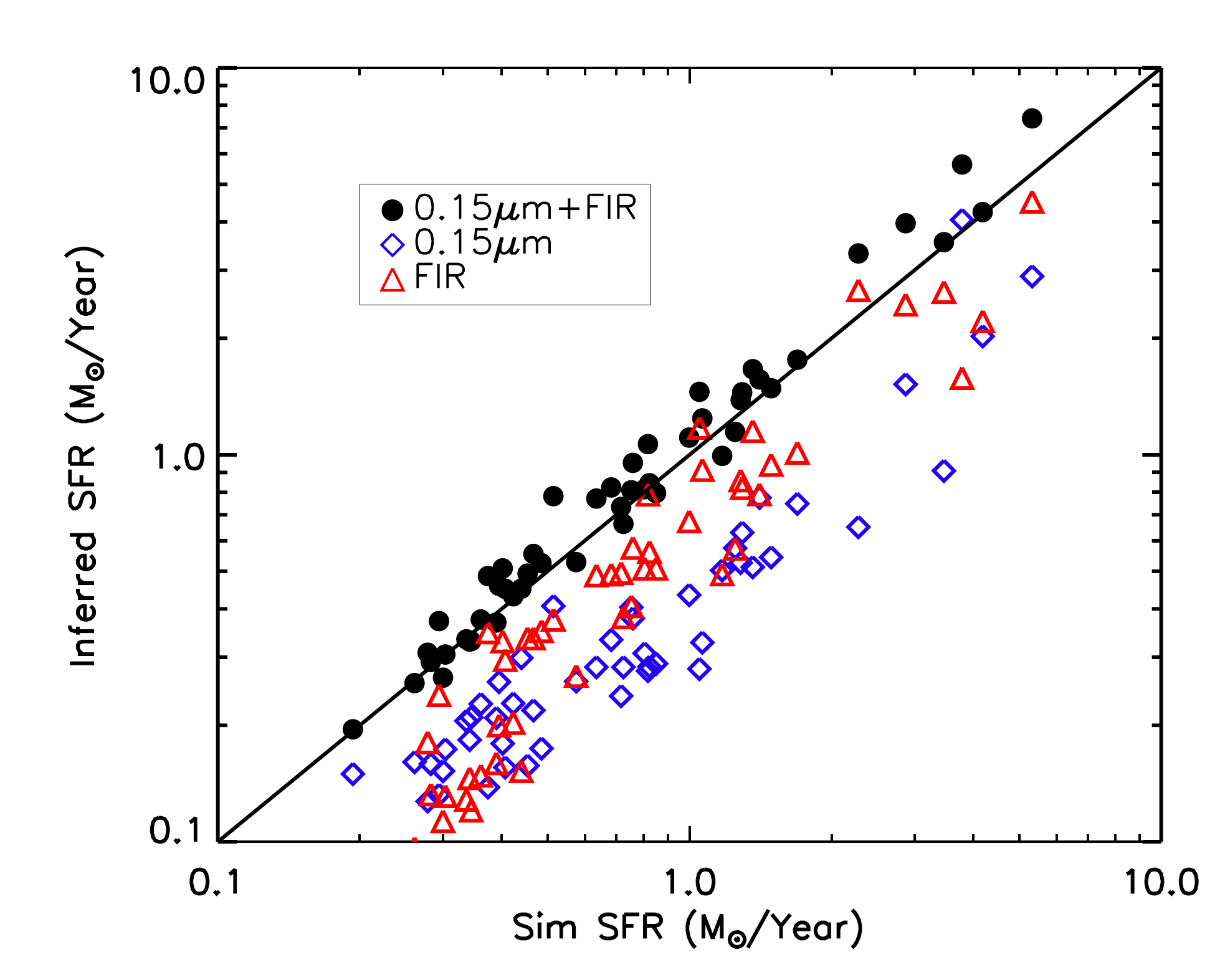}
\caption{Calibration of SFR indicators based on the UV (green diamond) and FIR (red triangle) luminosities of the 50 most massive galaxies at z=3.1 from the Milky Way Simulation, and a linear combination of the two (black dot). The black line is the diagonal line for a perfect match of SFR from the two indicators.}
\label{fig:sfr_uvir}
\end{figure}

\begin{figure}
\includegraphics[width=3.2in]{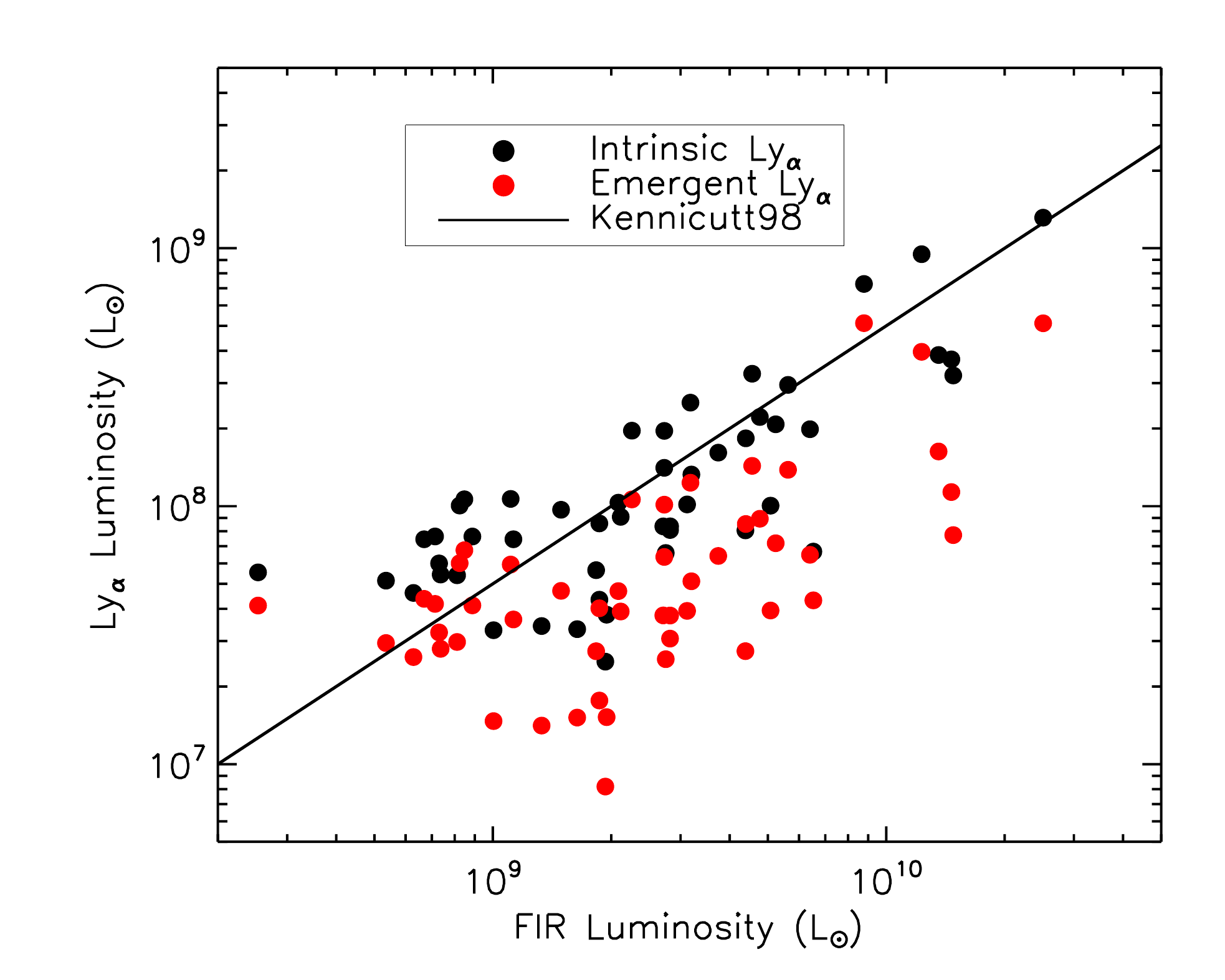}
\includegraphics[width=3.2in]{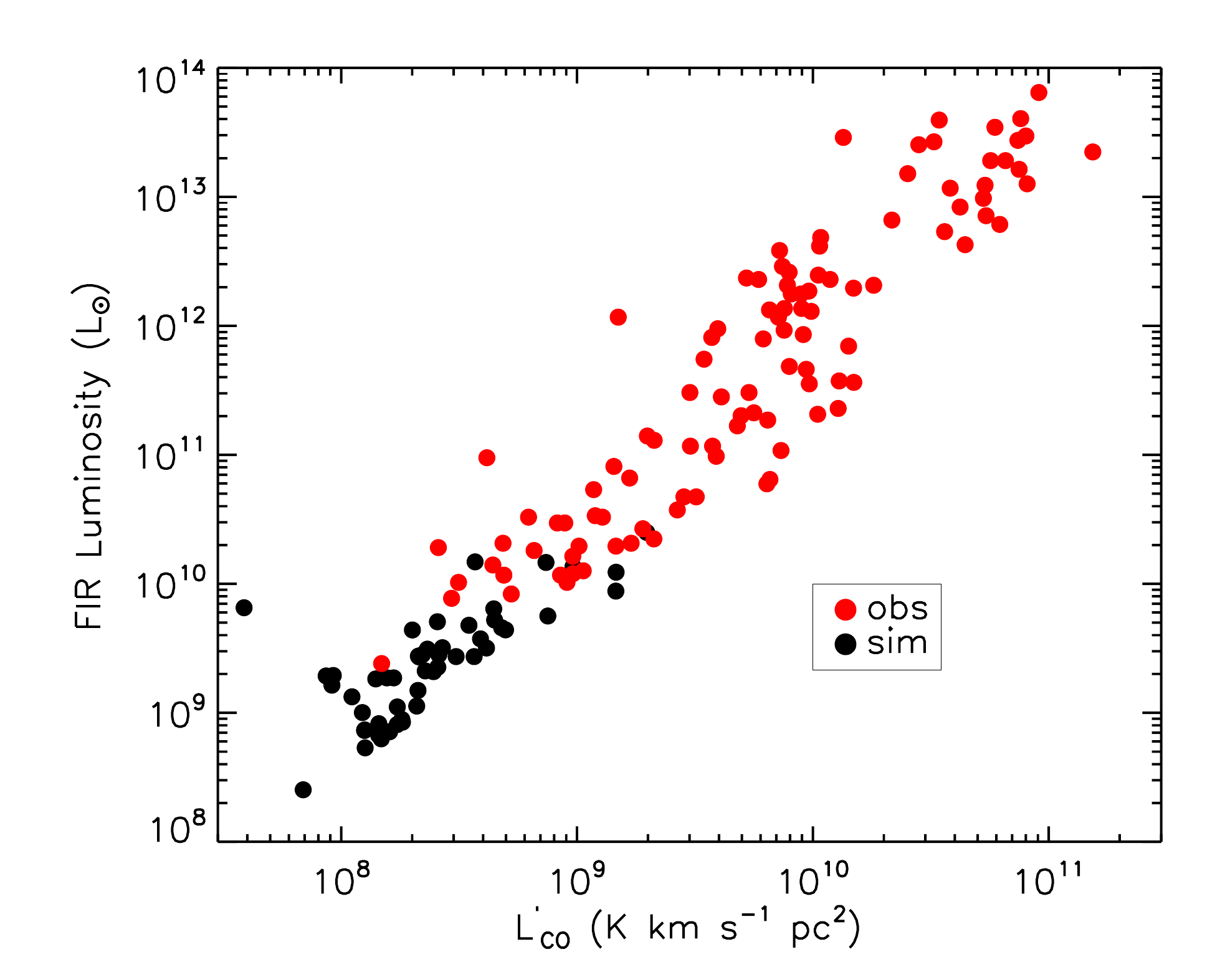}
\caption{Correlation between \lya and FIR luminosities (top), and  between CO  and FIR  luminosities (bottom) of the 50 most massive galaxies at z=3.1 from the Milky Way Simulation, compared with the scaling relation of  \citet{Kennicutt1998} (top), and the observations of \citet{Solomon2005} (bottom).}
\label{fig:lyairco}
\end{figure}

\begin{figure}
\includegraphics[width=3.2in]{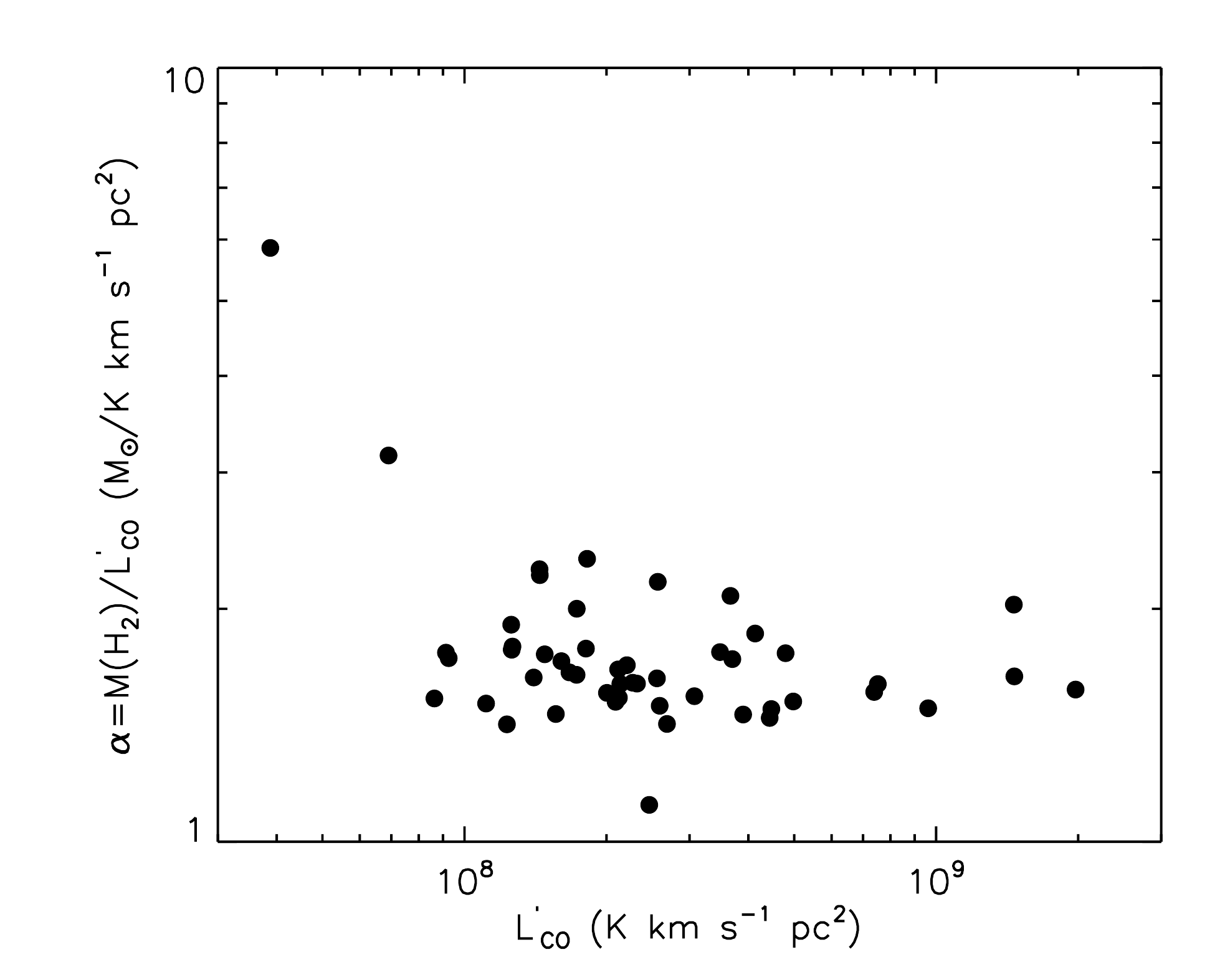}
\caption{The relation between CO-to-H$_2$ mass conversion factor $\alpha$  and CO luminosity $\rm{L^{\prime}_{\rm CO}(1-0)}$ of the 50
  most massive galaxies at z=3.1 from the Milky Way Simulation. }
\label{fig:aco}
\end{figure}

\begin{figure}
\includegraphics[width=3.2in]{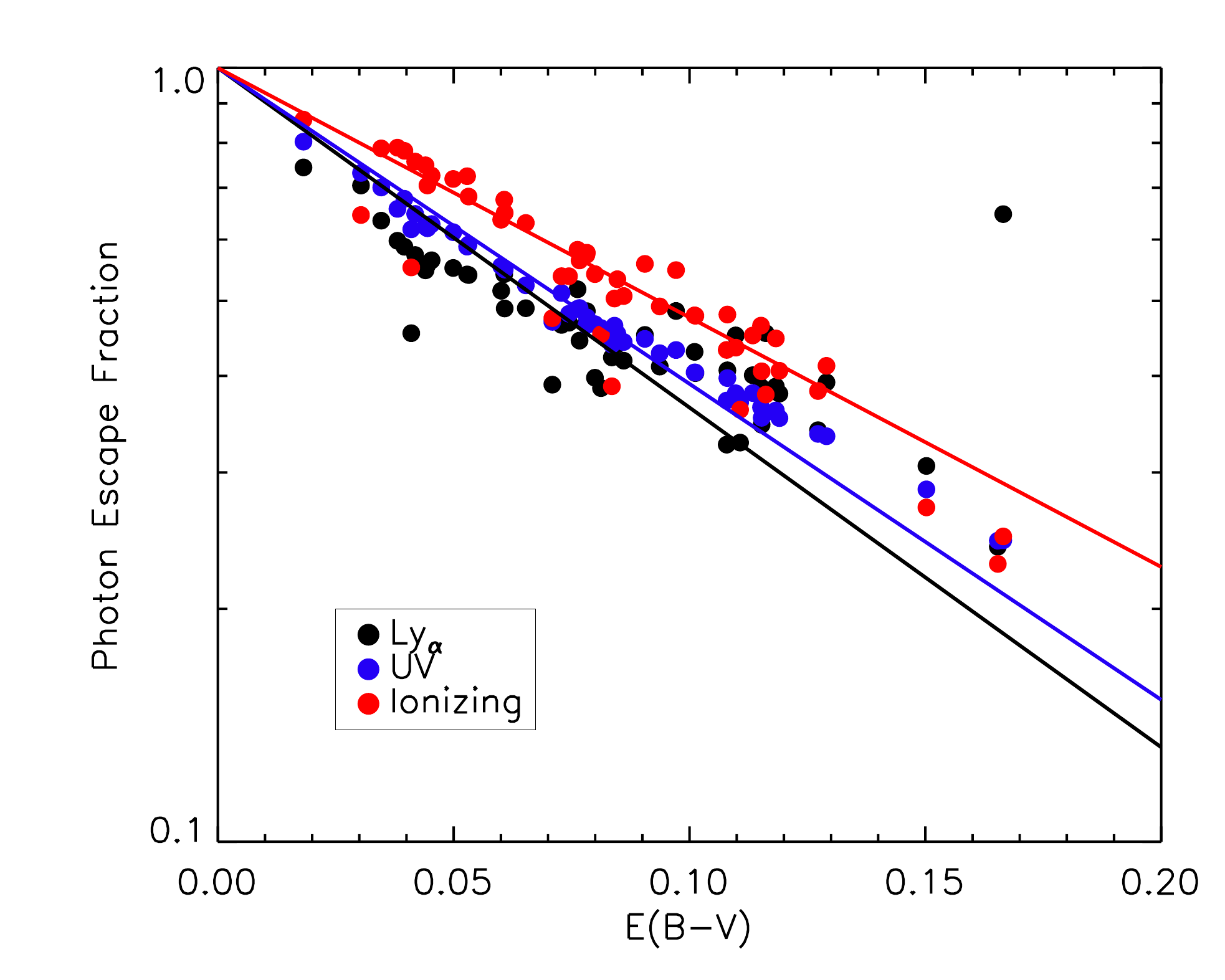}
\caption{Correlation between escape fraction of the ionizing, UV (0.15~$\mu$m),  
  and \lya photons from the 50 most massive galaxies at z=3.1 from the Milky Way Simulation and the $B-V$ color excess $E(B-V)$. The lines are fittings for the respective photons.}
\label{fig:fesc}
\end{figure}

\begin{figure}
\includegraphics[width=3.2in]{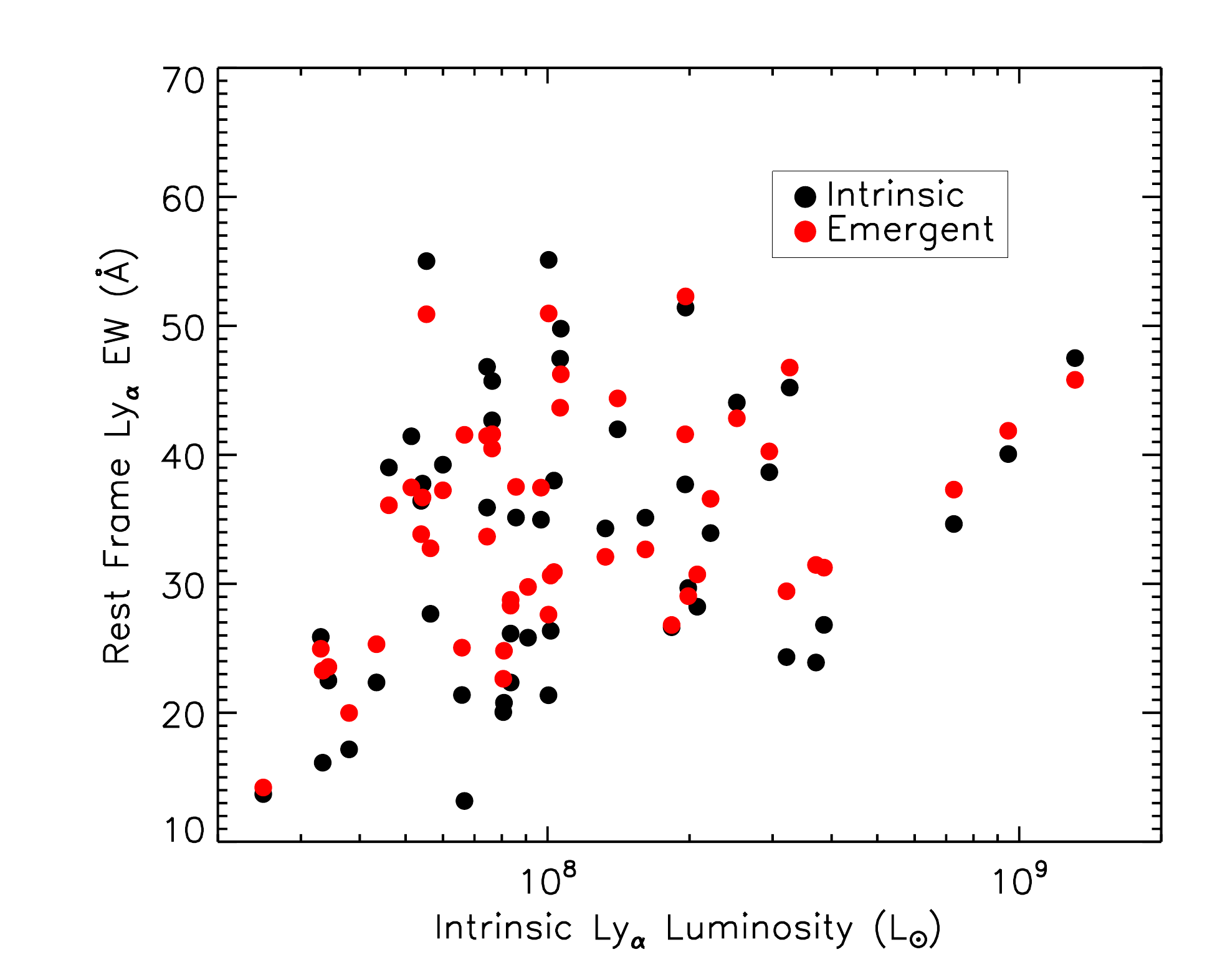}
\caption{Intrinsic and emergent rest-frame equivalent widths of \lya lines of the 50 most massive galaxies at z=3.1 from the Milky Way Simulation post processed by \art.}
\label{fig:lyaew}
\end{figure}

\begin{figure*}
\includegraphics[width=2.2in]{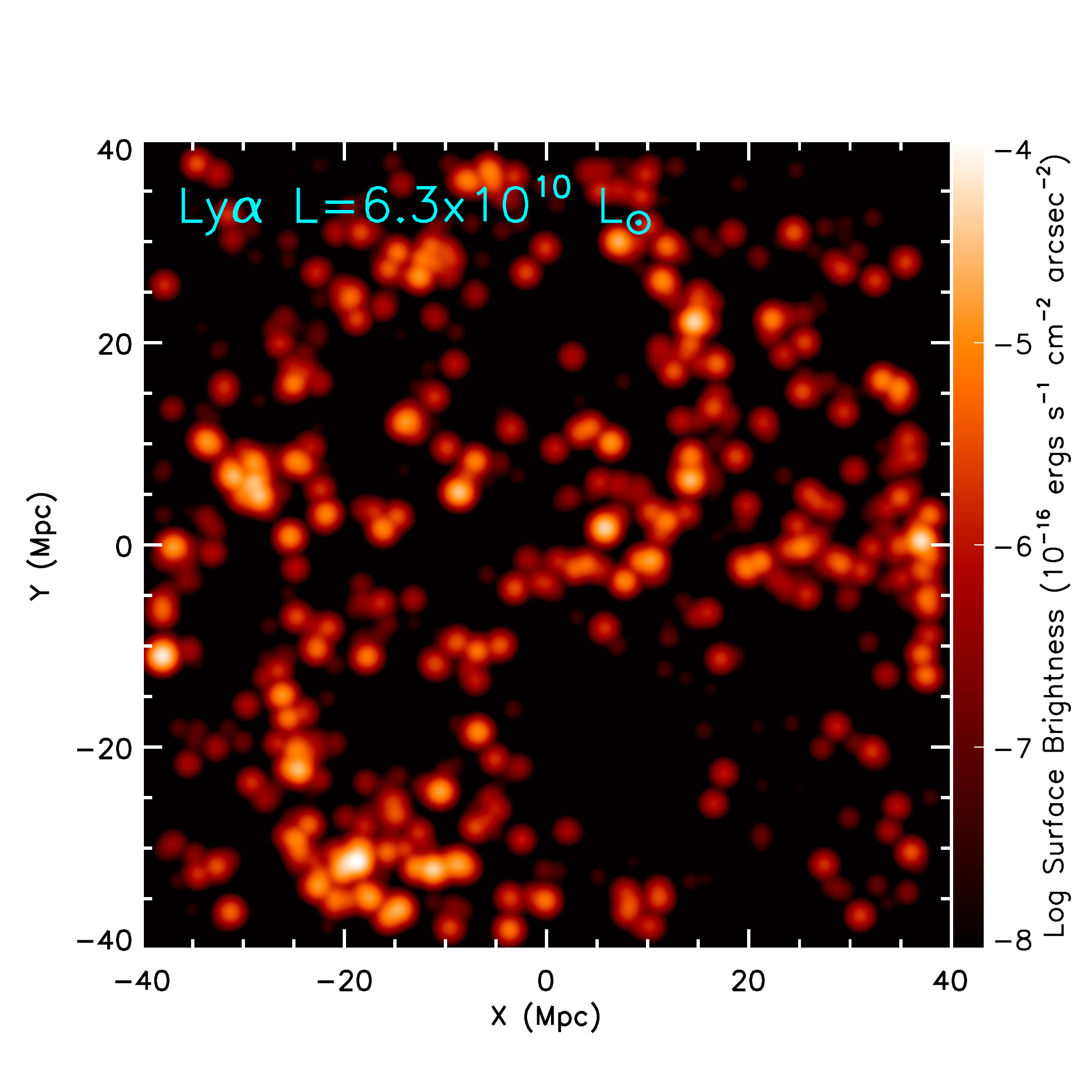}
\includegraphics[width=2.2in]{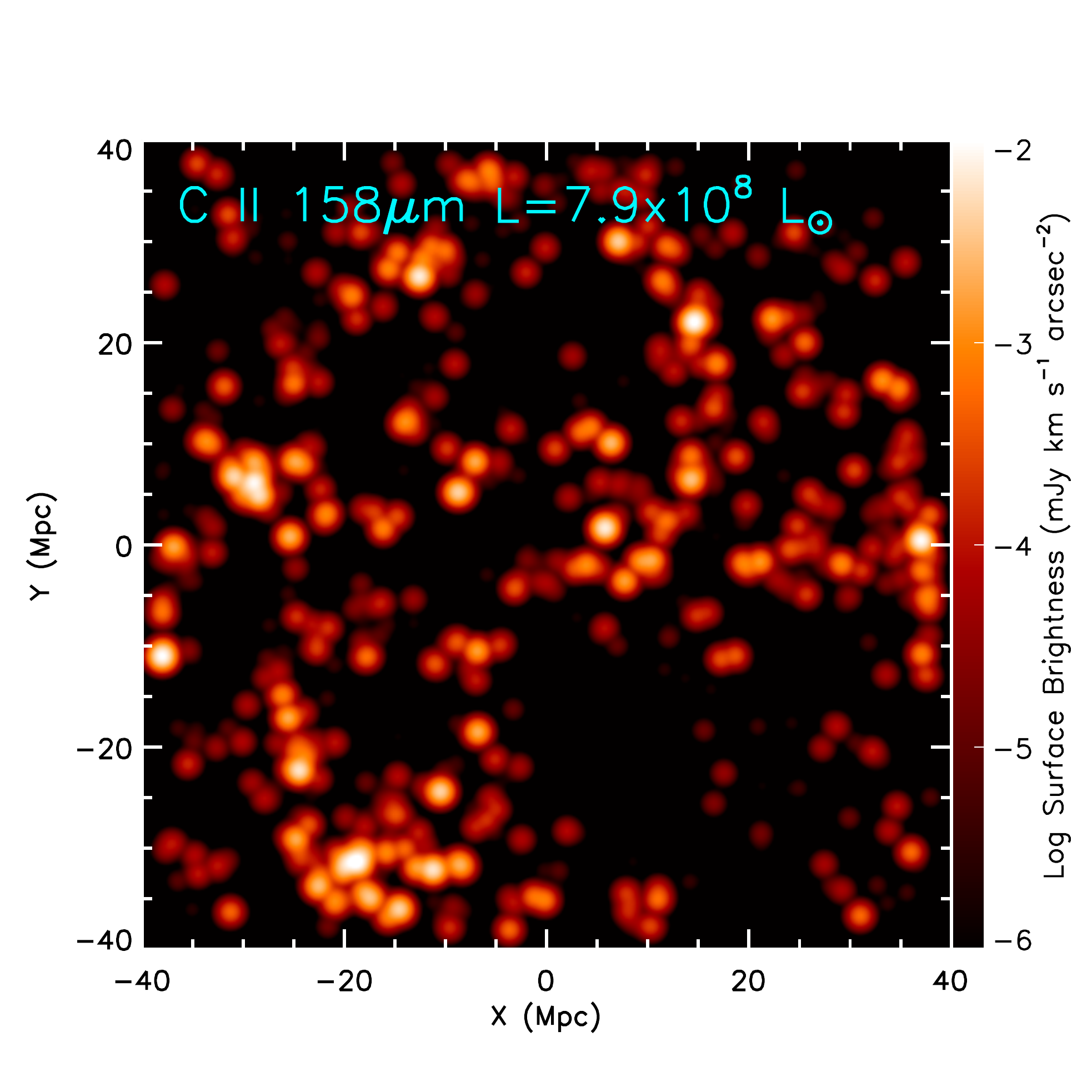} 
\includegraphics[width=2.2in]{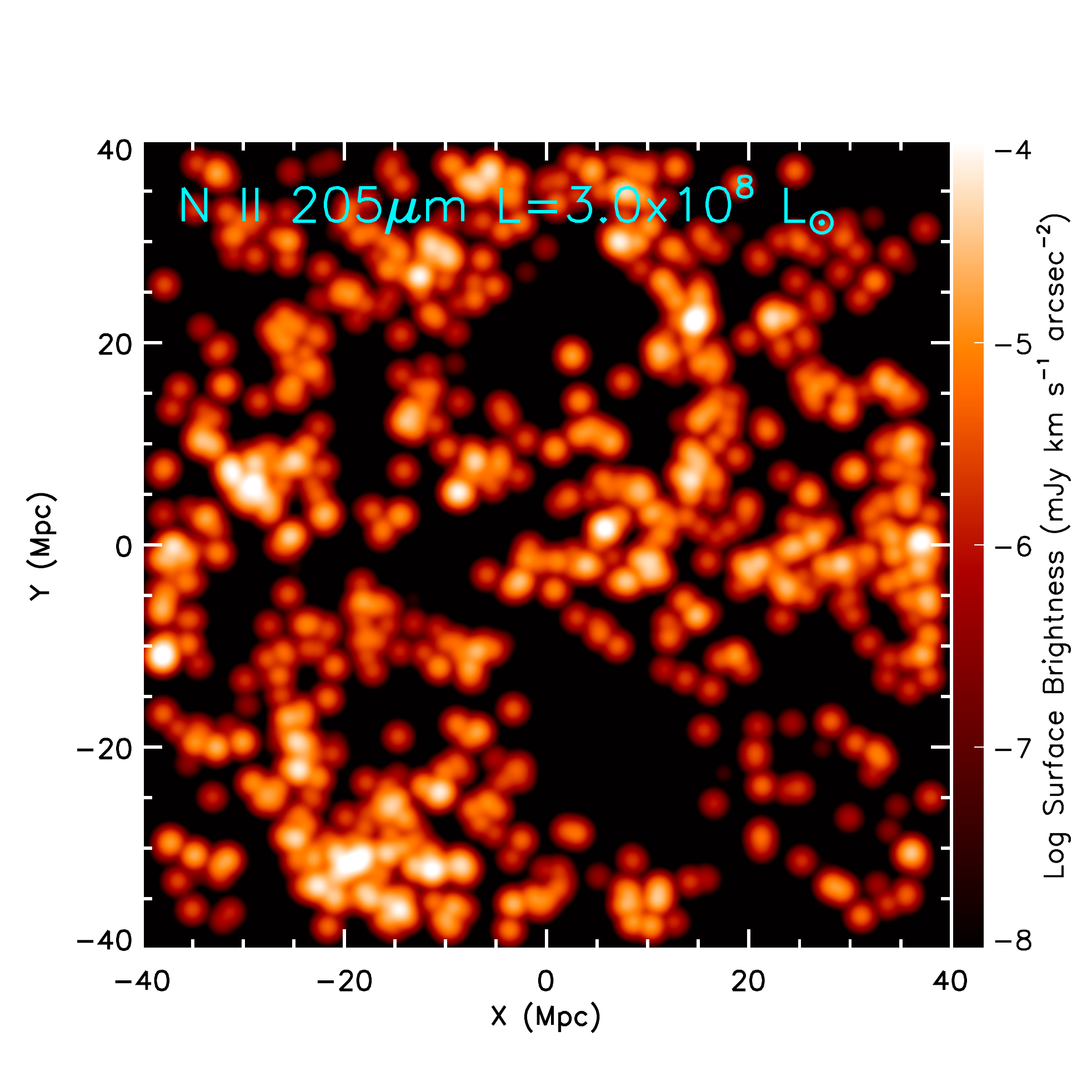} \\
\includegraphics[width=2.2in]{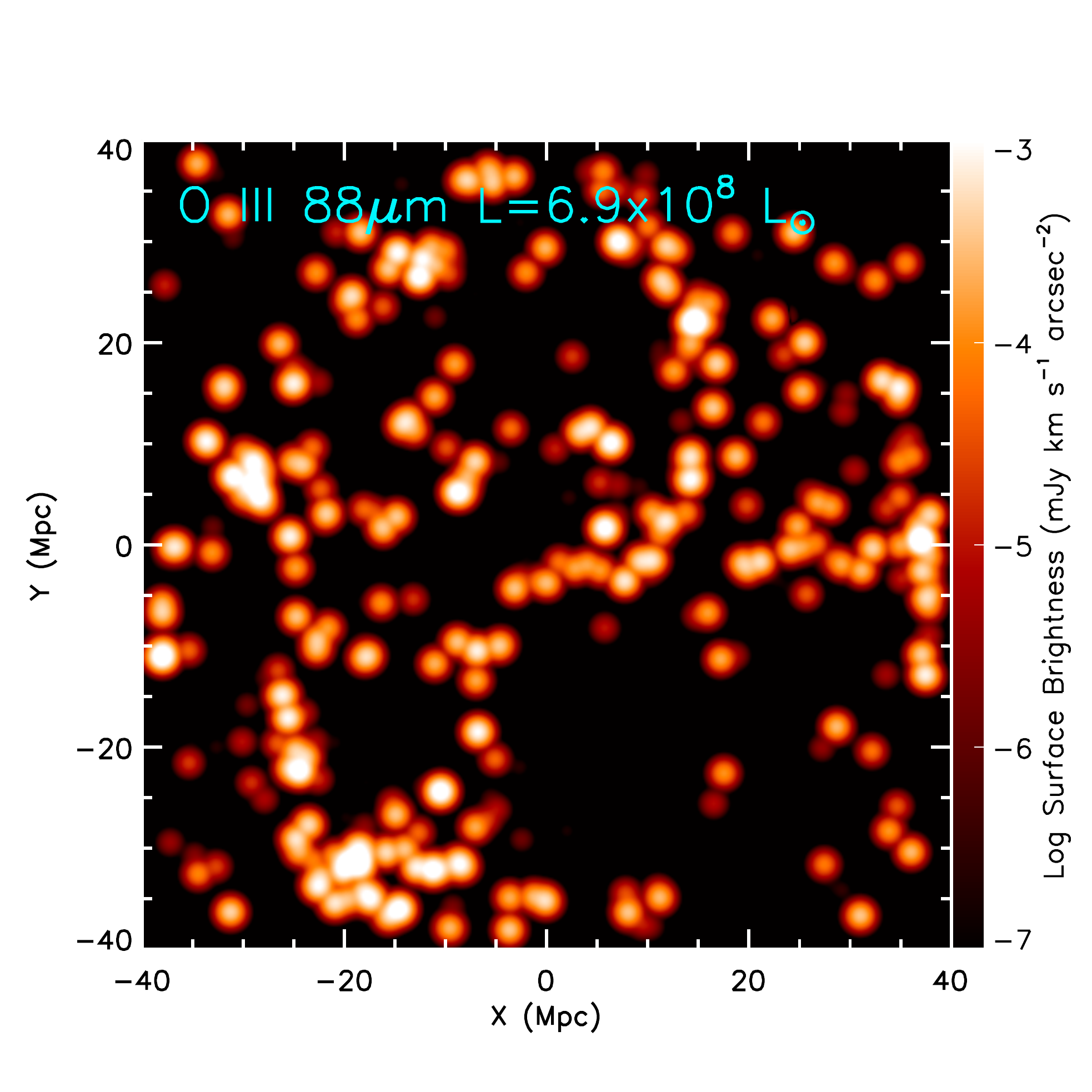} 
\includegraphics[width=2.2in]{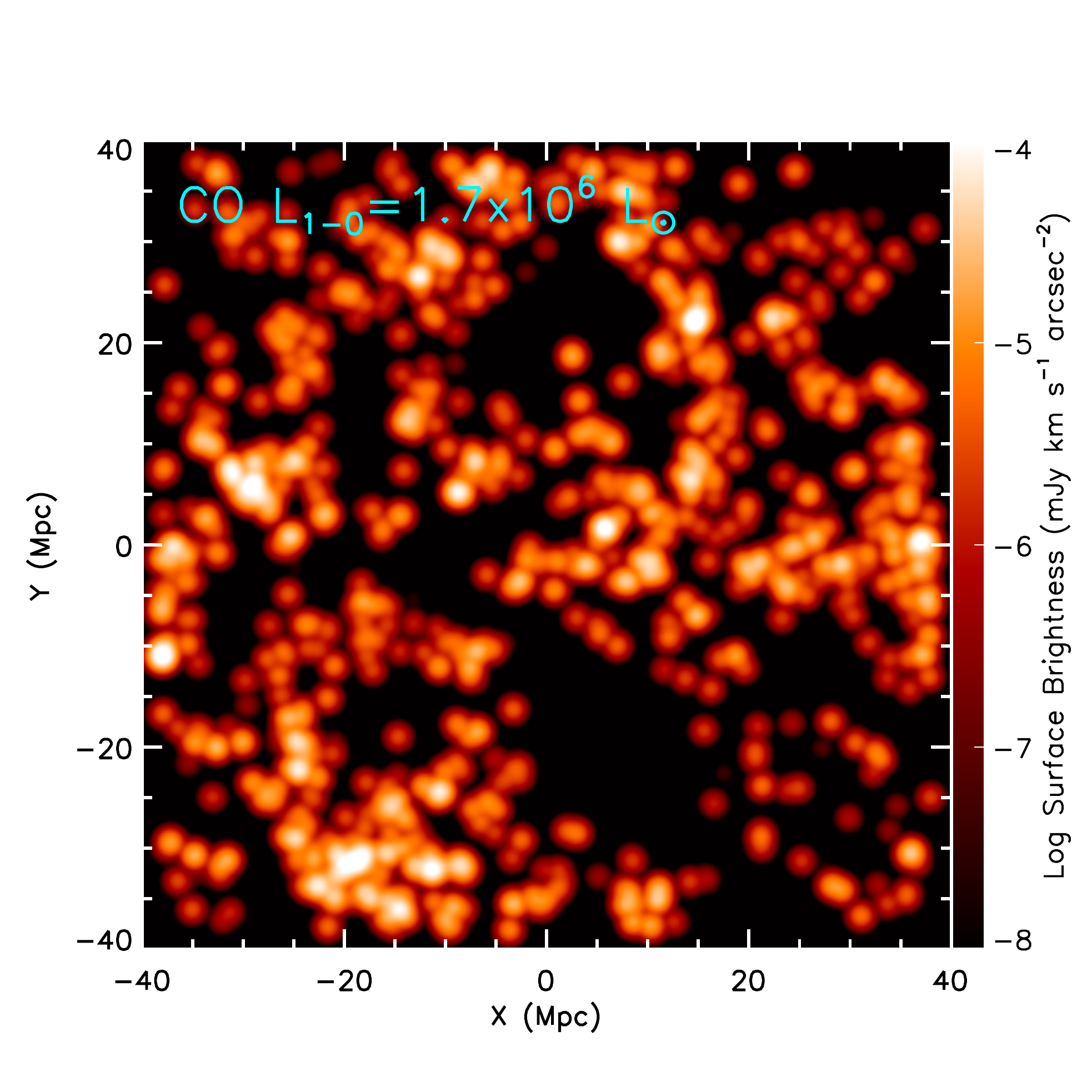}
\includegraphics[width=2.2in]{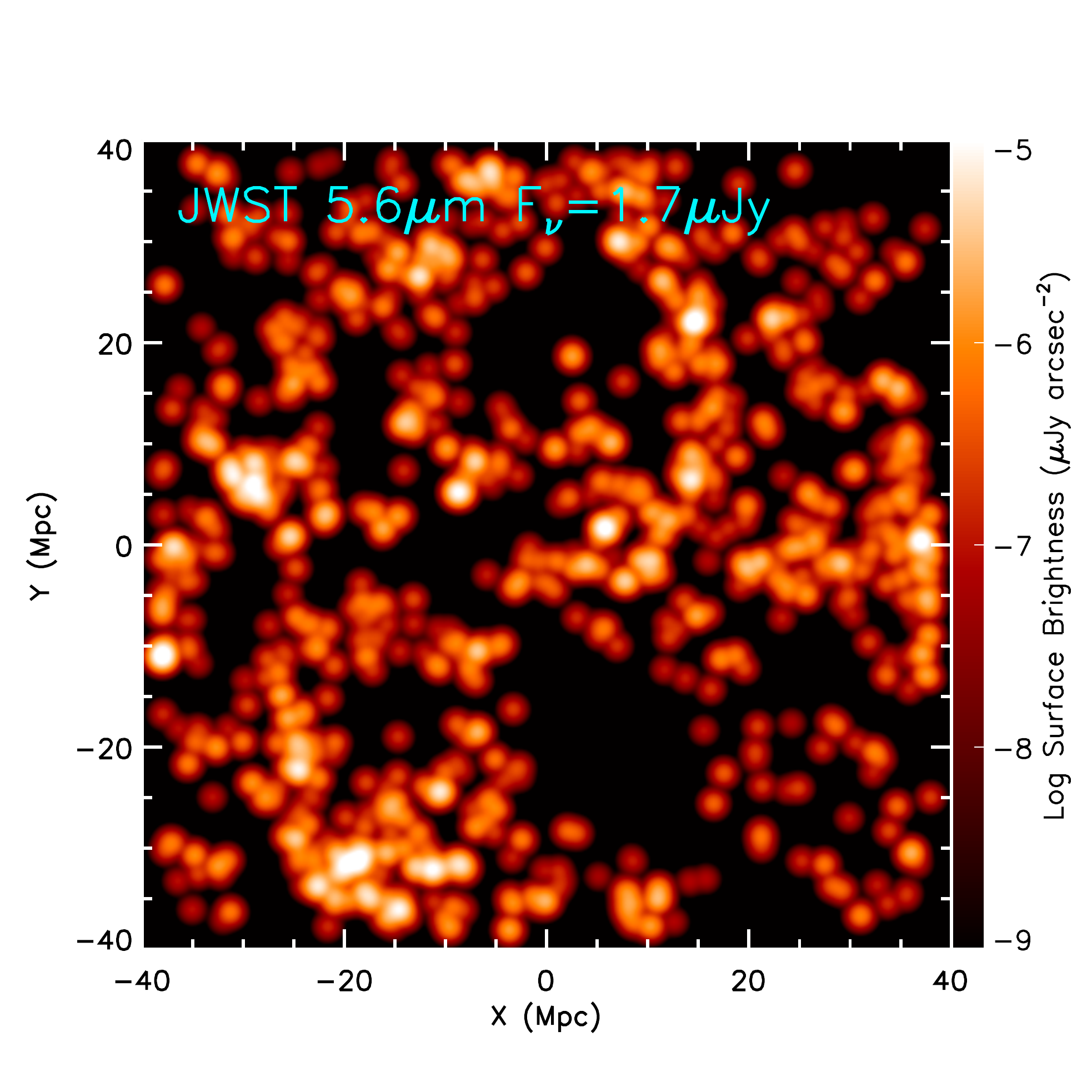}
\caption{Intensity maps of resonant scattering line \lya, atomic fine structure lines \cii, \nii and \oiii, molecular line CO (1-0), and continuum at JWST filter F560W, for  a 5~Mpc slice at $z=8$ from the IllustrisTNG100 Simulation post-processed by \art.}
\label{fig_lim}
\end{figure*}

The \art data is ideal to study not only individual properties as shown in the previous sections, but also statistical  properties of galaxies and ISM at different redshift, and when combined with data from the hydrodynamics simulation, we can study relations between photometric properties and galaxy properties such as galaxy mass, star formation rate and metallicity, as demonstrated in Figure~\ref{fig:lum_sfr}, which shows the correlations between luminosities of FIR, \lya, \cii and \oiii and SFR of the 50 most massive galaxies in the simulation at redshift z=3.1.

Figure~\ref{fig:sfr_uvir} shows the SFR indicator calibration using the UV (0.15~$\mu$mu) and FIR ($8 - 1000~\mu$m) luminosities. The inferred SFR from the FIR and UV, and a linear combination of the two are determined using the calibration given by \citet{Calzetti2012}. It is evident that the UV luminosity is a poor indicator
of the SFR due to the dust extinction, while the FIR indicator works much better. However, the linear combination of the two has an even tighter correlation with the SFR obtained directly from the simulation.

Figure~\ref{fig:lyairco} shows the relation between \lya and FIR (top), and between FIR and CO luminosity $\rm{L^{\prime}_{CO}(1-0)}$ (bottom). The intrinsic \lya luminosity correlates with FIR tightly and agrees with the scaling relation proposed by \citet{Kennicutt1998}. However, the emergent \lya luminosities are much lower due to the dust absorption. On the other hand, the correlation between CO  and FIR luminosity from our simulations smoothly matches the observations of \citet{Solomon2005},
although the galaxies in the present simulations are typically smaller and have lower luminosities than from the observation sample. 

The CO-to-H$_2$ mass conversion factor, $\alpha=\rm{M(H_2) /L^{\prime}_{CO}}$, is an important parameter to understand star formation and the ISM in galaxies. However, measurements of the $\alpha$ factor show large scatterings in different galaxies \citep{Bolatto2013}. We show in Figure~\ref{fig:aco} the relation between the $\alpha$ factor and CO luminosity $\rm{L^{\prime}_{CO}(1-0)}$ of the 50 most massive galaxies at z=3.1 in the Milky Way Simulation post processed by \art. The average  $\alpha$ factor from the simulation $\sim$ 2~M$_{\odot}$/K~km~s$^{-1}$~pc$^2$, which is consistent with the observed values of normal galaxies \citep{Bolatto2013}.

Figure~\ref{fig:fesc} shows the escape fraction of the \lya, UV at 0.15~$\mu$m, and ionizing photons from the galaxies as a function of $E(B-V)$, for the 50 most galaxies in this sample.

Finally, Figure~\ref{fig:lyaew} shows the intrinsic and emergent rest-frame equivalent widths of \lya lines from the simulation. The EW range of $10 - 60\, \AA$ is consistent with the EW range of star-forming galaxies at this redshift \citep{Yajima2012b}.

\subsection{Application to the Large-scale IllustrisTNG Simulation}
\label{sec:tng}

In this example, we apply \art to the IllustrisTNG Simulation to demonstrate the technique of making large-scale line intensity mappings of interested lines such as \lya, \cii, \oiii, and CO.  For each target redshift, we will divide the simulation snapshot into thin slices in the line of sight direction. The thickness of each slice corresponds to the frequency resolution element of the mapping experiment. The RT calculations are then performed for each slice, resulting the full spectrum including continuum and lines. The stacking of slices at different redshifts therefore results in a three-dimensional data cube similar to the observational data from intensity mapping experiments. Choices of different slices at the same redshift represent different realizations of the data cubes, and can be used to study the statistical variance. These simulated data cubes can be an important testbed for different analyses techniques, such as those for background and foreground contamination removal, and the models for the full-spectrum comic background radiation.  

The IllustrisTNG Simulations are the state-of-the-art cosmological simulations currently available~\citep[][]{Pillepich2018b, Springel2018, Marinacci2018, Naiman2018, Nelson2018a, Nelson2019a, Nelson2019b, Pillepich2019}. The simulations were carried out with the moving-mesh code {\sc Arepo} \citep{Springel2010, Springel2019}. The list of physical processes is similar to that of the Milky Way Simulation \citep{Zhu2016} with additional physics of black hole accretion and feedback, and magnetic fields~\citep[][]{Weinberger2017, Pillepich2018a}. 

We use the TNG100 Simulation in this work, which is publicly available~\citep{Nelson2019b}. TNG100 has a box size of 100 Mpc, and the mass resolutions are  $\rm{m_{b}} = 1.4 \times10^6\, \rm{M_{\odot}}$ for baryons  and  $\rm{m_{dm}} = 7.5 \times10^6\, \rm{M_{\odot}}$ for dark matter, and the minimum cell softening length is $\epsilon_{\rm gas}= 0.18\,\rm{kpc}$. It was run with the Planck cosmological parameters \cite{Planck2016}: $\Omega_{\Lambda}=0.6911$, $\Omega_{\rm m}=0.3089$, $\Omega_{\rm b}=0.0486$, $\sigma_8=0.8159$, $n_s=0.9667$ and $h=0.6774$. 

For the demonstration, we randomly choose the snapshot at redshift z=8, and a random slice of 5~Mpc in depth. We apply \art to the slice to calculate the total luminosity of  \lya, \cii, \oiii, and CO lines for the intensity maps, as well as the continuum at $5.6\, \mu\,m$ of JWST filter F560W for comparison. The resulting intensity maps are shown in Figure~\ref{fig_lim}.  These preliminary results show that \lya is the strongest line to study the high-redshift galaxies, while the CO 1-0 line is weaker than the FIR lines \cii, \nii and \oiii. A more comprehensive study of line intensity mappings using the IllustrisTNG Simulations will be presented in a future paper.

\section{Discussions and Conclusions}
\label{sec:summary}

The new \art  implements a number of novel methods to calculate the non-LTE molecular and atomic line transfer and to parallelize the code. These treatments ensure accurate, efficient and self-consistent calculations of the continuum from sub-millimeter to X-ray, and various lines from atoms and molecules, such as \lya, \cii, \oiii, and CO. In addition, \art  adopts a multi-phase ISM model, which ensures an appropriate prescription of the ISM physics in case hydrodynamics simulations have insufficient resolution to resolve the multi-phase ISM, and it employs an adaptive grid scheme, which can handle arbitrary geometry and cover a large dynamical range of gas densities in hydrodynamical simulations. These essential features make \art  a multi-purpose code to study the multi-wavelength properties of a wide range of astrophysical systems, from planetary disks, to star forming regions, to galaxies, and to large-scale structures of the Universe. 

To demonstrate the capability of the new \art, we applied it to two hydrodynamics cosmological simulations, the zoom-in Milky Way Simulation to obtain the multi-band properties of individual galaxies at different redshift, and the full-box IllustrisTNG100 Simulation to obtain line intensity maps of \lya, \cii, \nii, \oiii and CO lines and continuum JWST F560W. The RT outputs include multi-wavelength SEDs, lines and images, useful for an array of studies such as the correlations between physical and photometric properties, the physical conditions of line emission, the origin of emission line deficit observed in galaxies, the escape fraction of ionizing and UV photons, the luminosity functions, and the evolution of the multi-band properties of galaxies and gas with time. 

We will perform in-depth studies of the above topics with the comprehensive IllustrisTNG $+$ \art (ILART) Project, in which we will post process the IllustrisTNG100 Simulation with \art to derive multi-band properties of galaxies and ISM at different redshifts. In addition, we will also apply \art to IllustrisTNG50, which has the highest resolutions of all TNG runs ($\rm{m_{b}} = 8.5 \times10^4\, \rm{M}_{\odot},  \rm{m_{dm}} = 4.6 \times10^5\, \rm{M}_{\odot}, \epsilon_{\rm {gas}}= 0.18\,\rm{kpc}$), and which is expected to be released to the public sometime in the near future (private communication with Dr. Dylan Nelson). The ILART Project is expected to provide new insights into a wide array of studies, as well as effects of numerical resolutions on the RT results, in upcoming papers.

To conclude, \art is well-suited to study both individual objects and global properties of the entire system. It enables direct comparison between numerical simulations and multi-band observations, and it will provide a crucial theoretical framework for the understanding of existing multi-band astronomical observations, the  design and plan for future surveys, and the synergy between multi-band galaxy surveys and line intensity mappings to provide a full picture of the origin and evolution of the Universe. Therefore, \art can provide a powerful and versatile tool to bridge the gap between theories and observations of the cosmic structures.

\section*{ACKNOWLEDGEMENTS}
We thank the referee for a constructive report which has helped improve the manuscript. YL acknowledges support from NSF grants AST-1412719 and MRI-1626251, and thanks the Amaldi Research Center at the Sapienza University of Rome for hosting her sabbatical stay. HY acknowledges support from the MEXT/JSPS KAKENHI Grant Number JP17H04827 and 18H04570. We thank the IllustrisTNG Collaboration for making the TNG100 Simulation publicly available for the calculations presented in this work. The numerical computations and data analysis in this paper have been carried out on the CyberLAMP supercomputer cluster  at the Pennsylvania State University, which is funded by the MRI-1626251 award and managed by the Penn State Institute for CyberScience, as well as  the Odyssey cluster  supported by the FAS Division of Science, Research Computing Group at Harvard University. The Institute for Gravitation and the Cosmos is supported by the Eberly College of Science and the Office of the Senior Vice President for Research at the Pennsylvania State University. 


\label{lastpage}

\end{document}